\documentclass[twoside,12pt]{article}
\usepackage{amsmath,amssymb}
\usepackage{graphics,psfrag}
\usepackage{graphicx,psfrag}

\def\mpi{M_\pi}

\newcommand{\be}{\begin{equation}}
\newcommand{\ee}{\end{equation}}
\newcommand{\bea}{\begin{eqnarray}}
\newcommand{\eea}{\end{eqnarray}}

\newcommand{\Choose}[2]{{^{#1}C_{#2}}}
\newcommand{\lsim}{\raisebox{-0.7ex}{$\stackrel{\textstyle <}{\sim}$ }}
\newcommand{\gsim}{\raisebox{-0.7ex}{$\stackrel{\textstyle >}{\sim}$ }}

\def\siii{^3 \hskip -0.02in S _1}
\def\diii{^3 \hskip -0.04in D _1}
\def\abar{\overline{a}}

\begin{document}

\title{ \vspace{1cm} Nuclear Physics from Lattice QCD}
\author{S.R.~Beane$^{1}$,  W.~Detmold$^{2,3}$, K.~Orginos$^{2,3}$ and M.J.~Savage$^4$ \\
\\
$^1$ {\small Department of Physics, University of New Hampshire,
Durham, NH 03824-3568, USA.}\\
$^2$ {\small Department of Physics, College of William and Mary, Williamsburg, VA
23187-8795, USA.}\\
$^3$ {\small Jefferson Laboratory, 12000 Jefferson Avenue, Newport
  News, VA 23606, USA}\\
$^4$ {\small Department of Physics, University of Washington, 
Seattle, WA 98195-1560, USA.}}

\maketitle
\begin{abstract} 
We review recent progress 
toward establishing lattice Quantum Chromodynamics 
as a predictive calculational framework for nuclear physics.
A survey of the current 
techniques that are used to extract low-energy hadronic scattering amplitudes 
and interactions is followed by a review of recent two-body 
and few-body calculations by the NPLQCD collaboration and others. 
An outline of the nuclear physics that is expected to be 
accomplished with Lattice QCD
in the next decade, along with estimates of 
the required computational resources, 
is presented.
\end{abstract}
%\eject
%\tableofcontents

\section{Introduction}

\noindent It has been known for the past four decades that Quantum
Chromodynamics (QCD), together with the electroweak interactions,
underlies all of nuclear physics, from the s-wave nucleon-nucleon
scattering lengths to the mechanism of synthesis of heavy elements in
stars. 
Soon after the discovery of QCD it became apparent that
the complexities of the theory at strong coupling would hinder
analytic progress in understanding the properties of the simplest
hadrons, let alone the simplest features of the nuclear forces.
Wilson pointed the way to eventual
direct quantitative confirmation of the origins of nuclear physics by
formulating Lattice QCD~\cite{Wilson:1974sk}, a regularization 
and non-perturbative definition
of QCD
that is suitable for the intensive computational demands of solving
QCD in the infrared.

Only in the last five years or so has Lattice QCD emerged from a long
period of research and development ---where only qualitative agreement
between calculations and experiments could be claimed--- to the
present, where precise predictions for hadronic quantities are being
made. In particular, presently, fully-dynamical calculations with
near-exact chiral symmetry at finite lattice-spacing have become standard,
with lattice volumes of spatial extent $L \gsim 2.5~{\rm fm}$ and
with lattice spacings in the range $b\lsim 0.12~{\rm fm}$. Very
recently, preliminary calculations at the physical quark masses have
been carried out.  However, it is still the norm that the light-quark
masses, $m_q$, are larger than those of nature, with typical pion
masses $m_\pi\sim 300~{\rm MeV}$. It is expected that in the next five
years, calculations at the physical light-quark masses, $m_\pi\sim
140~{\rm MeV}$, in large volumes, $L\gsim 6~{\rm fm}$, and at small
lattice spacings, $b\lsim 0.06~{\rm fm}$ will become commonplace.

Nuclear physics is a vast, rich field, whose phenomenology has been
explored for decades through intense experimental and theoretical
effort. However, there is still little understanding of the connection
to QCD and the basic building blocks of nature, quarks and gluons.
For instance, as a first ``benchmarking'' step, Lattice QCD should
post-dict the large nucleon-nucleon scattering lengths and 
the existence of the deuteron, the simplest nucleus.  The
connection between QCD and nuclear physics will be firmly established
with Lattice QCD, and will allow for an exploration of how nuclei and
nuclear interactions depend upon the fundamental parameters of
nature. In particular, it is believed that an understanding of the
fine-tunings that permeate nuclear physics will finally be translated
into fine-tunings of the light-quark masses. While these
issues are of great interest, and it is important to recover what is
known experimentally to high precision, these goals are not the main
objective of the Lattice QCD effort in nuclear physics.  The primary
reason for investing resources in this area
is to be able to calculate physical quantities of importance that
cannot be accessed experimentally, or which can be measured with only
limited precision in the laboratory. Two important examples of how
Lattice QCD calculations can impact nuclear physics are in the
structure of nuclei and in the behavior of hadronic matter at 
densities beyond that of nuclear matter.

Refined many-body techniques for studying the
structure of nuclei, such as Greens function Monte-Carlo
(GFMC)~\cite{Pieper:2007ax}, already exist.
This method has led to calculations of
the ground states and excited states of light nuclei, with atomic
number $A\lsim 14$. Using only the modern nucleon-nucleon (NN)
potentials that reproduce all scattering data below inelastic
thresholds with $\chi^2/dof\sim 1$, such as ${\rm
AV}_{18}$~\cite{Wiringa:1994wb}, these models fail
to recover the observed structure of light nuclei. However, the inclusion of a
three-nucleon interaction greatly improves the predicted structure of
nuclei~\cite{Pieper:2007ax}. Lattice QCD will be able to calculate the
interactions of multiple nucleons, bound or unbound, in the same way that it
can be used to determine the two-body scattering parameters.  For
instance, a calculation of the three-neutron interactions will be
possible.

One of the great challenges facing nuclear physics is to determine the
behavior of hadronic matter at densities away from that of nuclear
matter, such as those that occur in the interior of neutron
stars~\cite{Page:2006ud}.  There are a number of possibilities for the
composition of the hadronic matter at the center of a neutron star.
One possibility is that it is composed of neutrons and protons.
Another possibility is that it is composed of neutrons, protons and a
kaon condensate, and a third possibility is that it is composed of
neutrons, protons and $\Sigma^-$'s.  At present, all of these
compositions (and other more exotic scenarios) are possible
theoretically, and not excluded experimentally or observationally.  On
the theoretical side, the main uncertainty in establishing the
composition of hadronic phases, 
and hence the equation of state, is the
interactions between the strange hadrons, such as the kaon and
$\Sigma^-$, and the protons and neutrons.  These interactions are
poorly known experimentally due to the short lifetime of the strange
hadrons, however Lattice QCD promises to reliably calculate these
interactions from QCD with quantifiable uncertainties. Indeed lattice
QCD calculations will provide the best determinations of
hyperon-nucleon and kaon-nucleon scattering amplitudes and hence will play
a crucial role in determining the role of hyperons and strange mesons
in neutron stars.

Lattice QCD is a technique in which Euclidean space correlation
functions are calculated by a Monte-Carlo evaluation of the Euclidean
space path integral~\cite{Wilson:1974sk}.  The calculations are
performed in Euclidean space so that 
field configurations
that have a large action are exponentially
suppressed.  This is in contrast with Minkowski space in which large
action contributions result in a complex phase which will average to
an exponentially small contribution with nearby configurations.  
In this approach, space-time is
discretized with the quarks residing on the lattice sites, and the
gluon fields residing on the links between lattice sites. 
The lattice spacing, $b$, the distance between adjacent lattice sites,
is required to be much smaller than the characteristic hadronic length
scale of the system under study. The effects of a finite lattice
spacing can be systematically removed by combining calculations of 
correlation functions at several lattice spacings with the low-energy
effective field theory (EFT) which explicitly includes  the 
discretization effects. 
This type of EFT is somewhat more complicated than its
continuum counterpart as it must reproduce matrix elements of the
Symanzik action constructed with higher dimension operators induced by
the discretization~\cite{Symanzik:1983gh}. While the action lacks
Lorentz invariance and rotational symmetry, it is constrained by
hypercubic symmetry.  As computers have finite memory and performance,
the lattice volumes are finite in all four space-time directions.
Generally, periodic boundary conditions (BC's) are imposed on the
fields in the space-directions (a three-dimensional torus), while
(anti-)periodic BC's are imposed on the (quark) gauge fields 
in the time-direction,
which in many cases is much larger than the space-directions 
(in order to approach the zero-temperature limit).

For the calculations we will be discussing in this review, the lattice
volumes are large compared with the Compton wavelength of the pion,
and deviations of single-particle properties from their infinite
volume values are exponentially small, generically $\sim e^{-m_\pi
L}$. These finite-volume effects may be removed by finite-volume EFT.
Perhaps the most important ``economic'' feature of Lattice QCD
calculations is that the 
computational resources required to perform
lattice calculations
increase with decreasing quark mass, and presently, 
no calculations exist at the physical quark masses, $m_\pi\sim
140~{\rm MeV}$, that have $m_\pi L \gg 1$.
It remains the case that lattice calculations are
performed at unphysical values of the quark masses, and the light
quark mass dependence of the observable of interest, which can be
determined perturbatively in the low-energy EFTs, is used to
extrapolate to the physical light quark masses.  Therefore, the
practical situation with current Lattice QCD calculations is that they
are performed at finite lattice spacing, within finite volumes and at
unphysical quark masses.  The appropriate EFT (e.g. chiral
perturbation theory ($\chi$-PT), heavy-baryon-$\chi$-PT (HB$\chi$-PT) ) 
is then used
to extrapolate to the infinite volume, continuum limit of QCD where
physical predictions can be made.

%%%%%%%%%%%%%%%%%%%%%%%%%%%
\section{Lattice QCD Technology}
\begin{figure}[tb]
\begin{center}
\begin{minipage}[t]{8 cm}
\centerline{\includegraphics[scale=0.45]{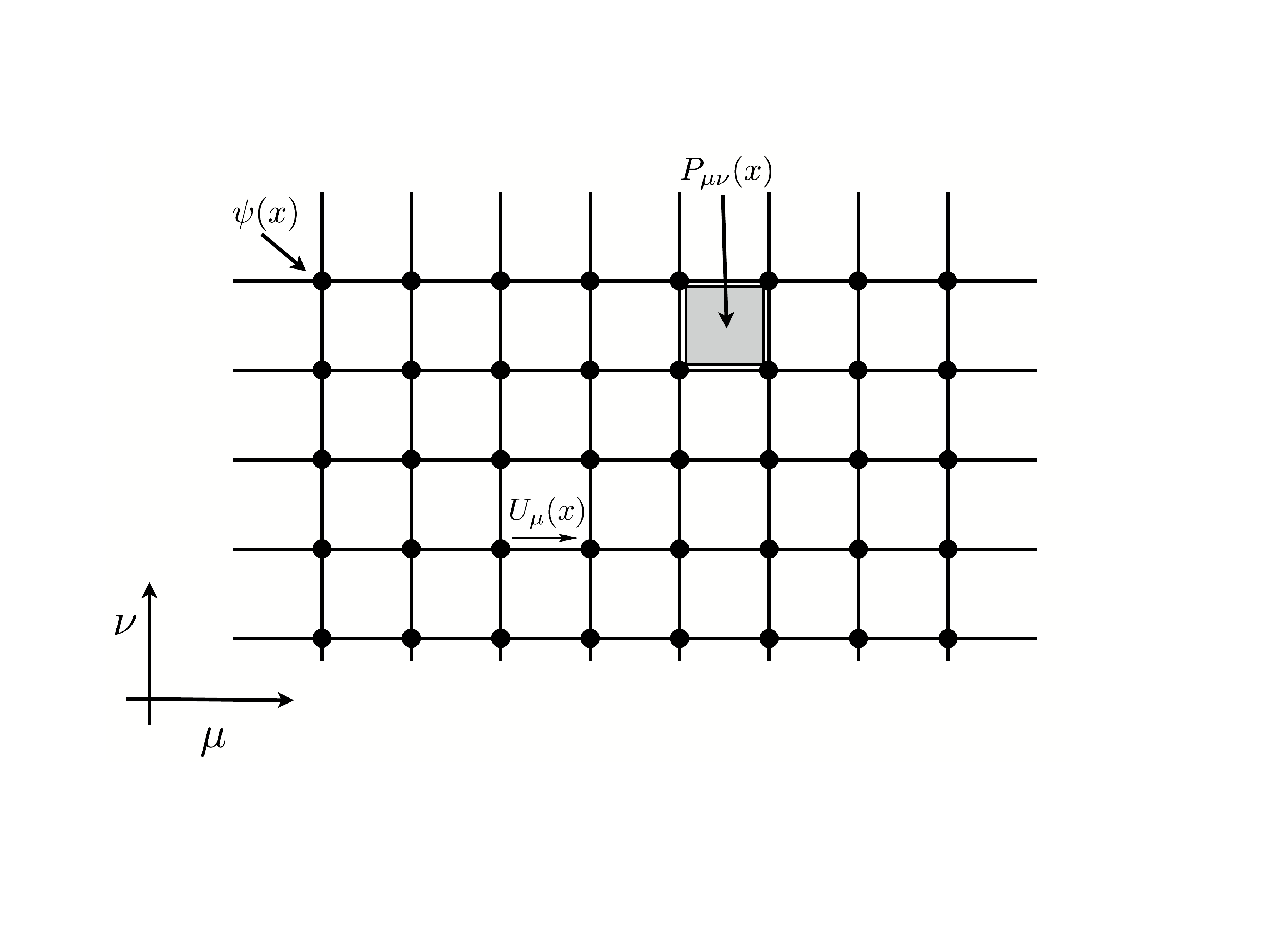}}
\end{minipage}
\begin{minipage}[t]{16.5 cm}
\caption{
A two dimensional slice of the four dimensional  space-time lattice.
$\mu$ and $\nu$ denote unit-vectors in the indicated directions.
$\psi(x)$ denotes a fermion-field at the lattice-site $x$, 
$U_\mu(x)$ denotes the gauge link from the lattice-site $x$ to the site
$x+b\mu$, and $P_{\mu\nu}(x)$ denotes the $1\times 1$ Wilson plaquette centered
at $x+b{\bf \mu}/2+b{\bf \nu}/2$.
\label{fig:lattice}}
\end{minipage}
\end{center}
\end{figure}

\noindent
Quantum Chromodynamics (QCD) can be defined non-perturbatively as the
continuum limit of a Lattice gauge theory. This approach provides both
an ultraviolet regulator of the continuum field theory and admits
numerical evaluation of the functional integrals required for
calculating physical observables. In the continuum, the QCD path
integral is~\footnote{A very nice introduction to Lattice QCD can be found in
  Ref.~\cite{Gupta:1997nd}.}
 \begin{eqnarray}
 {\cal Z} & = & \int {\cal D} A_\mu {\cal D} \bar \psi {\cal D}  \psi \; 
e^{-\int d^4 x \ {\cal L}_{\rm QCD} }
\nonumber\\
{\cal L}_{\rm QCD} & = &  
\sum_f \bar{\psi_f} \left[ D_\mu \gamma_\mu + m_f\right] \psi_f\ 
 + \ \frac{1}{4} \sum_a\ G^a_{\mu\nu} G^{a\mu\nu}
\ \ \ ,
\label{eq:Zqcd}
\end{eqnarray}
where ${\cal L}_{\rm QCD}$ is the QCD Lagrange density, 
$A_\mu$ is the QCD gauge field (describing the gluons),
$G^a_{\mu\nu}$ is the gauge field strength,
$m_f$ is the quark mass, and $\bar \psi_f$, $\psi_f$ are
the fermion fields representing the quark flavors.  $D_\mu$ is the covariant
derivative which ensures gauge invariance and $\gamma_\mu$ are
the Dirac-matrices.  
Explicitly, the covariant derivative acting on a quark field is given by 
\begin{eqnarray}
D_\mu\psi(x) & = & \partial_\mu\psi(x) + i g_s A_\mu(x)\psi(x)
\ \ ,\ \ A_\mu (x) \ = \ T^a\ A_\mu^a(x)
\ \ \ ,
\label{eq:covariant}
\end{eqnarray}
where $T^a=\lambda^a/2$ with the $\lambda^a$ being the Gell-Mann matrices.
The strong coupling $g_s$ appearing in eq.~(\ref{eq:covariant})  is related to
the strong fine-structure constant,
$\alpha_s$, via $\alpha_s = g_s^2/(4\pi)$.
The field strength tensor is defined in terms of the gluon field through
\begin{eqnarray}
G_{\mu\nu}(x) & = & \partial_\mu A_\nu (x) \ -\ \partial_\nu A_\mu (x)
\ +\ i g_s\ \left[\  A_\mu (x)\ ,\ A_\nu (x)\ \right]
\ \ ,\ \ 
G_{\mu\nu} (x)\ =\ G_{\mu\nu}^a (x)\ T^a
\ \ \ ,
\label{eq:Ddef}
\end{eqnarray}
Observables
(physical quantities) in this theory can be calculated from correlation
functions of operators $\cal O$ that are functions of the quantum
fields (quarks and gluons).
\begin{equation}
\langle {\cal O}\rangle = \frac{1}{\cal Z} \int {\cal D} A_\mu {\cal D} \bar
\psi {\cal D}  \psi \; {\cal O}\; e^{- \int d^4 x \ {\cal L}_{\rm QCD} } \ .
 \end{equation}
The path integral can be rigorously defined on a 
discrete space-time. In order to preserve gauge invariance the gauge
fields are discretized as special unitary matrices, in the group
$SU(3)$, on the links of the lattice (see
Figure~\ref{fig:lattice}).
The discrete gauge action is
the sum over all plaquettes $P_{\mu\nu}(x)$ which are the
product of the links $U_\mu(x) = \exp\left( i \int_x^{x+\hat\mu}\ dx^\prime
  A_\mu (x^\prime)\right)$ 
around the elementary plaquettes of the
lattice,
\begin{equation}
S_g(U) ={\beta} \sum_{x\mu\nu}\left( 1 - \frac{1}{3}
{\rm Re }\ {\rm  Tr}\ {P_{\mu\nu}(x)} \right)  
\ \ ,
 \label{eq:lattice_gauge_action}
\end{equation}
with
 \begin{equation}
 P_{\mu\nu} = U_\mu(x)U_\nu(x+\hat\mu)U^\dagger_\mu(x+\hat\nu)U^\dagger_\nu(x)
\, .
 \label{eq:plaquett}
 \end{equation}
$\beta$ is the lattice gauge coupling that is related to the strong coupling  
via $\beta = 2 N_C /g_s^2$ where $N_C$ is the number
of colors.
Taking the naive continuum limit, this action becomes the 
familiar  continuum gauge action,
 $\ -\int d^4 x \frac{1}{4}\left( G^a_{\mu\nu}(x)\right)^2 $. 
The action in Eq.~\ref{eq:lattice_gauge_action} is the
 well known Wilson gauge~\cite{Wilson:1974sk} action, and while 
this
 discretization is not unique,  it is the simplest. 
It can be modified by adding larger loops with coefficients
 appropriately chosen to achieve better convergence to the
 continuum, which is the ultimate goal of the calculation.
 
Inclusion of the quarks (fermions), which are 
defined on the vertices of the lattice, 
is a challenging problem. A naive discretization of a single 
continuum fermion field introduces   sixteen light 
lattice fermion flavors in the four dimensions.  In  QCD only three
light quarks (up, down and strange) and three heavy quarks (top,
bottom and charm) are needed. The doublers, i.e. the additional 
light lattice fermion degrees of freedom,  can be avoided by several ingenious
formulations of lattice fermions. 
Wilson fermions, which were introduced first~\cite{Wilson:1974sk}, 
eliminate the doublers by adding irrelevant dimension five operators 
in the action that lift the masses of the doublers, leaving only one 
light fermion in the spectrum.  However, this explicitly breaks chiral 
symmetry and introduces lattice artifacts that scale as $O(b)$. 
The Wilson action can be improved by the addition of one dimension-5 
bilinear operator, 
${\cal O}_{SW} = \overline{\psi}\sigma^{\mu\nu} G_{\mu\nu}\psi$, 
the Sheikholeslami-Wohlert term~\cite{Sheikholeslami:1985ij}, 
with a coefficient, $C_{SW}$,
that can be tuned so that the lattice artifacts are parametrically 
reduced to scale as $O(b^2)$.
In addition, the lattice spacing artifacts can be reduced to $O(b^2)$
by the addition of a twisted-mass term to the action~\cite{Frezzotti:2000nk}.
The isospin violation that the twisted-mass term introduces 
into the action can be removed using chiral perturbation theory.
Kogut-Susskind fermions~\cite{Kogut:1974ag} (staggered fermions) 
provide another way to remove some of the
doublers and re-interpret the remaining four as four degenerate
flavors.  In this approach a $U(1)$ chiral symmetry still remains
unbroken and lattice artifacts scale as $O(b^2)$.  Kogut-Susskind
fermions become problematic when the required number of flavors is not
a multiple of four (as is the case for QCD).  In addition, the broken
flavor and chiral symmetries introduce large lattice artifacts,
although they scale as $O(b^2)$. 
Finally, the so called domain wall
fermions~\cite{Kaplan:1992bt,Shamir:1993zy,Furman:1994ky} and overlap
fermions~\cite{Narayanan:1994gw,Neuberger:1997fp} are fermionic
actions that both preserve a lattice chiral symmetry at finite lattice spacing
(they satisfy the Ginsparg-Wilson relation~\cite{Ginsparg:1981bj}
$\gamma_5 D + D\gamma_5 = b\ D\gamma_5 D$)
and are doubler free. Unfortunately, such formulations are
computationally significantly more expensive.
It should also be stated that other actions are being explored that 
approximately satisfy the Ginsparg-Wilson relation, for instance the
fixed-point action~\cite{Hasenfratz:2006xi}, or the chirally-improved action,
e.g. Ref.~\cite{Gattringer:2000js,Gattringer:2003qx}.
In all cases the lattice fermion action is of the form
\begin{equation}
S_f = \bar \psi D(U) \psi
\end{equation}
where $\psi$ is the fermion ``vector'' and $D(U)$ is a sparse
matrix~\footnote{In certain cases, such as with overlap fermions, the
matrix is not sparse but has sparse like properties,  i.e. the matrix
vector multiplication is a computationally 
``cheap'' operation.}  acting on the fermion vector,
that depends on the gauge field $U$.
As an example of the form of $S_f$ for a lattice action, its explicit form for
the naive Wilson action is
\begin{eqnarray}
S_f^{\rm Wilson} & = & 
{1\over 2b}\sum_x\ 
\overline{\psi}(x) \gamma_\mu
\left[\ U_\mu(x)\psi(x+b\mu)
\ -\ U^\dagger_\mu(x-b\mu)\psi(x-b\mu)\ \right]
\ +\ 
m_f\ \sum_x\ \overline{\psi}(x) \psi(x) 
\ .
\label{eq:WilsonS}
\end{eqnarray}

The partition function in the case of two quark flavors is
\begin{eqnarray}
{\cal Z} &=& \int \prod_{\mu,x}dU_\mu(x)\prod_x 
d\bar{\psi}d\psi
\;\; e^{-S_g(U)-S_f(\bar{\psi},\psi,U)} \nonumber\\
&=&  \int \prod_{\mu,x}dU_\mu(x)\;\;
 {\rm det}\left(D(U)^\dagger D(U)\right)  \;\; e^{-S_g(U)}\,.
 \end{eqnarray}
The integration over the quark fields, which are represented by
Grassmann numbers, can be done exactly.  In addition, the quark matrix
$D(U)$ represents one flavor, however since ${\rm det} D(U)^\dagger =
{\rm det} D(U)$, the determinant ${\rm det}\left(D(U)^\dagger
D(U)\right) $ represents two flavors. In the case of correlation functions, 
integrating out
the quarks gives the following expression
\begin{equation}
\langle{\cal O}\rangle = \frac{1}{\cal Z}
 \int \prod_{\mu,x}dU_\mu(x)\;\; {\cal O}(\frac{1}{D(U)},U)\;\;
 {\rm det}\left(D(U)^\dagger D(U)\right)  \;\; e^{-S_g(U)}\, ,
 \label{eq:CorFunc}
 \end{equation}
where the quantities  ${\cal O}$  depend on the inverse of the
quark matrix and possibly explicitly on the gauge fields.  
The above expressions are only valid in the case of two
flavors of quarks (the up and the down) which both have the same mass,
which is a good approximation to the low energy physics of QCD. A
strange quark can be easily added by including ${\rm
det}\left(D(U)^\dagger D(U)\right)^{1/2}$ in the partition function.

The computation of Eq.~(\ref{eq:CorFunc}) is the main numerical task
faced in Lattice QCD calculations. The integral  
over the gauge fields in
Eq.~(\ref{eq:CorFunc}) is of extremely large
dimensionality. 
Given that QCD has a fundamental length 
scale of $\sim1~{\rm fm}$ ($10^{-13}~{\rm cm}$), 
calculations must be performed in lattice volumes that have a 
physical size much larger than $1~{\rm fm}$
in order to control finite volume effects, and with 
lattice spacings much smaller than $1~{\rm fm}$ in order to be close to the
continuum limit.  With moderate choices for the volume and the lattice
spacing, a lattice volume of $\gsim 32^3\times 256$ is currently practical.  
With the
color and spin degrees of freedom, such calculations involve
$\approx 10^9$ degrees of freedom. The only practical way that 
this type of computation can
be done is by using Monte-Carlo integration. Fortunately, the
combination of the quark determinant and the gauge action,
\begin{equation}
 {\cal P}(U) =\frac{1}{\cal Z}\  {\rm det}\left(D(U)^\dagger D(U)\right)  
\;\; e^{-S_g(U)}\,,
 \end{equation}
is a positive definite quantity which can be interpreted as a
probability and hence importance sampling can be employed.  The basic
algorithm is to produce $N$ gauge field configurations $\{U\}$ with
probability distribution ${\cal P}(U)$ and then evaluate
\begin{equation}
\langle {\cal O} \rangle = \lim_{N\rightarrow \infty} 
\frac{1}{N}\sum_{i=1}^N {\cal O}(\frac{1}{D(U_i)},U_i) \ .
\label{eq:average}
\end{equation}
At finite $N$, the estimate of ${\cal O}$ is approximate, with an uncertainty
that converges to zero as ${\cal O}(1/\sqrt{N})$.  Both for the gauge
field configuration generation and the evaluation of
Eq.~(\ref{eq:average}), the linear system of equations
\begin{equation}
 {D^\dagger(U)[m]D(U)[m]}\chi =  \phi\,,
 \label{eq:linearsyst}
\end{equation}
needs to be solved where the dependence of the quark matrix on the
quark mass $m$ is made explicit.  Since the quark matrix is
sparse, iterative solvers such as conjugate gradient (CG) can be used.  The
condition number of the quark matrix is inversely proportional to
the quark mass.  Since the physical quark masses for the up and down
quarks are quite small, the quark matrix has a large condition
number. With current computer resources this linear system cannot be
solved exactly at the physical quark mass. For that reason the
calculation is performed at heavier quark masses and then extrapolated
to the physical point.  The vast majority of the computer time used in
these calculations is devoted to the solution of this linear system
both in the context of gauge field generation and in the later stage
of the calculation of physical observables through
Eq.~(\ref{eq:average}).

Realistic lattice calculations require quark masses that result in
pion masses below 400 MeV, allowing chiral EFT's  to
be used with some reliability. In addition, a dynamical strange quark
is required in order to guarantee that the low energy constants of the
EFT 
determined with lattice QCD
match those of the physical theory.  Although this task seems
formidable, in the last several years there have been developments that
make phenomenologically interesting calculations now possible.

The works that we will be reviewing involves the computation of correlation
functions with fixed particle number, for instance, correlation functions
with the quantum numbers of the deuteron, or the correlation functions with the
quantum numbers of twelve charged kaons.  As the number of particles increases,
the complexity of forming the correlation function grows dramatically, due to
the need for the quarks to satisfy the Pauli principle.  An alternate way to
describe systems with large numbers of baryons, for instance, is to introduce a
baryon-number chemical potential into the partition function in
Eq.~(\ref{eq:Zqcd}).  However, the evaluation of the partition function with
such a chemical potential is complicated by the fact that the measure of the
integration is no longer positive definite, that is to say, it suffers from the
well-known sign-problem.  Incremental progress has been made in dealing with
the sign-problem, and a conceptual break through will be required in order to
calculate the properties of systems at high densities and low-temperatures.
Naively it appears that there should be a direct connection between the
sign-problem and the behavior of multi-baryon correlation functions.  However,
to date, little  connection between the two approaches to the numerical evaluation
of the properties of finite-density systems has been established.
A recent review of the status of calculations at finite density using
chemical potentials can be found in Ref.~\cite{deForcrand:2010ys}.

%%%%%%%%%%%%%%%%%%%%%%%%%
\subsection{\it Lattice Fermion Actions}
\noindent
The emergence of fermions that respect chiral
symmetry~\cite{Kaplan:1992bt,Shamir:1993zy,Furman:1994ky,Narayanan:1994gw,Neuberger:1997fp}
on the lattice was one of the major recent developments in Lattice
QCD. These formulations of lattice fermions allow for the reduction of
the lattice spacing errors and approach the continuum limit in a
smooth manner. However, the computational resources required to calculate
with these fermions
are an order of magnitude larger than any other variant of lattice
fermions.  In addition, the development of improved Kogut-Susskind
fermion actions~\cite{Orginos:1999cr,Orginos:1998ue} that
significantly reduce the $O(b^2)$ errors, allowed for ``cheap'' inclusion
of quark loop effects in the QCD correlation functions computed on the
lattice.  With this formulation, volumes with spatial extent as large
as $L\sim 5.8~{\rm fm}$ are being used with light-quark masses as low as
1/20th of the strange quark mass depending on available computing
resources. However the fact that Kogut-Susskind fermions represent
four flavors of quarks complicates calculations when two or one
flavors are needed. From the operational point of view the problem is
solved by introducing  the Kogut-Susskind
determinant raised to the $n_f/4$ power (rooted)
into the path integral, where $n_f$ is the
desired number of flavors. The non-integer power of the quark
determinant introduces non-localities in the lattice action. It has
been argued that the long distance physics that survives the
continuum limit is not affected by such
non-localities~\cite{Shamir:2006nj,Bernard:2006ee,Bernard:2007ma,Bernard:2007eh,Durr:2006ze,Durr:2004ta}. In
addition, at finite lattice spacing, the pathologies arising in the
Kogut-Susskind fermion formulation can be dealt with in staggered
$\chi$-PT~\cite{Bernard:2006ee,Bernard:2007ma,Bernard:2006vv}.
Although no rigorous proof exists, empirical evidence indicates that
Kogut-Susskind fermions do describe the correct physics as long as the
continuum limit is taken before the chiral 
limit~\cite{Durr:2004ta}~\footnote{It should be noted that there are 
some members of the lattice
community who believe that the rooted-staggered action is
fundamentally flawed and its continuum limit does not correspond to
QCD (for a summary of these arguments, see Ref.~\cite{Creutz:2007rk}).
We disagree with these arguments, however we acknowledge that there is
no proof that the continuum limit of the rooted-Kogut-Susskind action
corresponds to QCD.}. The results presented in this review that are
based upon mixed-action calculations on the MILC lattice ensembles
assumes that the continuum limit of such calculations is QCD.

Finally, significant simulation algorithm developments have been
achieved that have sped up the gauge field generation by orders of
magnitude.  Developments in the integrators needed for the molecular
dynamics~\cite{Takaishi:2005tz} that form the core of the Hybrid Monte-Carlo 
algorithm~\cite{Duane:1987de,Gottlieb:1987mq} are one important
component of the improvements. A second important
component of the new algorithms is preconditioned Hybrid 
Monte-Carlo~\cite{Luscher:2005rx,Kamleh:2005wg,Hasenbusch:2001ne} 
and rational hybrid Monte-Carlo (RHMC)~\cite{Clark:2006fx} 
for calculations involving an odd number of flavors.
Together
with multiple time scales~\cite{Weingarten:1991ra,Peardon:2002wb} for
evolving different parts of the molecular dynamics
Hamiltonian, calculations close to or at the physical quark masses became
possible for a variety of lattice fermion 
actions including Wilson,  Kogut-Susskind and
domain-wall.
For a status review of the
current dynamical simulation algorithms the reader is referred
to~\cite{Jung:2010jt}. The result of these algorithmic developments
was that calculations with Wilson or improved Wilson fermions again became
feasible. In addition dynamical simulations with domain wall
fermions are also possible at relatively small additional ``cost''
compared to ``cheap'' fermion variants at the same physical parameters. 
The choice of the computational
approach is made based on the resources each collaboration has available as
well as the physics objectives.

%%%%%%%%%%%%%%%%%%%%%%%%%
\subsubsection{\it Mixed Action Calculations}
\noindent
The mixed-action calculations discussed in this review employed 
Kogut-Susskind fermions to represent
the QCD vacuum polarization effects associated with the two light
flavors (up/down quarks) and the somewhat heavier strange quark. This
is done by using gauge configurations generated with the appropriate
Kogut-Susskind fermion determinants incorporated into the probability
distribution that enters the path integral. 
Since this part of the
computation is separated from the calculation of
correlation functions, gauge fields generated by other
collaborations can be used. 
NPLQCD made use of a number of ensembles of gauge
configurations generated by the MILC
collaboration~\cite{Bernard:2001av}. 
Domain wall fermions were used to describe all external (valence) quarks.
Because of the chiral symmetry that domain wall
fermions satisfy, all correlation functions satisfy chiral Ward
identities, ensuring that the leading order (LO) chiral behavior is
continuum-like.  The small corrections appearing due to Kogut-Susskind
fermions in the vacuum loops can be taken care of systematically in
$\chi$-PT~\cite{Chen:2007ug,Chen:2005ab,Chen:2006wf}.  Compared to
calculations with Kogut-Susskind fermions in the valence sector, this
formulation results in better control of the chiral behavior and
possibly smaller discretization errors and also simplifies 
calculations with baryons. This approach was first
introduced by the LHP collaboration for the study of nucleon
structure~\cite{Edwards:2006zza,Renner:2007pb,Hagler:2007xi,Edwards:2006qx,Edwards:2005ym}.
Because the valence and sea quark actions are different, such
calculations are inherently partially quenched and are not unitary. 
This type of calculation 
is sometimes referred
to as the mixed action scheme.  Unlike conventional partially quenched
calculations, which become unitary when the valence quark mass is
tuned to the sea quark mass, unitarity cannot be restored by tuning
the valence quark mass.  The next best option is to tune the valence
quark mass in such a way that the resulting pions have the same mass
as those made of the sea Kogut-Susskind fermions. 
In this case
unitarity should be restored in the continuum limit, where the $n_f=2+1$
staggered action has an $SU(12)_L\otimes SU(12)_R\otimes U(1)_V$ chiral
symmetry due to the four-fold taste degeneracy of each flavor, and
each $\pi$, K or $\eta$
has 15 degenerate additional partners.  At finite lattice
spacing this symmetry is broken and the taste multiplets are no longer
degenerate, but have splittings that are ${\cal O}(\alpha^2
b^2)$~\cite{Orginos:1999cr,Orginos:1998ue,Toussaint:1998sa,Orginos:1999kg,Lee:1999zxa}.
The domain wall fermion mass is tuned to give valence pions (kaons) 
that match
the Goldstone Kogut-Susskind pion (kaon)~\footnote{This is the only Goldstone
boson that becomes massless in the chiral limit at finite lattice
spacing.}.  This choice gives pions that are as light as possible,
resulting in better convergence of the $\chi$-PT extrapolation of
the lattice results to the physical quark mass point.  This tuning was
used by the LHPC
collaboration~\cite{Edwards:2006zza,Renner:2007pb,Hagler:2007xi,Edwards:2006qx,Renner:2004ck,Edwards:2005kw}.

%%%%%%%%%%%%%%%%%%%%%%%%%
\subsubsection{\it Anisotropic Clover Wilson Fermions}

Recently anisotropic lattices have proven useful for spectroscopy
projects, and as the calculations needed for studying multi-hadron
systems rely on spectroscopy, NPLQCD has recently adopted
clover-improved Wilson fermion actions.  
In particular, the $n_f=2+1$ flavor anisotropic Clover Wilson
action~\cite{Okamoto:2001jb,Chen:2000ej} with stout-link
smearing~\cite{Morningstar:2003gk} of the spatial gauge fields
in the fermion action with a smearing weight of $\rho=0.14$ has been used.  
The
gauge fields entering the fermion action are not smeared in the time
direction, thus preserving the ultra-locality of the action in the time
direction.  Further,
a tree-level tadpole-improved Symanzik gauge action
with no $1\times 2$ rectangle in the time direction is used. Anisotropy
allows for a better extraction of the excited states as well as
additional confidence that plateaus in the effective mass plots formed from
the correlation functions (discussed in Section~\ref{sec:ESCF})
have been observed, significantly
reducing the systematic errors in observables due to fitting.  
The gauge field generation is done by the Hadron Spectrum Collaboration 
(HSC),
and these gauge field configurations have been used
for excited hadron spectrum calculations by
HSC~\cite{Dudek:2009qf,Bulava:2009jb,Lin:2008pr,Edwards:2008ja}.

%%%%%%%%%%%%%%%%%%%%%%%%%%%%%%%
\subsection{\it Euclidean Space Correlation Functions}
\label{sec:ESCF}
\noindent
Most Euclidean space correlation functions computed in LQCD calculations 
(suitably Fourier transformed) are the sums of exponential functions.
The arguments of the exponentials 
are the product of Euclidean time with the eigenvalues
of the Hamiltonian associated with eigenstates in the finite-volume 
that couple to the hadronic sources and sinks.  
For a lattice that has infinite extent in the time-direction, the
correlation function at large times becomes a single exponential dictated by the
ground state energy and the overlap of the source and sink with the
ground state.  As an example, consider the pion two-point function,
$C_{\pi^+}(t)$, generated by a source (and sink) of the form
$\pi^+({\bf x},t)=\overline{u}({\bf x},t)\gamma_5 d({\bf x},t)$,
\begin{eqnarray}
C_{\pi^+}(t) & = & 
\sum_{\bf x}\ \langle 0|\ \pi^- ({\bf x},t)\ \pi^+ ({\bf 0},0)\ |0\rangle
\ \ \ ,
\label{eq:singlepioncorrelator}
\end{eqnarray}
where the sum over all lattice sites at each time-slice, $t$, projects onto the
 ${\bf p}={\bf 0}$ spatial momentum states.
The source $\pi^+({\bf x},t)$ not only produces single pion states, but also
all states with the quantum numbers of the  pion.
More generally, the source and sink are smeared over lattice sites in the
vicinity of $({\bf x},t)$ to increase the overlap onto the ground state and
lowest-lying excited states.
Translating the sink operator in time via $\pi^+({\bf x},t)=e^{\hat H t}
\pi^+({\bf x},0)e^{-\hat H t}$,
and inserting a complete set of states,
gives~\footnote{We assume the absence of external electroweak fields that
  may exert forces on hadrons in the lattice volume.}
\begin{eqnarray}
C_{\pi^+}(t) & = & 
\sum_n\ {e^{-E_n t}\over 2 E_n}\  \sum_{\bf x}\ \langle 0|\ \pi^- ({\bf
  x},0) |n\rangle 
\langle n|\pi^+ ({\bf 0},0) |0\rangle
\ \rightarrow\ A_0\ {e^{-m_\pi t}\over 2 m_\pi}
\ \ \ .
\label{eq:singlepioncorrelatorASYMP}
\end{eqnarray}
At finite lattice spacing, the correlation functions for Wilson
fermions remain sums of exponential functions, but for particular
choices of parameters used in the domain-wall discretization, the
correlation functions exhibit additional sinusoidally modulated
exponential behavior at short-times with a period set by the lattice
spacing~\cite{Syritsyn:2007mp}.

It is straightforward to show that the lowest energy eigenvalue
extracted from the correlation function in
Eqs.~(\ref{eq:singlepioncorrelator}) and
(\ref{eq:singlepioncorrelatorASYMP}) corresponds to the mass of the
$\pi^+$ (and, more generally, the mass of the lightest hadronic state
that couples to the source and sink) in the finite volume.  The masses
of stable single particle states can be extracted from a Lattice QCD
calculation with high accuracy as long as the lattice spatial extent
is large compared to the pion Compton-wavelength~\footnote{Finite-volume
effects are exponentially suppressed~\cite{Luscher:1985dn} by factors
of $e^{-m_\pi L}$.}.

Once a correlation function is calculated, a
common objective is to extract the argument of the exponential function that
persists at large times. One way to do this is to simply fit the function
over a finite number of time-slices to a single exponential function.  
A second method, that is somewhat more 
useful in visually assessing the quality of the calculation, 
is to form the effective mass (EM) function, e.g
\begin{eqnarray}
M_{\rm eff.}(t; t_J) & = &
{1\over t_J}\  
\log\left({ C_{\pi^+}(t)\over  C_{\pi^+}(t+t_J)}\right)\ \rightarrow\ m_{\pi}
\ \ \ ,
\label{eq:effectivemassfunction}
\end{eqnarray}
where both $t$ and $M_{\rm eff.}(t; t_J)$ are in lattice units.
At large times, $M_{\rm eff.}(t; t_J)$ becomes  a constant equal to the mass of the
lightest state contributing to the correlation function~\footnote{This is
  obviously the most simplistic approach to this problem.  
One well-known method to extract the ground state and excited state energies is
that of L\"uscher and Wolff~\cite{Luscher:1990ck, Michael:1985ne} 
in which the correlation functions 
resulting from different
sources and sinks are calculated.   The resulting matrix of correlation
functions is diagonalized, and the EM function for each resulting
eigenvalue can be used to extract the spectrum.}. 
The anti-periodic boundary-conditions in the time-direction, imposed on the
quark-fields in order to recover the correct fermionic partition function, 
result in the single meson correlation functions being periodic in the time
direction and symmetric about the mid-point in the time direction.  They are
the sum of $\cosh$-functions, and 
for such correlation functions it is useful to
construct the EM using
\begin{eqnarray}
M_{\rm eff.}(t; t_J) & = & {1\over t_J}\ 
\ \cosh^{-1}\left({ C_{\pi^+}(t+t_J) +  C_{\pi^+}(t-t_J)\over  2 \ C_{\pi^+}(t)}\right)\
\rightarrow\ m_{\pi}
\ \ \ ,
\label{eq:effectivemassfunctionCOSH}
\end{eqnarray}
which becomes a constant value for a single exponential 
or a single hyperbolic-cosine 
(near the mid-point of the time-direction of the lattice).

%%%%%%%%%%%%%%%
\subsection{\it Statistical Analysis Methods}
\label{sec:StatAn}
 Since Monte-Carlo integration is used to compute the relevant
 correlation functions, the statistical uncertainty must be carefully
 determined.  The main observables extracted from the  calculations
 presented in this review are energy eigenvalues and their
 differences, which contain information about phase shifts, scattering
 lengths and the three body interaction.  The extraction of energy
 eigenvalues is done by fitting the relevant correlation functions to
 a sum of exponentials (or hyperbolic cosine functions).  The
 optimal values for the energy are extracted from correlated
 $\chi^2$-minimization fits that take into account the time
 correlations in the lattice calculations.  In particular, the
 relevant parameters, such as the energies and the amplitude of each
 state that contributes to the correlation function, are determined as
 those that minimize
 \begin{equation}
\chi^2(A) = \sum_{ij} \left[\bar G(t_i) - F(t_i, A)\right]C^{-1}_{ij} 
\left[\bar G(t_j) - F(t_j, A)\right]
\ \ \ ,
\end{equation}
where $\bar G(t)$ are the lattice two point correlation functions,
$F(t,A)$ are the fitting functions used, $A$ denotes the set of
fitting parameters over which $\chi^2(A)$ is minimized, and $C_{ij}$
is the covariance matrix.  The lattice two point correlation functions
are determined as averages over $N$ Monte-Carlo samples of the correlation
function, $G_k(t)$:
\begin{equation}
\bar G(t) = \frac{1}{N} \sum_{k=1}^N G_k(t)
\ \ \ , 
\end{equation}
and 
\begin{equation}
C_{ij} = \frac{1}{N(N-1)}  \sum_{k=1}^N \left[G_k(t_i) - \bar G(t_i)\right] 
\left[G_k(t_j) - \bar G(t_j)\right] \ .
\end{equation}
The (standard) errors on the fitted parameters are determined by the 
boundaries of the
error ellipsoid~\footnote{For a pedagogical presentation of fitting 
see the
TASI lectures by D. Toussaint~\cite{DeGrand:1990ss}.}.

In computing scattering parameters, the procedure for determining the 
statistical uncertainties is somewhat more involved due to the highly
non-linear relation between the scattering amplitude and the energy
levels of the two-hadron system. First, one is interested in the
energy differences between the energy levels of the two-hadron
system and the sum of the masses of the two free hadrons (similarly
for the case of three or more hadrons).  These energy differences
can be determined in two ways. The simplest is 
where ratios of correlators are constructed in
such a way so that they are a sum over exponentials parametrized
by the desired energy splittings. 
In this case Jackknife is used to determine
the covariance matrix and then a correlated $\chi^2$-fit is performed. 
For a  single elimination Jackknife, the
covariance matrix of a ratio of correlation functions is
\begin{equation}
C_{ij} = \frac{N-1}{N}  \sum_{k=1}^N \left[R_k(t_i) - \bar R(t_i)\right] 
\left[R_k(t_j) - \bar R(t_j)\right]
\end{equation}
where $R_k$ is the desired ratio computed with the $k$th sample omitted from
the full ensemble and $\bar R$ is the ratio computed on the full ensemble.

Fitting correlation functions to the sum of $p$ exponential functions
to extract the ground state energy
requires fitting ranges that start at time separations from the source
that are large enough so that the $p+1^{\rm th}$ and higher 
excited states have negligible
contributions. The determination of the minimum time separation that
can be included in the fit is sometimes subjective. Hence a systematic
uncertainty due to the choice of the minimum time separation in the fit is
included. This uncertainty is determined by observing the variation of the
extracted results as a function of the choice of fitting interval.  The
final uncertainties  include both systematic and statistical uncertainties 
combined
in quadrature.

%%%%%%%%%%%%%%%%%%%%%%%%%
\subsection{\it Developments in Fitting Methodology}

Fitting correlation functions to a sum of exponentials is a
notoriously hard problem.  However, the fitting is greatly simplified
and more robust 
if only the lowest energy eigenvalue is to be extracted.  With this
restriction, only a limited amount of information about the
spectral properties of the theory can be extracted from the computed
correlation functions and, in addition, control of the systematic
uncertainties of the extracted ground state energy eigenvalue is
somewhat limited.  Because the statistical uncertainties in baryon
correlation functions grow exponentially with Euclidean time at large
times, extracting the lowest energy eigenvalue at large time
separations (in order to reduce the systematic uncertainty due to the 
contamination
from excited states) typically results in large statistical
uncertainties.  In other words one can trade statistical uncertainty  
growth for
systematic uncertainty reduction. There are two approaches in resolving this
problem. One is to develop better sampling methods to reduce
statistical uncertainties 
in the correlation functions.  Such techniques have been
developed in simple lattice field theories such as spin systems
(cluster algorithms). However, of the general correlation functions 
computed in Lattice QCD only one approach seems viable at this
point. This is to extract as much information from the correlation functions 
at
short time separations where the statistical noise does not overwhelm
the signal. In this region, multiple  exponentials contribute to the
correlation functions, and although the general multi-exponential fit
problem is difficult and not well behaved, 
correlation functions can be designed 
so that such fits are facilitated. 
Variational analysis on symmetric positive
definite matrices of correlation functions has been successfully used
in the
Lattice QCD community 
to extract  the energy eigenvalues 
contributing to the correlation functions (the
L\"uscher-Wolff method). These methods were originally introduced
in Refs.~\cite{Luscher:1990ck, Michael:1985ne}, and have been subsequently
developed~\cite{Dudek:2009qf,Foley:2007ui,Basak:2007kj,Blossier:2009kd}.

%%%%%%%%%%%%%%%%%%%%%%%%%%%%%%
\subsubsection{\it The Prony Method}
A simple and widely used method of estimating the mass of the ground state 
energy contributing to a correlation function
is the EM, 
as given in Eq.~(\ref{eq:effectivemassfunction}).
It is conventional to define the EM from the
logarithm of the ratio of the correlation function on adjacent
time-slices. It is also possible \footnote{This was suggested by
  K.~Juge in a talk at Lattice 2008, see Ref.~\cite{JugeAllHands2008},
  but may have been used earlier.} to form a more general EM from
time-slices separated by $t_J>1$.
For exponentially decreasing signals with time-independent noise, this
will naturally reduce the statistical uncertainty in the EM and
improve the extraction of energy eigenvalues as it increases the
``lever-arm'' of the exponential.  In such a case, the uncertainty in
$M_{\rm eff.}(t;t_J)$ in Eq.~(\ref{eq:effectivemassfunction}) will decrease as $1/t_J$.
Simple correlation functions involving pions have time-independent
uncertainties, but this is not the case for baryonic correlation
functions, whose relative uncertainties grow exponentially with time at large times.
Improvements to baryon EMs, and ultimately the
extraction of baryon masses and the energy eigenvalues in the volume,
that result from $t_J>1$ have been explored~\cite{Beane:2009kya}. 
In fitting an energy to an EM (and other
generalizations), either the Bootstrap or Jackknife procedures are
used to generate the covariance matrix associated with the time-slices
in the range of the fit.
This covariance matrix is then used to form the $\chi^2/{\rm dof}$ and 
extract the
best estimate of the mass as well as it's uncertainty by using methods 
described in Sec.~\ref{sec:StatAn}.

Based on the findings of NMR spectroscopists,
the EM method has been generalized
to two or more exponential 
functions~\cite{Fleming:2004hs}~\footnote{ The method is more
  generally referred to as Prony's method~\cite{Hamming} after Gaspard
  Riche de Prony who first constructed it in 1795~\cite{Prony}. These
  techniques and other related methods are known as linear prediction
  theory in the signal analysis community.},
and is found to compare favorably~\cite{Lin:2007iq} to the
variational approach.
A detailed study of this approach in the analysis of correlation 
functions
with small statistical uncertainties  
was performed in Ref.~\cite{Beane:2009kya}.
The method was found to be quite stable and to provide a simple and reliable
method of determining the lowest energy eigenvalue.  
This is  in contrast to an analysis of low-statistics correlation functions
for which the method is quite unstable.

%%%%%%%%%%%%%%%%%%%%%%%%%%%%%%%%%%
\subsubsection{\it The Matrix-Prony Method}
In Lattice QCD, it is simple and ``cost'' effective to construct a set of correlation 
functions with the same source (creation operator) interpolating field and 
several interpolating fields at the sink (annihilation operator). 
Multiple sinks and one source have been used 
for achieving the high statistics required to study 
multi-hadron systems. 
In this case, it is straightforward to generalize Prony's method
to include all correlation functions with the same source, and  we call 
this the matrix-Prony method.  This
method leads to a further reduction in the uncertainty of the extraction
of the energy eigenvalues. A similar approach has been briefly
discussed in~Ref.~\cite{Fleming:2006zz}.
While the matrix-Prony method provides a computational simplification 
in the analysis of multiple correlation functions, 
and provides a tool with which to extract multiple energy eigenvalues and
eigenstates from a single source,
it should not be considered
to be more reliable than the L\"uscher-Wolff method~\cite{Luscher:1990ck, Michael:1985ne}.

Assume there are $N$ 
correlation functions~\footnote{We have applied the method for $N=2$ and $3$.} 
from
which the lowest lying  energy eigenvalues are desired. 
If these correlation
functions are a sum of exponentials they satisfy the following
recursion relation,
\begin{equation}
  M y(\tau+t_J) - V y(\tau) = 0
  \ \ \ ,
  \label{eq:recursion}
\end{equation}
where $M$ and $V$ are $N\times N$ matrices and $y(t)$ is a column
vector of $N$ components corresponding to the $N$ correlation
functions.  Eq.~(\ref{eq:recursion}) implies that the correlation
functions are
\begin{equation}
  y(t) = \sum_{n=1}^N  \ C_n\  q_n\  \lambda_n^{t}
  \ \ \ ,
  \label{eq:signal}
\end{equation}
where $q_n$ and $\lambda_n=\exp(m_n t_J)$  are the eigenvectors and
eigenvalues of the following generalized eigenvalue problem
\begin{equation}
  M q = \lambda V q
\ \ \ .
  \label{eq:gev}
\end{equation}
Given the $N$ sets of correlation functions, the energy eigenvalues can be found
by determining the matrices $M$ and $V$ that are needed in order for
the signal to satisfy Eq.~(\ref{eq:recursion}). Solving
Eq.~(\ref{eq:gev}) leads to the eigenvalues $\lambda_n=\exp(m_n
t_J)$ and the eigenvectors $q_n$ that are needed to reconstruct the amplitudes
with which each exponential enters the correlation functions.  A
simple solution can be constructed as follows. First note that
\begin{equation}
  M \sum_{\tau=t}^{t+t_W} y(\tau+t_J)y(\tau)^T - V  \sum_{\tau=t}^{t+t_W}
  y(\tau)y(\tau)^T = 0
  \ \ \ .
  \label{eq:recursion3}
\end{equation}
Clearly, a solution for $M$ and $V$ is
\begin{eqnarray}
  M &=& \left[\
    \sum_{\tau=t}^{t+t_W} y(\tau+t_J)y(\tau)^T\
  \right]^{-1}
  \ \ \ \ ,\
  \ \ \
  V \ =\  \left[\ \sum_{\tau=t}^{t+t_W} y(\tau)y(\tau)^T \ \right]^{-1}
  \ \ \ ,
  \label{eq:solution}
\end{eqnarray}
where these inverses exist provided that the range, $t_W$, is large
enough to make the matrices in the brackets full rank ($t_W\ge N -
1$).    Once the eigenvalues, $\lambda_n$ and
eigenvectors $q_n$ are determined, the amplitudes, $C_n$, can be
reconstructed using $t$ as a normalization point.  The shift parameter
$t_J$ can be used to improve stability and
understand systematic effects.  The above solution is equivalent 
to determining $M$ and
$V$ by requiring that
\begin{eqnarray}
  \Psi^2 &=& \sum_{\tau=t}^{t+t_W}\left[ M y(\tau+t_J) - V
    y(\tau)\right]^T  \left[ M y(\tau+t_J) - V y(\tau)\right] 
  \label{eq:chi2}
\end{eqnarray}
is minimized.

To go beyond extracting $N$ states, a
second order recursion relation can be constructed and solved. 
The minimization condition of
Eq.~(\ref{eq:chi2}), augmented to contain the second order terms in the
recursion, can be used to determine the unknown matrices.  The
resulting eigenvalue problem is a second order nonlinear generalized
eigenvalue problem which is straightforward to solve.  However, 
tests show that
this approach is unstable and hence 
such analyses are restricted to extract,
at most, $N$ energies from $N$ correlation functions.

%%%%%%%%%%%%%%%%%%%%%%%%%%%%%%%%%%%%%%%%%%%%%%%%%%%
%
% FIGURE: Matrix prony tJ=tW=10
%
%%%%%%%%%%%%%%%%%%%%%%%%%%%%%%%%%%%%%%%%%%%%%%%%%%%
%
\begin{figure}[!htb]
\begin{center}
\begin{minipage}[t]{8 cm}
\centerline{\includegraphics[scale=0.7]{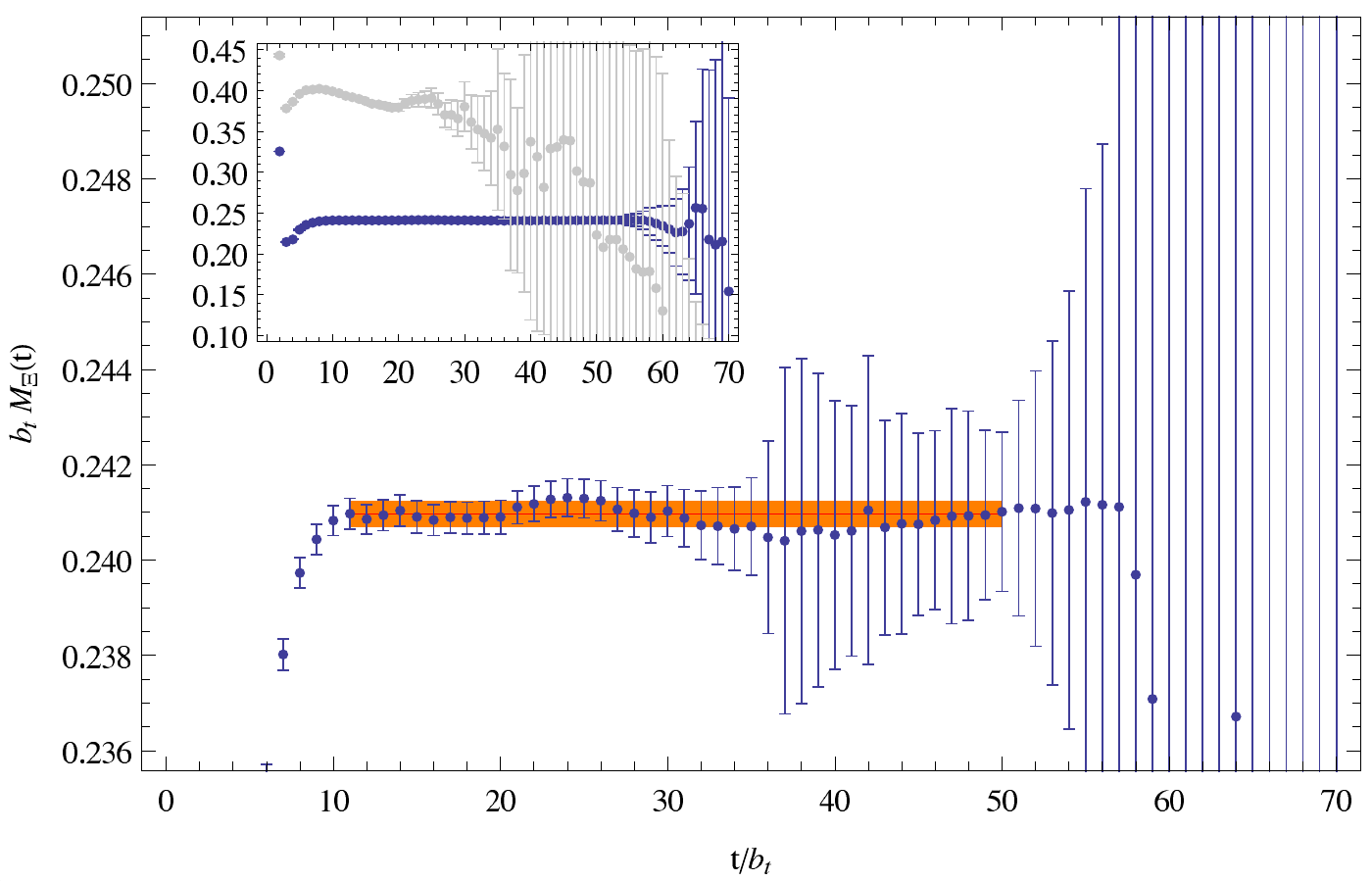}}
\end{minipage}
\begin{minipage}[t]{16.5 cm}
\caption{
The generalized EMP for the mass
    of the $\Xi$ using a Matrix-Prony analysis~\cite{Beane:2009kya}.
The correlation function was calculated on the $20^3\times 128$ anisotropic
clover gauge field configurations with $m_\pi\sim 390~{\rm MeV}$.
The inner (darker) region corresponds to the
    statistical uncertainty, while the outer (lighter) region
    corresponds to the statistical and fitting systematic
    uncertainties combined in quadrature. The inset shows 
the energies of both states
    extracted with the matrix-Prony method.
\label{fig:2x2MatrixexpXi}}
\end{minipage}
\end{center}
\end{figure}
To demonstrate how this method works, results from  the $\Xi$ 
baryon correlation
function calculated on the $20^3\times 128$ anisotropic
clover gauge field configurations with $m_\pi\sim 390~{\rm MeV}$ 
are presented.
Figure~\ref{fig:2x2MatrixexpXi} shows the generalized effective mass 
plot (EMP) for the
$\Xi$ mass as a function of time determined with a $N=2$ matrix-Prony
extraction, using both the smeared-smeared (SS) and smeared-point (SP)
correlation functions.  The inset shows the second extracted state in
addition to the ground state. The extracted value of the $\Xi$ mass,
determined by fitting in the time interval $t=11$ to $t=50$, is
\begin{eqnarray}
  M_\Xi & = & 0.24097\pm 0.00025\pm 0.00003
  \ \ ,\ \
  \chi^2/{\rm dof} \ =\ 0.81
  \ \ \ .
  \label{eq:Neq2expFits}
\end{eqnarray}
The energy of the dominant state in Figure~\ref{fig:2x2MatrixexpXi} plateaus
around time-slice $t=10$, and is well-defined over a large interval.

%%%%%%%%%%%%%%%%%%%%%%%%%%%%%%%%%%%%%%%%%%%%%%%%%%%
%
% FIGURE: Matrix prony varying tJ=tW
%
%%%%%%%%%%%%%%%%%%%%%%%%%%%%%%%%%%%%%%%%%%%%%%%%%%%
%
\begin{figure}[!htb]
\begin{center}
\begin{minipage}[t]{8 cm}
\centerline{\includegraphics[scale=0.7]{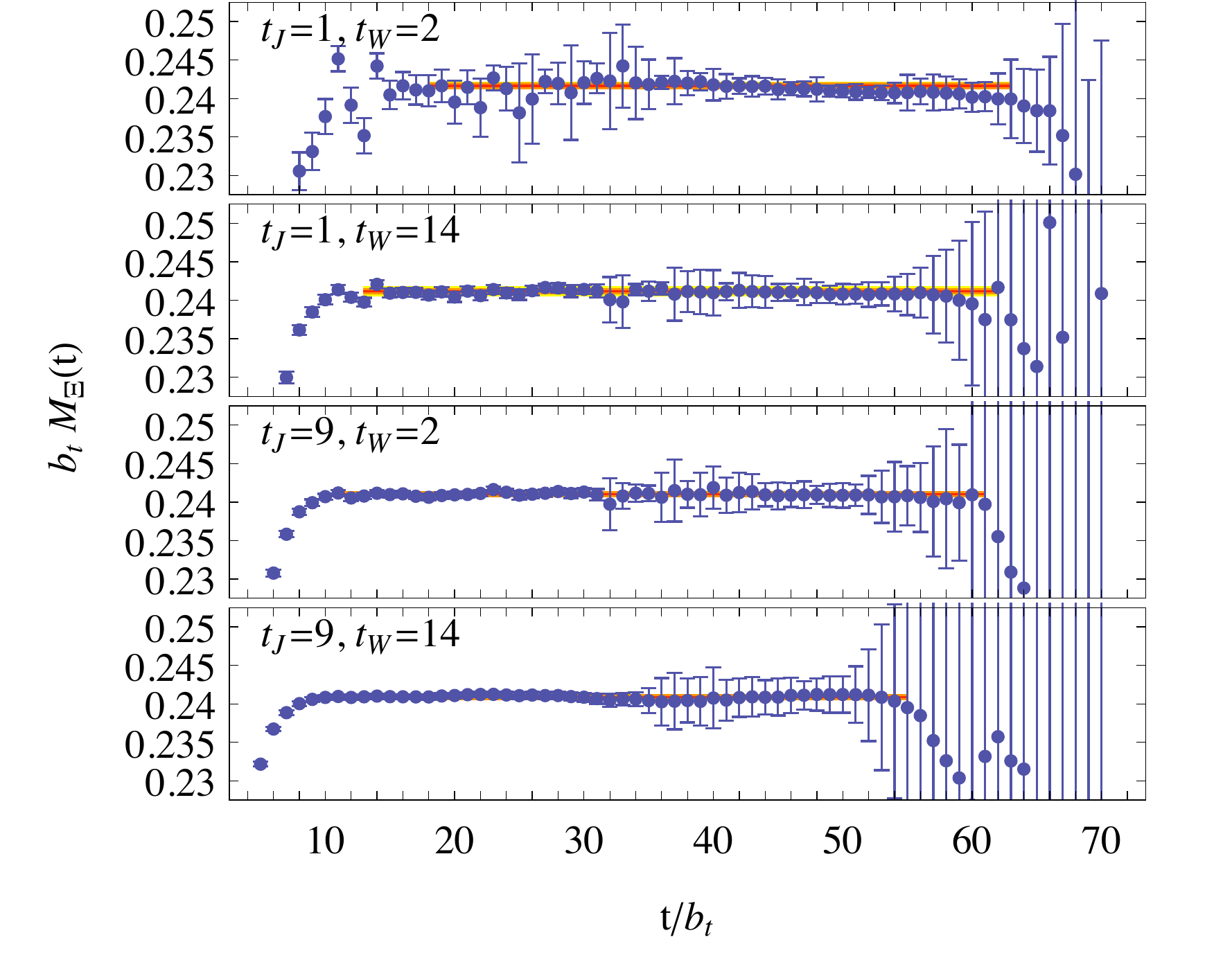}}
\end{minipage}
\begin{minipage}[t]{16.5 cm}
\caption{
The generalized EMP for the
    mass of the $\Xi$ using a Matrix-Prony analysis for a variety of
    values for $t_J$ and $t_W$.
\label{fig:2x2MatrixexpXiMULTI}}
\end{minipage}
\end{center}
\end{figure}
\begin{figure}[tb]
\begin{center}
\begin{minipage}[t]{8 cm}
\centerline{\includegraphics[scale=0.65]{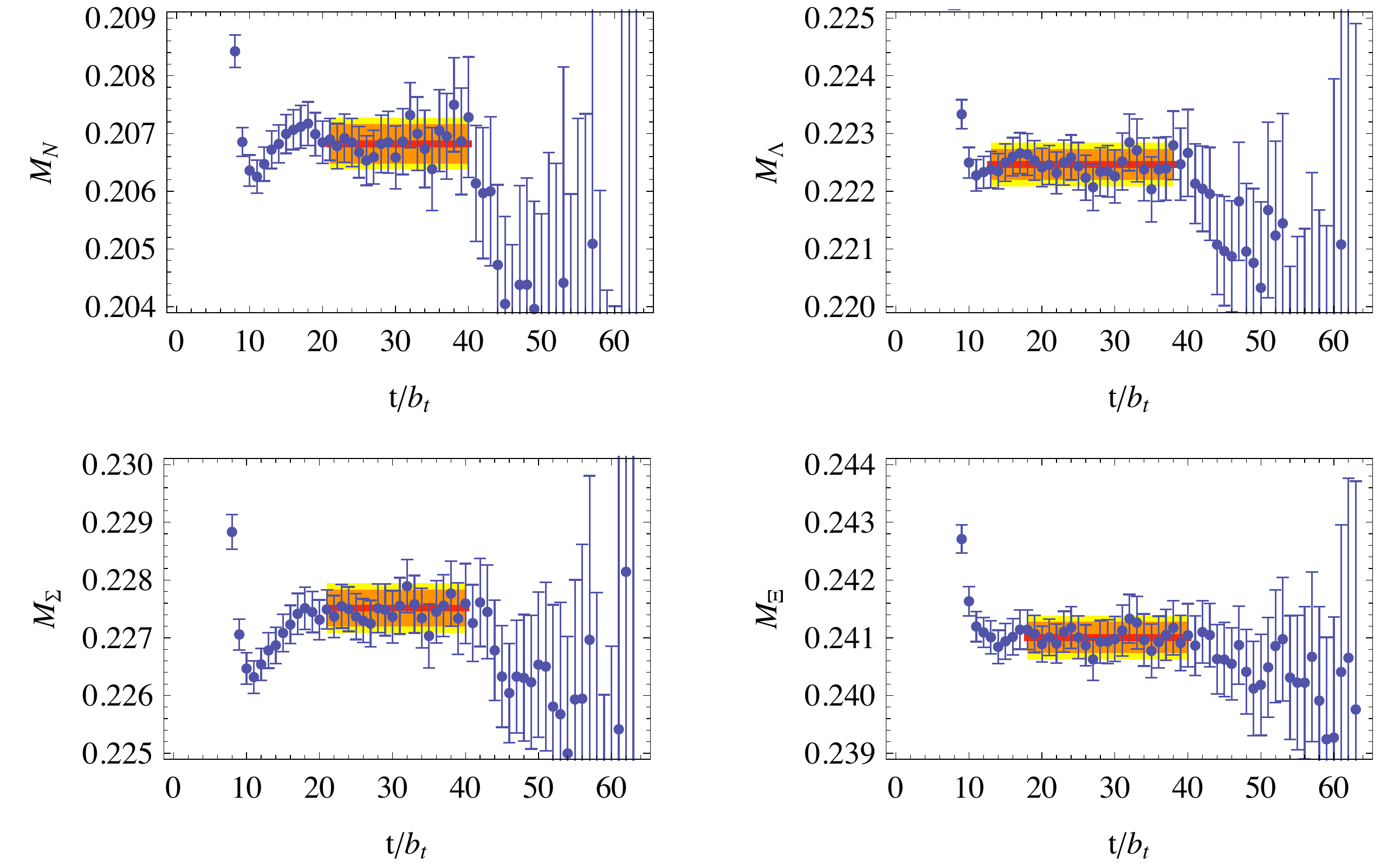}}
\end{minipage}
\begin{minipage}[t]{16.5 cm}
\caption{
The EMPs for the masses of the 
octet-baryons from the eigenstates of a $2\times 2$
matrix-Prony
analysis of the SS and SP correlation 
functions~\protect\cite{Beane:2009kya}
calculated on the $20^3\times 128$ anisotropic
clover gauge field configurations with $m_\pi\sim 390~{\rm MeV}$.
\label{fig:2x2SINGLE}}
\end{minipage}
\end{center}
\end{figure}

The matrix-Prony method is currently our preferred approach in analyzing 
correlation functions.
Detailed tests of this method were presented in Ref.~\cite{Beane:2009kya},
which yield ground state energies that
are in agreement with those from the other methods.
The procedure for fitting parameters
and determining their statistical uncertainty has been described in
Section~\ref{sec:StatAn}.  Systematic uncertainties can be  calculated by
performing fits over rolling windows of time-slices within the quoted
overall range and considering  the standard deviation of the central
values of those fits. This is combined in quadrature with a further
systematic uncertainty that is generated by sampling a large range of
possible values of $t_J$ and $t_W$ and taking the standard deviation
of the central values of the resulting fits.
The generalized EMP for the $\Xi$ extracted with the matrix-Prony
method for a variety of values of $t_W$ and $t_J$ can be seen in
Figure~\ref{fig:2x2MatrixexpXiMULTI}.
One further aspect of this method, that is perhaps the most appealing, is that
the correlation functions corresponding to the eigenstates of the matrix-Prony
matrix with lowest energy eigenvalue are dominated by the ground-state in the
lattice volume, and hence have the longest plateau in the EMP.
The EM corresponding to the eigenstates of the octet-baryons calculated on the 
$20^3\times 128$ anisotropic
clover gauge field configurations with $m_\pi\sim 390~{\rm MeV}$ 
are shown in
Figure~\ref{fig:2x2SINGLE}
(compared with  
Figure~\ref{fig:2x2MatrixexpXi} and Figure~\ref{fig:2x2MatrixexpXiMULTI}).
%%%%%%%%%%%%%%%%%%%%%%%%%%%%%%%%%%%%%%%%%%%%%%%%%%%
%
% FIGURE: Matrix prony estates
%
%%%%%%%%%%%%%%%%%%%%%%%%%%%%%%%%%%%%%%%%%%%%%%%%%%%
%
%

%%%%%%%%%%%%%%%%%%%%%%%%%%%%%%%%%%%%%%%%%
\subsection{\it Hadronic Interactions, the Maiani-Testa Theorem and L\"uscher's
  Method}
\noindent
Extracting hadronic interactions from Lattice QCD calculations is far
more complicated than the determination of the spectrum of stable particles.
This is encapsulated in the
Maiani-Testa theorem~\cite{Maiani:1990ca}, which states that S-matrix elements
cannot be extracted from infinite-volume Euclidean-space Green functions except
at kinematic thresholds.
This is clearly problematic from the nuclear physics perspective, as a main
motivation for pursuing Lattice QCD is to be able to compute nuclear reactions
involving multiple nucleons.
Of course, it is clear from the statement of this theorem how it can be evaded,
Euclidean-space correlation functions are calculated at finite volume to extract
S-matrix elements, the formulation of which was known for decades in the
context of non-relativistic quantum mechanics~\cite{Huang:1957im} and extended
to 
quantum field theory
by L\"uscher~\cite{Luscher:1986pf,Luscher:1990ux}.
The energy of two particles in a  finite volume depends
in a calculable way upon their elastic scattering amplitude and their masses
for energies below the inelastic threshold. 
As a concrete example consider $\pi^+\pi^+$ scattering.
A $\pi^+\pi^+$ correlation function 
in the $A_1$ representation of the cubic group~\cite{Mandula:ut} 
(that projects onto the continuum s-wave state amongst others) is
\begin{eqnarray}
C_{\pi^+\pi^+}(p, t) & = & 
\sum_{|{\bf p}|=p}\ 
\sum_{\bf x , y}
e^{i{\bf p}\cdot({\bf x}-{\bf y})} 
\langle \pi^-(t,{\bf x})\ \pi^-(t, {\bf y})\ \pi^+(0, {\bf 0})\ \pi^+(0, {\bf 0})
\rangle
\ \ \ . 
\label{pipi_correlator} 
\end{eqnarray}
In relatively large lattice volumes, the energy
difference between the interacting and non-interacting two-meson states
is a small fraction of the total energy, which is dominated by the
masses of the mesons.  
This energy difference can be extracted from
the ratio of correlation functions, $G_{\pi^+ \pi^+}(p, t)$,
where
\begin{eqnarray}
G_{\pi^+ \pi^+}(p, t) & \equiv &
\frac{C_{\pi^+\pi^+}(p, t)}{C_{\pi^+}(t) C_{\pi^+}(t)} 
\ \rightarrow \ {\cal B}_0\ e^{-\Delta E_0\ t} 
\  \ ,
\label{ratio_correlator} 
\end{eqnarray}
and where the arrow denotes the large-time behavior of $G_{\pi^+ \pi^+}$. 
For calculations performed with $p=0$,
the energy eigenvalue,  $E_n$, and its deviation from the sum of the rest
masses of the particle, $\Delta E_n$, are related to a
momentum magnitude $p_n$ by
\begin{eqnarray}
\Delta E_n \ & \equiv & 
E_n\ -\  2m_\pi\ =\ 
\ 2\sqrt{\ p_n^2\ +\ m_\pi^2\ } 
\ -\ 2m_\pi \ .
\label{eq:energieshift}
\end{eqnarray}
To obtain $k\cot\delta(k)$, where $\delta(k)$ is the phase shift, the
square of $p_n$ is extracted from the
energy shift and inserted
into~\cite{Huang:1957im,Luscher:1986pf,Luscher:1990ux,Hamber:1983vu}
\begin{eqnarray}
k\cot\delta(k) \ =\ {1\over \pi L}\ {\bf
  S}\left(\,\left(\frac{k L}{2\pi}\right)^2\,\right)
\ \ ,\ \ 
{\bf S}\left(\, x \, \right)\ \equiv \ \sum_{\bf j}^{ |{\bf j}|<\Lambda}
{1\over |{\bf j}|^2-x}\ -\  {4 \pi \Lambda}
\ \ ,
\label{eq:energies}
\end{eqnarray}
where $k=p_n$, and 
which is only valid below the inelastic threshold. 
The regulated three-dimensional sum~\cite{Beane:2003da}
extends over all triplets of integers ${\bf j}$ such that 
$|{\bf j}| < \Lambda$ and the
limit $\Lambda\rightarrow\infty$ is implicit.
Therefore, by calculating the energy-shift, $\Delta E_n$,  of the two 
particles in the finite lattice
volume, the scattering phase-shift is determined at $\Delta E_n$.
In the absence of interactions between the particles 
the energy eigenstates in the finite volume  
occur at momenta ${\bf p} =2\pi{\bf j}/L$.  
It is important to re-emphasize that the relation 
in Eq.~(\ref{eq:energies})
is valid relativistically~\cite{Luscher:1986pf,Luscher:1990ux}.
Perhaps most important for nuclear physics is that this expression is valid
for large and  even infinite scattering lengths~\cite{Beane:2003da}.  
The only restriction is that the lattice volume be much larger than the
range of the interaction between the hadrons, which for two nucleons, is set by
the mass of the pion.

For the scattering of two nucleons, the scattering length is known to
be unnaturally large at the physical pion mass, and therefore, the
relation in Eq.~(\ref{eq:energies}) will have to be used to extract
the scattering parameters.  For systems that are not finely-tuned,
such as the $\pi^+\pi^+$ system, an expansion in the volume can be
used.  In the large volume limit ($L\gg |a|$) the energy of the two
lowest-lying continuum states in the $A_1$ representation of the cubic
group~\cite{Mandula:ut} are~\cite{Luscher:1986pf,Luscher:1990ux}
\begin{eqnarray}
\Delta E_0 & = & + {4\pi a\over M L^3}\left[\ 1\ -\ c_1 {a\over L}\ 
+\ c_2 \left({a\over L}\right)^2\ +\ ...\right]
\ +\ {\cal O}(L^{-6})
\nonumber\\
\Delta E_1 & = & {4\pi^2\over M L^2} - 
{12\tan\delta_0\over M L^2}\left[\ 
1 + c_1^\prime\tan\delta_0 + c_2^\prime \tan^2\delta_0\ +\ ...\ \right]
\ +\ {\cal O}(L^{-6})
\ \ \ ,
\label{eq:e0}
\end{eqnarray}
where $\delta_0$ is the s-wave phase-shift evaluated at $p=2\pi/L$.
The coefficients, which result from sums over the allowed
momenta~\cite{Luscher:1986pf,Luscher:1990ux} in the finite cubic volume,
are $c_1=-2.837297$, $c_2=+6.375183$, $c_1^\prime=-0.061367 $,  and
$c_2^\prime=-0.354156$~\footnote{We use the nuclear physics 
sign convention for the scattering length.}.
In addition, 
for $a>0$ with an attractive interaction
a bound state exists with energy~\cite{Beane:2003da} (in the large volume limit)
\begin{eqnarray}
\Delta E_{-1} & = & -{\gamma^2\over M}\left[\ 
1\ +\ {12\over \gamma L}\  {1\over 1-2\gamma (p\cot\delta)^\prime}\ 
e^{-\gamma L}\ +\ ...
\right]
\ \ \ ,
\label{eq:eb}
\end{eqnarray}
where $(p\cot\delta)^\prime={d\over dp^2}\ p\cot\delta$ evaluated at
$p^2=-\gamma^2$. The quantity $\gamma$ is the solution of
\begin{eqnarray}
\gamma\  +\  p\cot\delta |_{p^2=-\gamma^2} \ & = & 0
\ \ \ ,
\label{eq:pctdg}
\end{eqnarray}
which yields the bound-state binding energy in the infinite-volume limit.
As expected, the finite volume corrections are exponentially suppressed by the
binding momentum~\footnote{The finite volume dependence
of bound states has been explored numerically in Ref.~\cite{Sasaki:2006jn}.}.
This is consistent with the corrections to a single particle state where the
lightest hadronic excitation is the zero-momentum two-particle continuum state,
as opposed to a state containing an additional pion for, say, the finite volume
corrections to the single nucleon mass.

In the limit where $L\ll |(p\cot\delta)^{-1}|$, which is a useful
limit to consider when systems have unnaturally-large scattering
lengths, the solution of Eq.~(\ref{eq:energies}) gives the energy of
the lowest-lying states:
\begin{eqnarray}
\Delta\tilde E_0 & = & 
{4\pi^2 \over M L^2}\left[\ d_1 \ +\   d_2\  L p\cot\delta_0\  + ...\
  \right]\ \ \ ,\ \ \ 
\Delta\tilde E_1 \ = \ 
{4\pi^2 \over M L^2}\left[\  d_1^\prime \ +\  d_2^\prime \  L
  p\cot\delta_0
\  + ...\  \right]
\ \ \ ,
\label{eq:usE0}
\end{eqnarray}
where the coefficients are $d_1 = -0.095901$, $d_2 = +0.0253716$,
$d_1^\prime = +0.472895$, $d_2^\prime = +0.0790234$
and where $p\cot\delta_0$ in 
the expression for 
$\Delta\tilde E_0$ is evaluated at an energy
$\Delta E={4\pi^2\over M L^2}\ d_1$, while 
$p\cot\delta_0$ in the expression for $\Delta\tilde E_1$ is evaluated at an energy
$\Delta E={4\pi^2\over M L^2}\ d_1^\prime$.
The values of the $d_i^{(\prime)}$ are determined by zeros of 
${\bf S}(x)$, and 
the expressions for $\Delta E_i$ and $\Delta\tilde E_i$, excluding 
$\Delta E_{-1}$,
are valid for both positive and negative scattering lengths.

%%%%%%%%%%%%%%%%%%%%%%%%%%%%%%%%%%%%%%%%%%%%%%%%%%
\subsection{\it Bound-States Versus Scattering-States}
\label{sec:BSorSS}
\noindent
It is important to understand what 
can be extracted from finite-volume calculations.  One important
question that arises is: if a negative energy-shifted state is
calculated on the lattice at finite-volume, does it correspond to a
bound state or to a scattering state?  Clearly, calculations of the
same correlation function in multiple volumes will allow for the
exponential volume dependence of a bound 
state
to be distinguished
from the power-law volume dependence of a scattering state.  However,
one can make an educated guess about the nature of the state by the
magnitude of the energy shift.  Consider a simple system whose
scattering amplitude is dominated at low-energies by the scattering
length and effective range, as is the case for the scattering of two
nucleons. 
\begin{figure}[tb]
\begin{center}
\begin{minipage}[t]{8 cm}
\centerline{\includegraphics[scale=0.31]{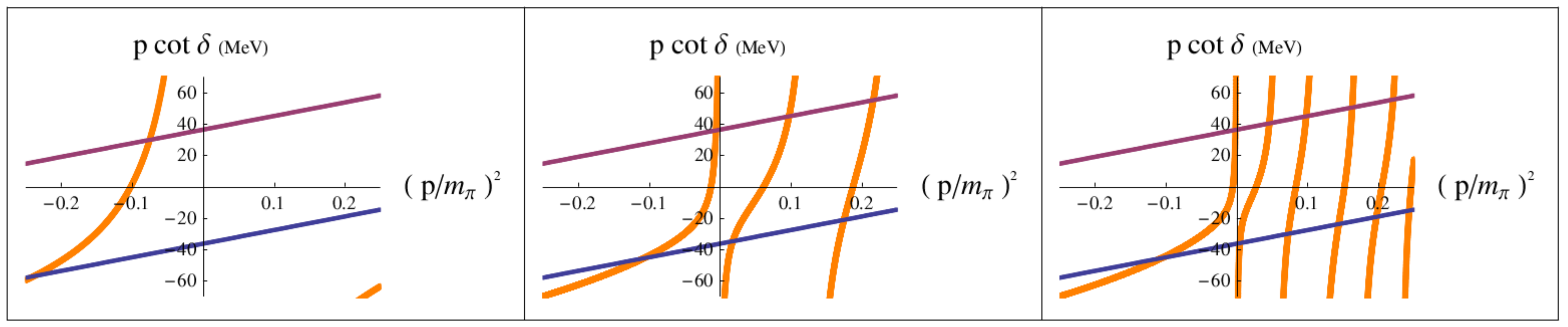}}
\end{minipage}
\begin{minipage}[t]{16.5 cm}
\caption{
The functions ${\bf S}(x)$ (curved orange lines) and 
$p\cot\delta$ (straight red and blue lines) versus $(p/m_\pi)^2$ 
at the physical pion mass.
The lower (blue) straight line 
corresponds to using the experimentally determined 
$\siii$-channel nucleon-nucleon scattering length and
effective range, while the upper (red) straight line corresponds to using a
scattering length of opposite sign but equal magnitude.
The left, center and right panels corresponds to lattices with spatial extent
of $L = 8.5~{\rm fm}, 24.5~{\rm fm}$ and $ 36.8~{\rm fm}$, respectively.
The intercepts of the curves corresponds to the 
energy eigenvalues in the finite volume.
\label{fig:Sfunpcots}}
\end{minipage}
\end{center}
\end{figure}
The location of the states in the lattice volume are
determined by the solution of Eq.~(\ref{eq:energies}).
In Figure~\ref{fig:Sfunpcots} we show the graphical solution to
Eq.~(\ref{eq:energies}) for two systems, one with 
$a=+5.425~{\rm fm}$ and $r=1.749~{\rm fm}$ (blue line) and the other with
$a=-5.425~{\rm fm}$ and $r=1.749~{\rm fm}$ (red line).
One finds that states with $E<0$ and $p\cot\delta<0$ (which occur for
$x=\left({pL\over 2\pi}\right)^2 < d_1$)
likely correspond to a bound-state, while states with $E<0$ and $p\cot\delta>0$
($x>d_1$)
likely correspond to continuum states.
However, one can imagine scattering parameters in the momentum 
expansion of  $p\cot\delta$ that modify these rules.

\begin{figure}[tb]
\begin{center}
\begin{minipage}[t]{8 cm}
\centerline{\includegraphics[scale=0.7]{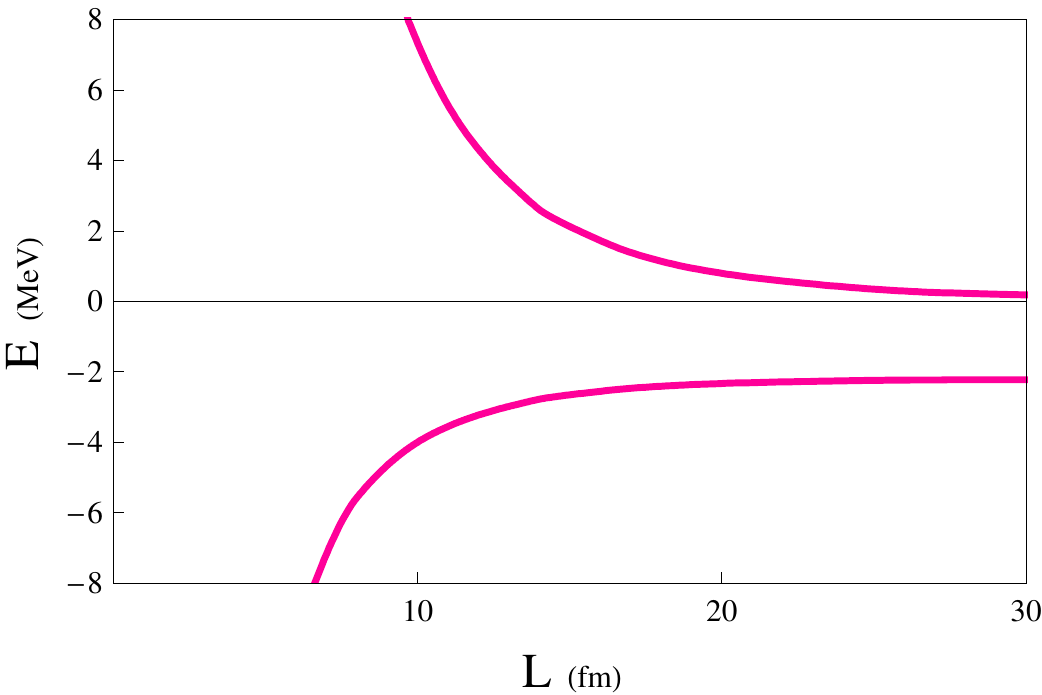}}
\end{minipage}
\begin{minipage}[t]{16.5 cm}
\caption{
The energy of the bound-state and first continuum state of 
two nucleons in
the $^3S_1$ channel (neglecting D-wave interactions and mixing) 
as a function of the spatial extent of the lattice (fm).
\label{fig:levels}}
\end{minipage}
\end{center}
\end{figure}

The location of the bound-state and 
the first
continuum state for two  nucleons in
the $^3S_1$ channel (neglecting D-wave interactions and mixing) as a 
function of
lattice volume are shown in Figure~\ref{fig:levels}.  
In order for the bound-state energy to be very close to that of the deuteron
binding energy, the lattice volumes must be very large.  However, as
shown in Section~\ref{sec:fakedata}, a single calculation of the lowest energy in the
lattice volume is not the best way to determine the deuteron binding energy.
The deuteron binding energy will best be determined from the calculation of 
ground-state and continuum state energies in one or more lattice volumes.

%%%%%%%%%%%%%%%%%%%%%%%%
\subsection{\it Scattering Parameters from Wavefunctions.}

\noindent
One remarkable feature of nuclear physics is that one can understand
and compute (to reasonable accuracy) the properties and interactions of
nuclei working with an energy-independent two-nucleon potential alone.
Phenomenologically, one finds that the three-nucleon interaction, and,
in the case of electroweak matrix elements, meson exchange currents, are
required to improve agreement with experiment.
But the
fact remains that the two-nucleon potential is the dominant
interaction in nuclei. There is a  desire to construct such 
nucleon-nucleon
potentials~\footnote{In this context, the word {\it potential} means an
energy-independent potential.}  directly from QCD, and hence from
Lattice QCD.
Nucleon-nucleon potentials may be defined from Lattice QCD
calculations in the same way that phenomenological
potentials are determined from
experimental measurements of the elastic scattering cross-section.  
A
large number of Lattice QCD calculations will be performed, 
producing values for the
phase shift, along with an uncertainty, over a wide range of
low-energies.
Potentials can be defined that minimize the
$\chi^2/dof$ in a global fit to the calculated phase-shifts.  At present,
there is no practical program underway to perform such an analysis due
to limited computational power.

The utility of two particle wavefunctions and potentials in LQCD
calculations was first addressed in a substantive way by
L\"uscher~\cite{Luscher:1990ux}.  The methodology outlined by
L\"uscher has been used by the CP-PACS collaboration in early work
that determined scattering parameters and phase shift of the 
$\pi^+\pi^+$ system from quenched calculations at relatively large
pion masses with Wilson fermions~\cite{Aoki:2004wq,Aoki:2005uf}.
Consider an interpolating operator for $\pi^+\pi^+$ states of the
form
\begin{eqnarray}
\hat\theta({\bf x},{\bf y}; t) = \overline{u}({\bf x},t)\gamma_5 d({\bf x},t)
 \overline{u}({\bf y},t)\gamma_5 d({\bf y},t)
\ \ ,
\label{eq:pipiinterpolator}
\end{eqnarray}
where the quark-field operators $q(x)$ may be smeared about the
point $x$ to enhance the overlap of the single-$\pi^+$ interpolating operator,
$\overline{u}({\bf x},t)\gamma_5 d({\bf x},t)$, onto the single-$\pi^+$ state.
A $\pi^+\pi^+$ correlation function 
\begin{eqnarray}
G(|{\bf x}|,t) & = & 
{1\over 48\ L^3}\ 
\sum_{{\bf R},{\bf y}}\ 
\langle\ 
\hat\theta({\bf R}[{\bf x}] + {\bf y},{\bf y}; t)\ 
\hat\theta^\dagger({\bf 0},{\bf 0}; 0)\ \rangle
\nonumber\\
& \rightarrow & 
Z_{\pi\pi}(0;k)\ Z_{\pi\pi}(|{\bf x}|;k)\ e^{-E_{\pi\pi}^{(0)} t}\ 
\psi_{\pi\pi}(|{\bf x}|;k)
\ \ ,
\label{eq:Gpipi}
\end{eqnarray}
can be constructed, where ${\bf R}$ represents an element of the cubic
group.  
The magnitude of
momentum, $k$, is defined 
through $E_{\pi\pi}^{(0)} = 2\sqrt{k^2+m_\pi^2}$, as in
Eq.~(\ref{eq:energieshift}).  The summation over ${\bf R}$ and ${\bf y}$
projects onto the $A_1$ representation of the cubic-group with zero
total momentum.  We have performed a separation~\footnote{Alternatively, we
  could have simply indicated that the wavefunction depends upon the sink by
$\psi_{\pi\pi}(|{\bf x}|;k)\rightarrow \psi_{\pi\pi}^{(\theta)}(|{\bf x}|;k)$.
}
 into the product of a
sink-dependent overlap factor, $Z_{\pi\pi}(|{\bf x}|;k)$, and a
hadronic wavefunction, $\psi_{\pi\pi}(|{\bf x}|;k)$.  
The
spatially-dependent overlap factor $Z_{\pi\pi}(|{\bf x}|;k)$ 
is introduced in order to stress the fact that the 
overlap (onto a given eigenstate in the lattice volume)
of the 
composite sink-operator $\hat\theta({\bf R}[{\bf x}] + {\bf y},{\bf y}; t)$ 
is not the square of 
the overlap of the 
interpolating operator for the single $\pi^+$
when $|{\bf x}-{\bf y}| \lsim R$ where $R$ is a typical strong
interaction length scale set by the mass of the pion (i.e. it is set
by the range of the interaction between two $\pi^+$'s.).  The
wavefunction given in Eq.~(\ref{eq:Gpipi}), $\phi(|{\bf x}|;k)\sim
Z_{\pi\pi}(|{\bf x}|;k)\ \psi_{\pi\pi}(|{\bf x}|;k)$ satisfies the
Schrodinger equation~\cite{Luscher:1990ux}
\begin{eqnarray}
\left(\ \nabla^2 \ +\ k^2\ \right)\phi(|{\bf x}|;k) 
 & = & 
\int d^3{\bf y}\ U_k({\bf x},{\bf y}) \phi(|{\bf y}|;k) 
\ \ ,
\label{eq:Upipidef}
\end{eqnarray}
where $U_k({\bf x},{\bf y})$ is the Fourier transform of the modified
Bethe-Salpeter kernel for the $\pi^+\pi^+$ interaction, which is
generally nonlocal and
energy-dependent~\cite{Luscher:1990ux,Aoki:2004wq,Aoki:2005uf}.  
An investigation of this construction under the assumption that $U_k({\bf x},{\bf y})$
is independent of energy can be found in Ref.~\cite{Aoki:2009ji}.
In
the limit where the separation between the two $\pi^+$'s is large
compared to the range of the interaction (the inverse pion mass),
$|{\bf x}|\gg R$, the normalization factor tends to constant value
that is the product of the single-$\pi^+$ overlap factor,
$Z_{\pi\pi}(|{\bf x}|;k)\rightarrow Z_\pi(k)^2$, and the non-local,
energy-dependent kernel tends to zero, $ U_k({\bf x},{\bf
y})\rightarrow 0$, leaving the hadronic wavefunction,
$\psi_{\pi\pi}(|{\bf x}|;k)$, to satisfy the Helmholtz equation,
$\left(\ \nabla^2 \ +\ k^2\ \right) \psi_{\pi\pi}(|{\bf x}|;k)=0$.  As
the $A_1$ representation of the cubic-group projects onto angular
momentum $l=0,4,6,8,..$, at low-momentum only s-wave scattering is
significant and outside the range of the interaction the wavefunction
becomes $\psi_{\pi\pi}(|{\bf x}|;k) \ =\ A_0\ j_0(k |{\bf x}|) + B_0\
n_l (k |{\bf x}|)$.  The scattering phase shift is simply determined
from $\tan\delta_0(k) \ =\ -{B_0/A_0}$.  The CP-PACS collaboration has
claimed that the uncertainty in the scattering length extracted from
LQCD calculations of the asymptotic behavior of the $\pi^+\pi^+$
wavefunction is approximately two-thirds that of the extraction from
the energy eigenvalue via the usual
L\"uscher-method~\cite{Aoki:2004wq,Aoki:2005uf}.
 
One can define an energy- and sink-dependent potential,
defined at one energy (the energy of the two $\pi^+$'s determined in
the finite volume) through an uncontrolled modification to
Eq.~(\ref{eq:Upipidef}),
\begin{eqnarray}
{
\left(\ \nabla^2 \ +\ k^2\ \right)\phi(|{\bf x}|;k) 
\over
m_\pi\ \phi(|{\bf x}|;k) }
 & = & 
V_k(|{\bf x}|)
\ \ ,
\label{eq:Vpipi}
\end{eqnarray}
but it is clear that, except in the
case of infinitely massive hadrons~\cite{Detmold:2007wk}, 
$V_k(|{\bf x}|)$ contains no
more information than the phase shift at the energy determined with
L\"uscher's method~\footnote{
The phase-shift is recovered by solving
\begin{eqnarray}
\left(\ -\nabla^2 + m_\pi\ V_k(|{\bf x}|)\ \right)\phi(|{\bf x}|;k) \ =\ 
k^2\ \phi(|{\bf x}|;k) 
\ \ ,
\label{eq:Vsolve}
\end{eqnarray}
at the calculated value of $k^2$.}.  
One such potential for
nucleon-nucleon scattering was calculated in quenched QCD in
Ref~\cite{Ishii:2006ec} but, for the reasons detailed above, it cannot
be used as an input in nuclear calculations and it cannot be
meaningfully compared to phenomenological nucleon-nucleon potentials~\footnote{
The information content of the potentials derived from the analysis
of hadron-hadron wavefunctions calculated with Lattice QCD is displayed by the
following modifications to the method of 
Ref~\cite{Ishii:2006ec}.  One could use wall-sinks for the two hadrons, 
$\overline{\cal J}^\dagger (0,t)$, by 
projecting each quark to have zero momentum.
The interpolating operator has the same quantum numbers as that used 
in Ref~\cite{Ishii:2006ec} and the associated correlation function
\begin{eqnarray}
\overline{G}({\bf x},t)
& = & 
\langle 0|\  
\overline{\cal J}^\dagger (0,t) \overline{\cal J} (0,0) \ |0\rangle
\ \ \ ,
\end{eqnarray}
is uniform over the spatial volume at each time-slice, and gives rise to
a potential that is also uniform on each time-slice:
$V_k(|{\bf x}|)= k^2/m$.  
This provides a clear demonstration of the fact that potentials extracted by
such methods depend upon the choice of sink operator and are not unique
predictions of QCD.
}.  
An investigation of the sink dependence of these potentials 
has been performed and presented in a set of 
lectures~\cite{AokiSINK}. In this study, block spin 
transformations were applied to the quark fields used in the 
calculations and ${\cal O}(1)$ modifications to the potentials were found 
out to a distance of $\sim 0.5$~fm. 
Analogous energy- and sink-dependent potentials have been
constructed for hyperon-nucleon scattering as well~\cite{Nemura:2008sp}.

%%%%%%%%%%%%%%%%%%%%%%%%%%%%%%%%%%%%%
\section{Two-Body Physics}

%%%%%%%%%%%%%%%%
\subsection{\it Meson-Meson Interactions}

\noindent The low-energy scattering of pions and kaons, the
pseudo-Goldstone bosons of spontaneous chiral symmetry breaking,
provides a perfect testing ground for Lattice QCD calculations of
scattering parameters. There is little or no signal-to-noise problem 
in such calculations
and
therefore highly accurate Lattice QCD calculations can be performed
with moderate resources. Moreover, the EFTs which describe the
low-energy interactions of pions and kaons, including lattice-spacing
and finite-volume effects, have been developed to non-trivial orders
in the chiral expansion. 
\begin{figure}[tb]
\begin{center}
\begin{minipage}[t]{8 cm}
\centerline{\includegraphics[scale=0.5]{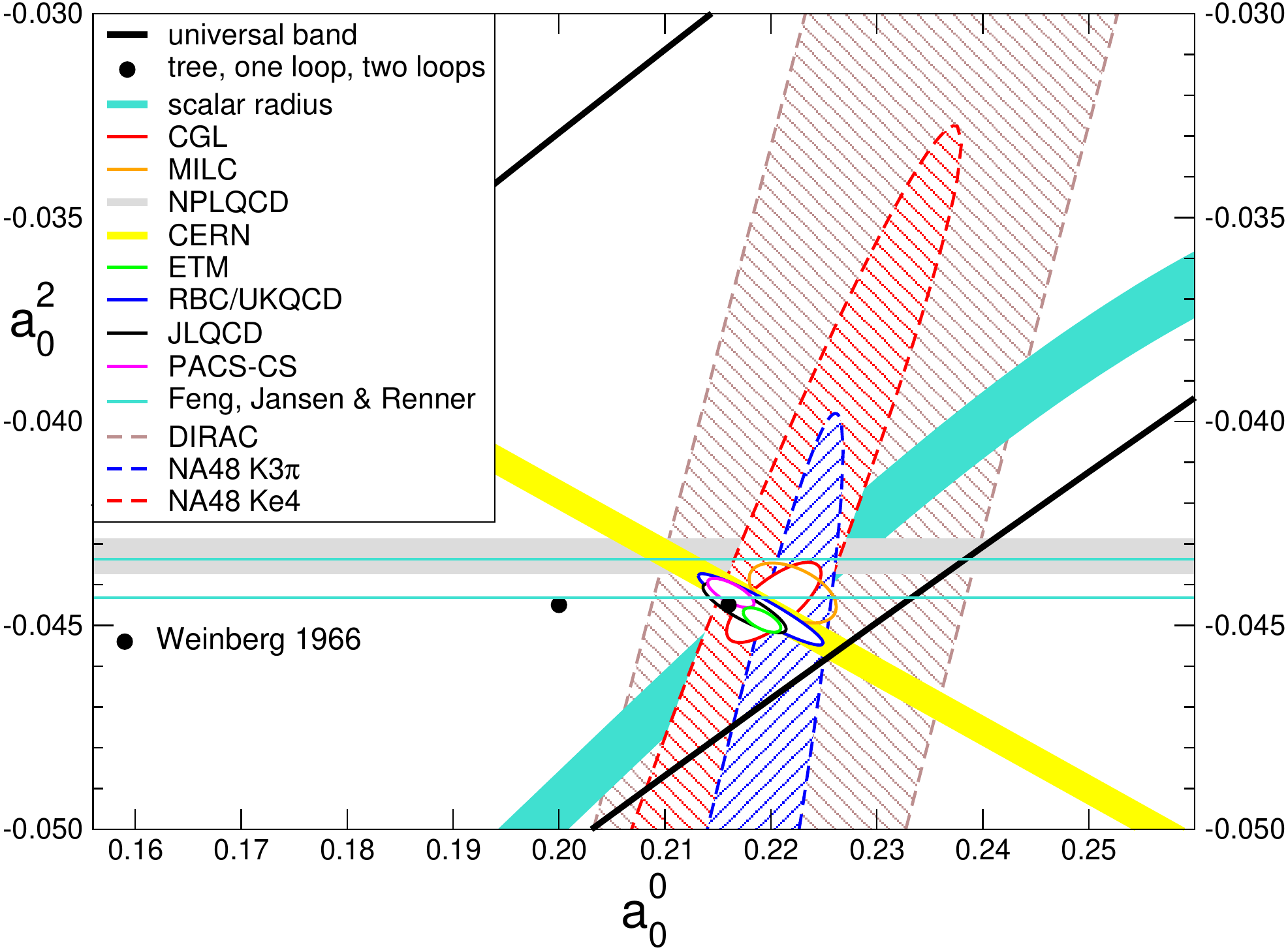}\hspace{0.4cm}\includegraphics[scale=0.31]{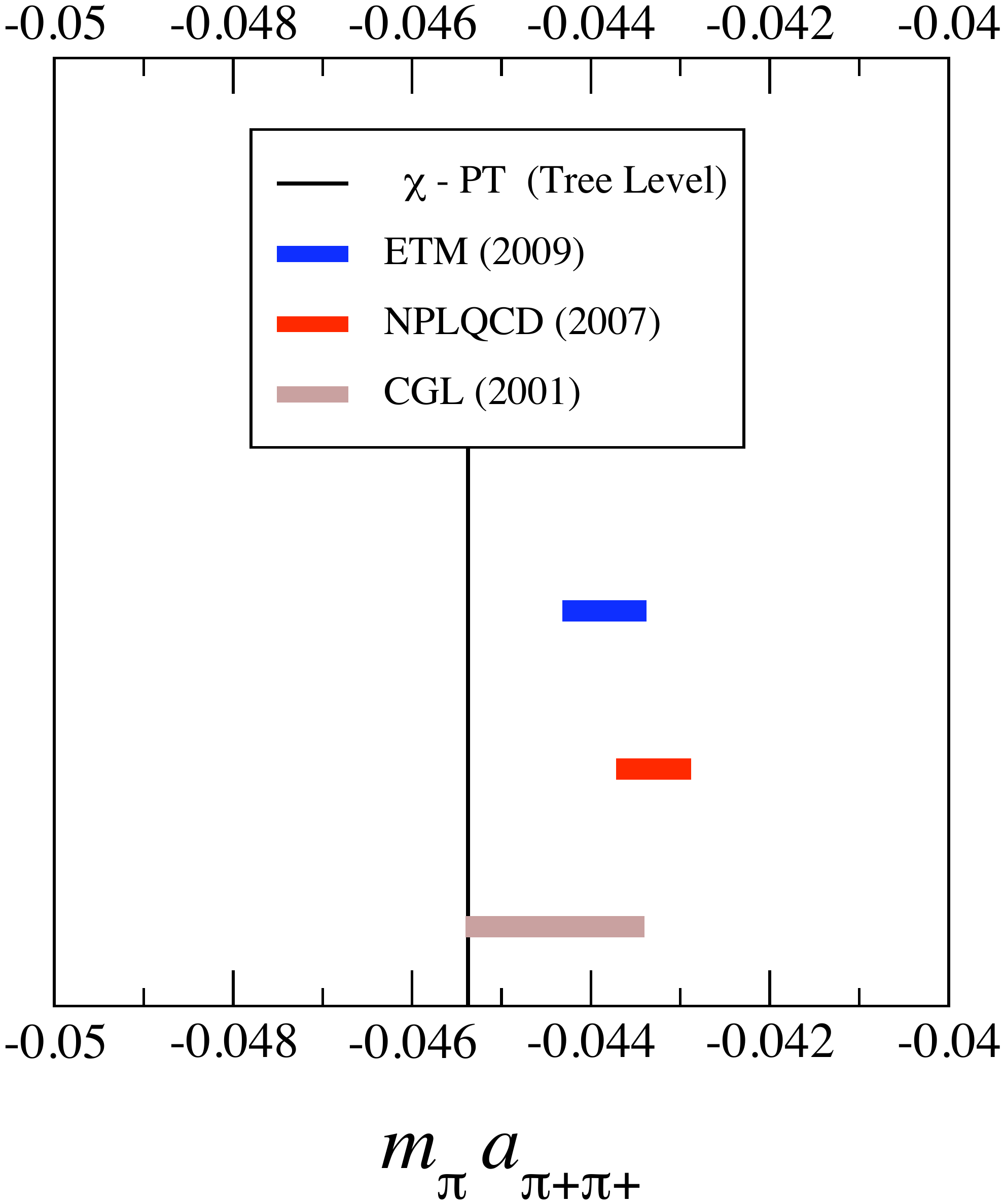}}
\end{minipage}
\begin{minipage}[t]{16.5 cm}
\caption{
Present constraints on threshold s-wave $\pi\pi$ scattering. 
Noteworthy in the left panel~\protect\cite{Leutwyler:2008fi} are
the red hatched ellipse from the Roy equation analysis and 
the grey band from the direct Lattice QCD calculation of
the $\pi^+\pi^+$ scattering length, as discussed in the text.
The right panel shows the $\pi^+\pi^+$ scattering length results only 
(the ETM result is at the top, 
the NPLQCD result is in the middle and the CGL result is at the 
bottom).
\label{fig:pipi2}}
\end{minipage}
\end{center}
\end{figure}

The $I=2$ pion-pion ($\pi^+\pi^+$) scattering length serves as a
benchmark calculation with an accuracy that can only be aspired to at
present for other systems.  Furthermore, due to the chiral symmetry of
QCD, $\pi\pi$ scattering at low energies is the simplest and
best-understood of the hadron-hadron scattering processes.  The scattering
lengths for $\pi\pi$ scattering in the s-wave are uniquely predicted
at LO in $\chi$-PT~\cite{Weinberg:1966kf}:
\begin{eqnarray}
m_\pi a_{\pi\pi}^{I=0} \ = \ 0.1588 \ \ ; \ \ m_\pi a_{\pi\pi}^{I=2} \ = \
-0.04537 
\ \ \ ,
\label{eq:CA}
\end{eqnarray}
when $m_\pi$ is set equal to the charged pion mass. While experiments 
do not directly provide stringent
constraints on the scattering lengths, a determination of s-wave
$\pi\pi$ scattering lengths using the Roy equations has reached a
remarkable level of
precision~\cite{Colangelo:2001df,Leutwyler:2008fi}:
\begin{eqnarray}
m_\pi a_{\pi\pi}^{I=0} \ = \ 0.220\pm 0.005 \ \ ; \ \ m_\pi a_{\pi\pi}^{I=2} \ = \ -0.0444\pm 0.0010
\ \ \ .
\label{eq:roy}
\end{eqnarray}
The Roy equations~\cite{Roy:1971tc} use dispersion theory to relate
scattering data at high energies to the scattering amplitude near
threshold. At present, Lattice QCD can compute $\pi\pi$ scattering
only in the $I=2$ channel with precision 
as the $I=0$ channel contains disconnected
diagrams which require large computational resources. 
It is of great interest to compare the precise Roy
equation predictions with Lattice QCD
calculations. Figure~\ref{fig:pipi2} summarizes theoretical and
experimental constraints on the s-wave $\pi\pi$ scattering
lengths~\cite{Leutwyler:2008fi}. It is clearly a strong-interaction
process where theory has somewhat out-paced the very challenging experimental
measurements.

The only existing $n_f=2+1$ Lattice QCD prediction of the $I=2$
$\pi\pi$ scattering length involves a mixed-action Lattice QCD scheme
of domain-wall valence quarks on a rooted staggered sea. Details of
the lattice calculation can be found in Ref.~\cite{Beane:2007xs}.
The scattering length was computed at pion masses, $m_\pi\sim 290~{\rm
  MeV}$, $350~{\rm MeV}$, $490~{\rm MeV}$ and $590~{\rm MeV}$, and at
a single lattice spacing, $b\sim 0.125~{\rm fm}$ and lattice size
$L\sim 2.5~{\rm fm}$~\cite{Beane:2007xs}. The physical value of the
scattering length was obtained using two-flavor mixed-action $\chi$-PT
which includes the effect of finite lattice-spacing
artifacts to $\mathcal{O}(m_\pi^2 b^2)$ and
$\mathcal{O}(b^4)$~\cite{Chen:2006wf}.  The final result is:
\begin{eqnarray}
m_\pi a_{\pi\pi}^{I=2} & = &  -0.04330 \pm 0.00042
\ \ \ ,
\label{eq:nplqcd2}
\end{eqnarray}
where the statistical and systematic uncertainties have been combined
in quadrature.  
Notice that this is a $1\%$ calculation, but it does rely on the assumption that
the rooting procedure used in the generation of the staggered gauge field
configurations is valid, and also relies on mixed-action chiral
perturbation theory at NLO to describe the lattice spacing artifacts. 
The agreement between this result and the Roy equation
determination is a striking confirmation of the lattice methodology,
and a powerful demonstration of the constraining power of chiral
symmetry in the meson sector. 
However, lattice calculations at one or more smaller lattice spacings are
required to verify and further refine this calculation.
It is also desirable that the scattering amplitude be calculated with 
mixed-action chiral perturbation theory to one higher order in the lattice
spacing in order to explore the convergence of this expansion.

\begin{figure}[tb]
\begin{center}
\begin{minipage}[t]{8 cm}
\centerline{\includegraphics[scale=0.35]{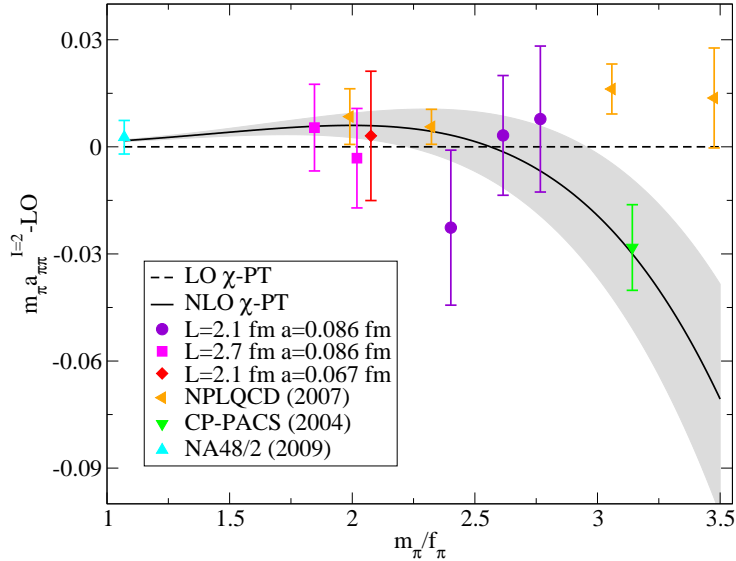}}
\end{minipage}
\begin{minipage}[t]{16.5 cm}
\caption{Lattice QCD calculations of the $\pi^+\pi^+$ scattering length
with the LO $\chi$-PT prediction 
removed~\protect\cite{Feng:2009ij}.
The shaded band
corresponds to the NLO $\chi$-PT fit to the lattice calculations.
The calculations with $L=2.1~{\rm fm}$ and $2.7~{\rm fm}$ were performed by the
ETM collaboration~\protect\cite{Feng:2009ij}. 
\label{fig:ETM-Tree-diff}}
\end{minipage}
\end{center}
\end{figure}

It is of great importance to have other Lattice QCD
determinations of the s-wave meson-meson scattering lengths which use
different lattice discretizations. 
Very recently, the ETM
collaboration has 
performed a $n_f=2$ calculation of
the $I=2$ $\pi\pi$ scattering length at
pion masses ranging from $m_\pi\sim 270~{\rm MeV}$ to $485~{\rm MeV}$,
at two lattice spacings ($b\sim 0.086~{\rm fm}$ and $b\sim 0.067~{\rm
fm}$) and at two lattice sizes ($L\sim 2.1~{\rm fm}$ and $L\sim
2.7~{\rm fm}$)~\cite{Feng:2009ij}. 
The result extrapolated to 
the physical pion mass is:
\begin{eqnarray}
m_\pi a_{\pi\pi}^{I=2} & = &  -0.04385 \pm 0.00028 \pm 0.00038
\ \ \ ,
\label{eq:TMpipi}
\end{eqnarray}
where the first uncertainty is statistical and the second is an estimate of
systematic effects. The agreement among the Lattice QCD calculations
and the Roy equation determination is striking.
In Figure~\ref{fig:ETM-Tree-diff}, 
the CP-PACS, NPLQCD and ETM Lattice QCD calculations 
of the $\pi^+\pi^+$ scattering length are
shown with the LO $\chi$-PT prediction subtracted. The shaded band
corresponds to the NLO $\chi$-PT fit to the lattice calculations.

\begin{figure}[!htb]
\begin{center}
\begin{minipage}[t]{8 cm}
\centerline{\includegraphics[scale=0.35,angle=0]{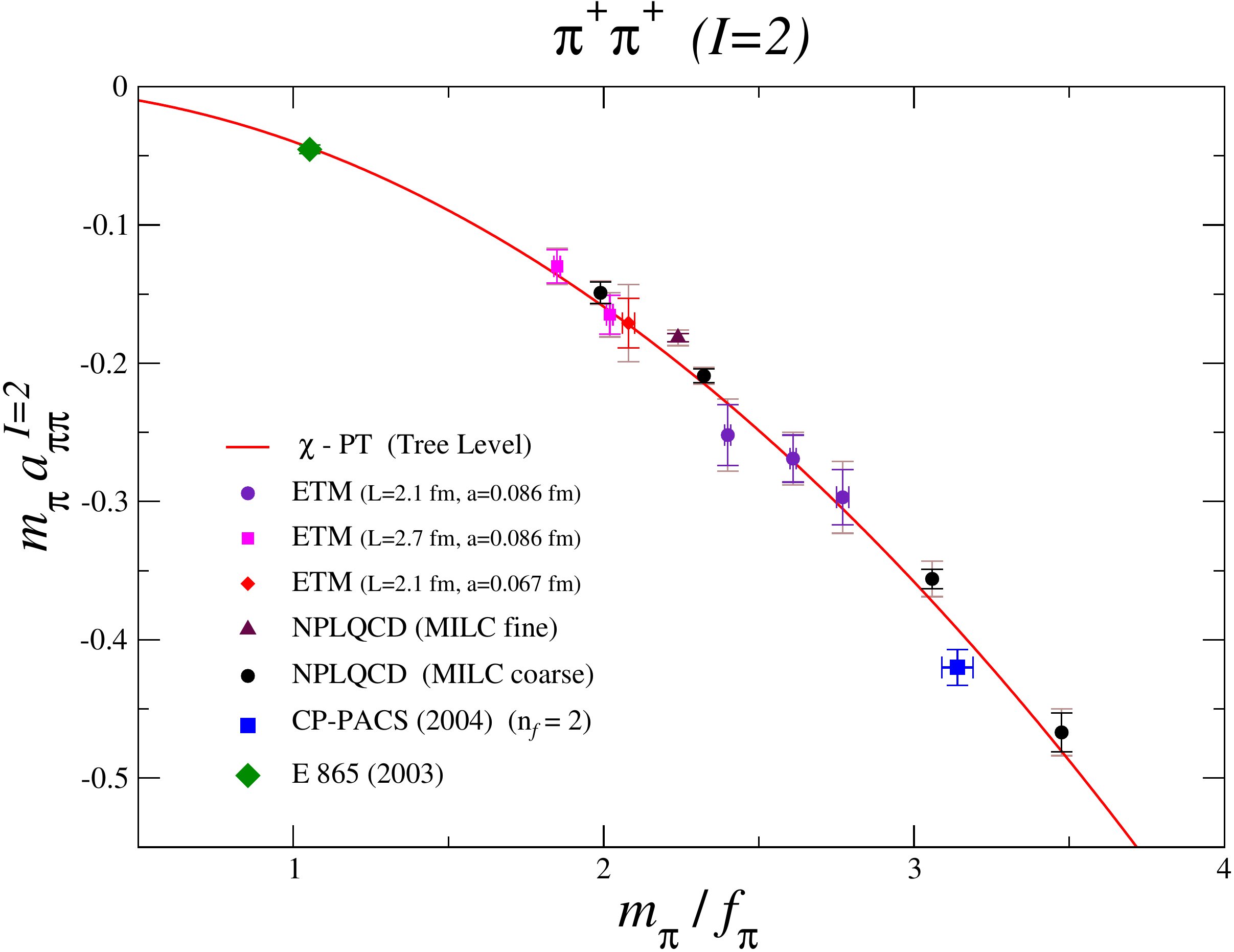}\hspace{0.54cm}\includegraphics[scale=0.35,angle=360]{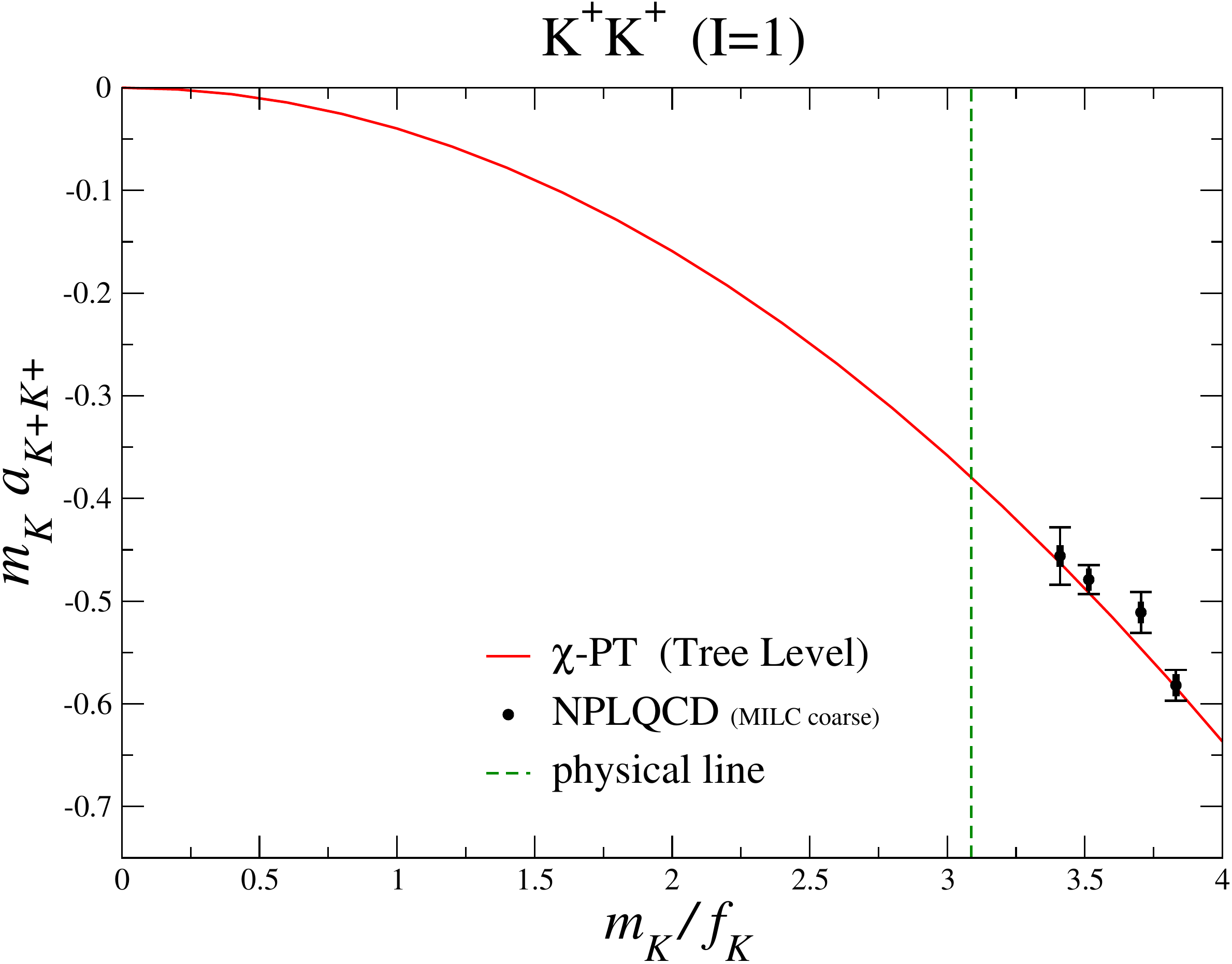}}
\end{minipage}
\begin{minipage}[t]{16.5 cm}
\caption{$m_\pi a_{\pi^+\pi^+}$ vs. $m_\pi/f_\pi$ (left panel)
and $m_k a_{K^+K^+}$ vs. $m_k/f_k$ (right panel). 
The black (darker) circles and dark-brown triangles
are the results of  Lattice QCD calculations by 
the NPLQCD collaboration, 
the blue (darker) square (left panel) is 
from the CP-PACS collaboration,
the purple (lighter) circles, 
red diamonds and magenta (lighter) squares 
(left panel) 
are from the ETM collaboration.
The solid (red) lines are the current algebra predictions.
\label{fig:CAplots}}
\end{minipage}
\end{center}
\end{figure}

\begin{figure}[!htb]
\begin{center}
\begin{minipage}[t]{8 cm}
\centerline{ \includegraphics[scale=0.35]{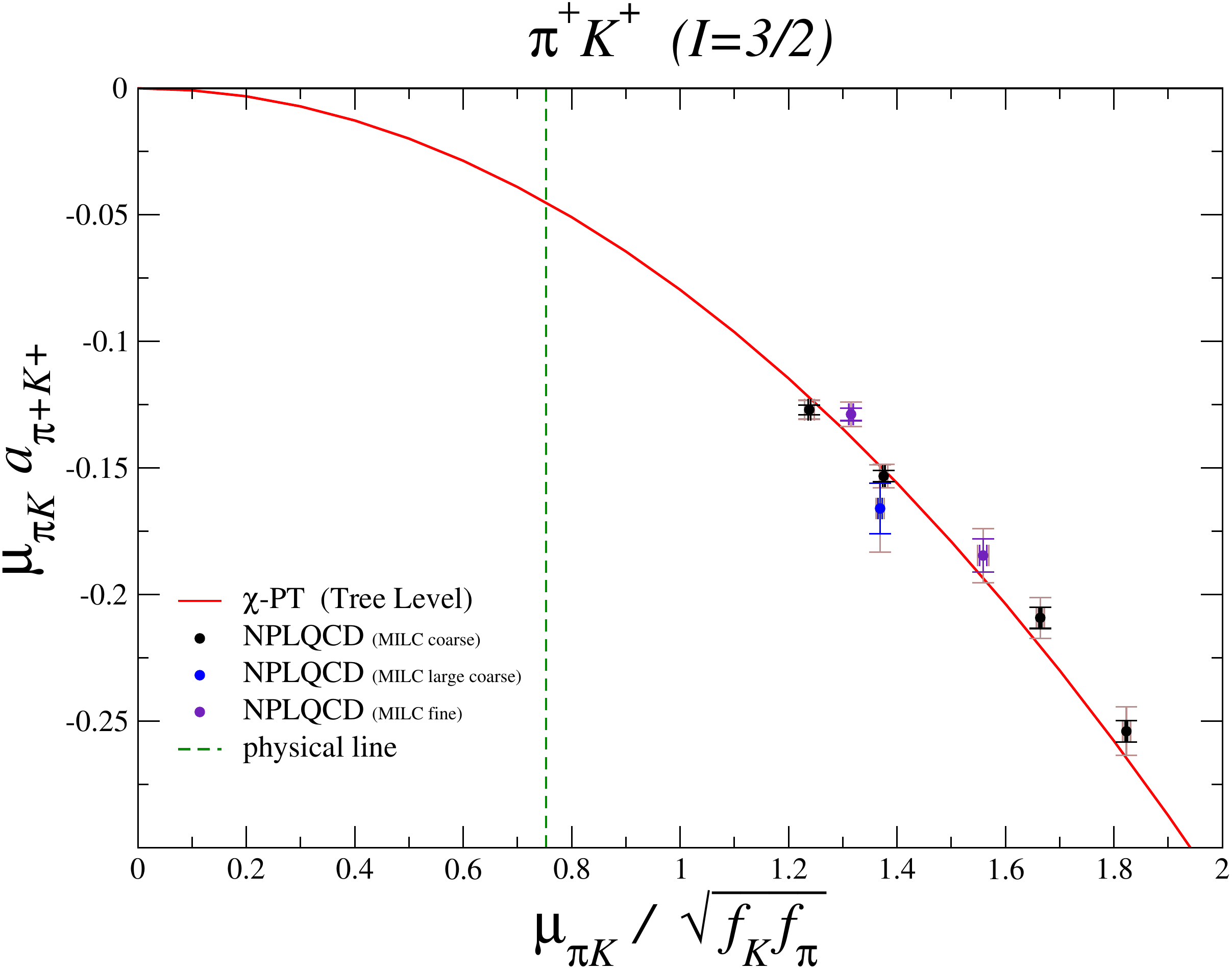} }
\end{minipage}
\begin{minipage}[t]{16.5 cm}
\caption{
$\mu_{K\pi}   a_{K^+\pi^+}$ vs. $\mu_{K\pi}/\sqrt{f_K f_\pi}$. The points
are the results of Lattice QCD calculations by the NPLQCD collaboration and the
solid (red) line is the current algebra prediction.
\label{fig:CAplotsKPi}}
\end{minipage}
\end{center}
\end{figure}

It is interesting to consider the pion mass dependence of the
meson-meson scattering lengths~\footnote{
The $K^+K^+$ and $\pi^+K^+$ scattering
lengths have also been computed by the NPLQCD collaboration. We refer
the interested reader to Figure~\ref{fig:CAplots} and
Figure~\ref{fig:CAplotsKPi}, 
and to Ref.~\cite{Beane:2008dv} for details.
Exploratory calculations of $\pi K$ scattering in both isospin channels have
been recently performed using $n_f=2+1$
improved Wilson quarks~\cite{Sasaki:2009cz}, and two quenched studies
have also been performed~\cite{Miao:2004gy,Nagata:2008wk}.
} 
as compared to the current algebra
predictions.  In Figure~\ref{fig:CAplots} (left panel) one sees that the
$I=2$ $\pi\pi$ scattering length is consistent with 
the current algebra result up
to pion masses that are expected to be at the edge of the chiral
regime in the two-flavor sector. While in the two flavor theory one
expects fairly good convergence of the chiral expansion and, moreover,
one expects that the effective expansion parameter is small in the
channel with maximal isospin, the lattice calculations clearly imply a
cancellation between chiral logs and counterterms (evaluated at a
given scale). However, as one sees in Figure~\ref{fig:CAplots} (right
panel), the same phenomenon occurs in $K^+K^+$ where the chiral
expansion is governed by the strange quark mass and is therefore
expected to be much more slowly converging.  The $\pi^+K^+$ scattering
length exhibits similar behavior when $\mu_{K\pi}\ a_{K^+\pi^+}$ is
plotted against $\mu_{K\pi}/\sqrt{f_K f_\pi}$ as shown in 
Figure~\ref{fig:CAplotsKPi}.  This remarkable cancellation between
chiral logs and counterterms for the meson-meson scattering lengths is
quite mysterious.

One (very) naive interpretation of these results is that the
contributions to these quantities from higher order terms in the chiral
expansion are much smaller than one would naively anticipate, and this
is not just a result of cancellation between terms.  This leaves one
with the task of understanding the origin of this suppression of the
higher order
terms.  
The totality of precision results may
suggest that there is a dynamically induced length scale governing
the size of contributions beyond tree-level that has yet to be
understood.

%%%%%%%%%%%%%%%%%%%%%%%%
\subsection{\it Meson-Baryon Interactions}

\noindent Pion-nucleon scattering has long been considered a
paradigmatic process for the comparison of $\chi$-PT and experiment.
To this day, controversy surrounds determinations of the pion-nucleon
coupling constant and the pion-nucleon sigma term. The $K^- n$
interaction is important for the description of kaon condensation in
the interior of neutron stars~\cite{KaplanNelson}, and meson-baryon
interactions are essential input in determining the final-state
interactions of various decays that are interesting for standard model
phenomenology (see Ref.~\cite{Lu:1994ex} for an example). In
determining baryon excited states on the lattice, it is clear that the
energy levels that represent meson-baryon scattering on the
finite-volume lattice must be resolved before progress can be made
regarding the extraction of single-particle excitations.  

\begin{figure}[hb!]
\begin{center}
\centerline{\includegraphics[width=7 cm]{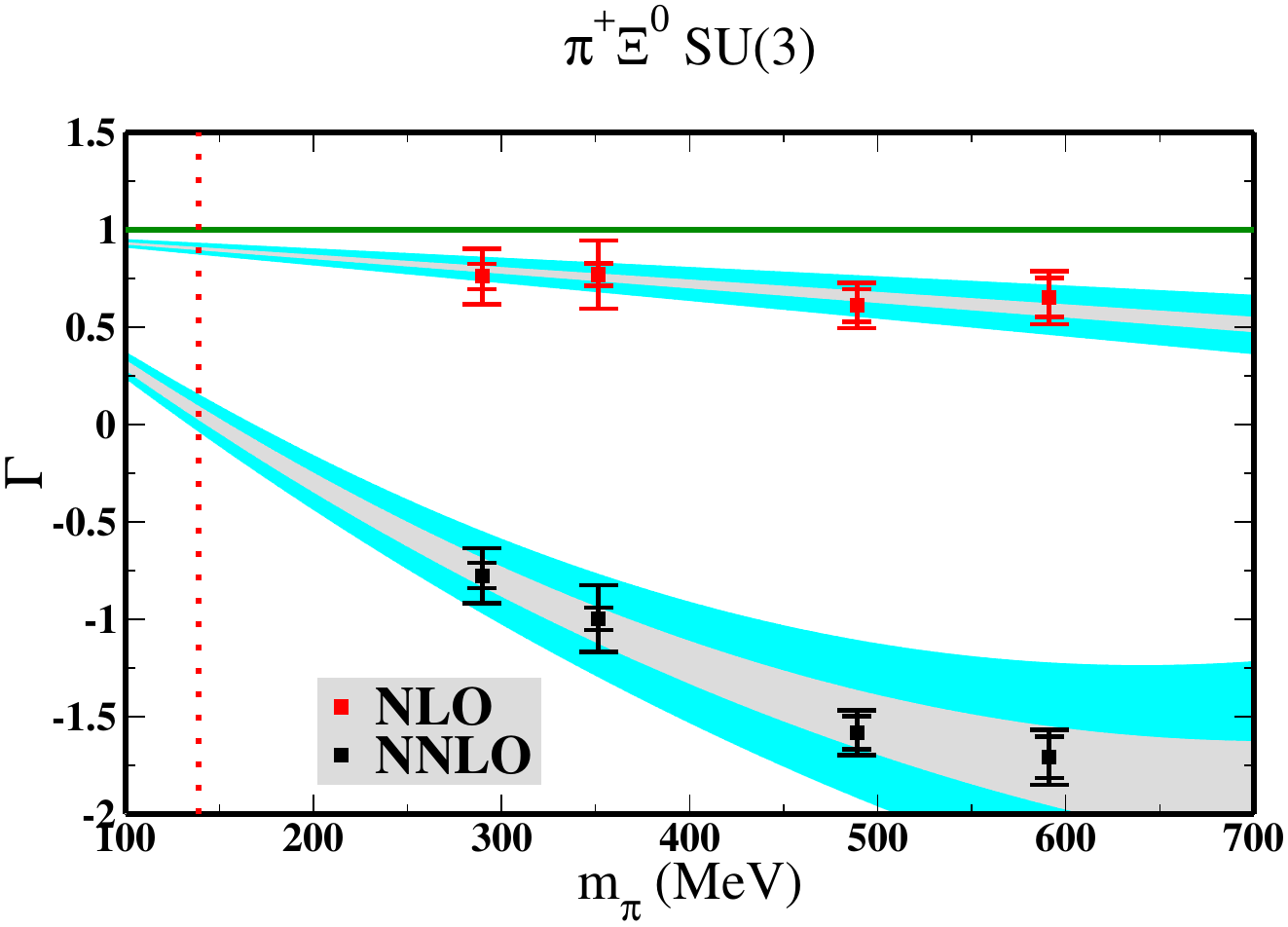}\ \ \ \ \includegraphics[width=7 cm]{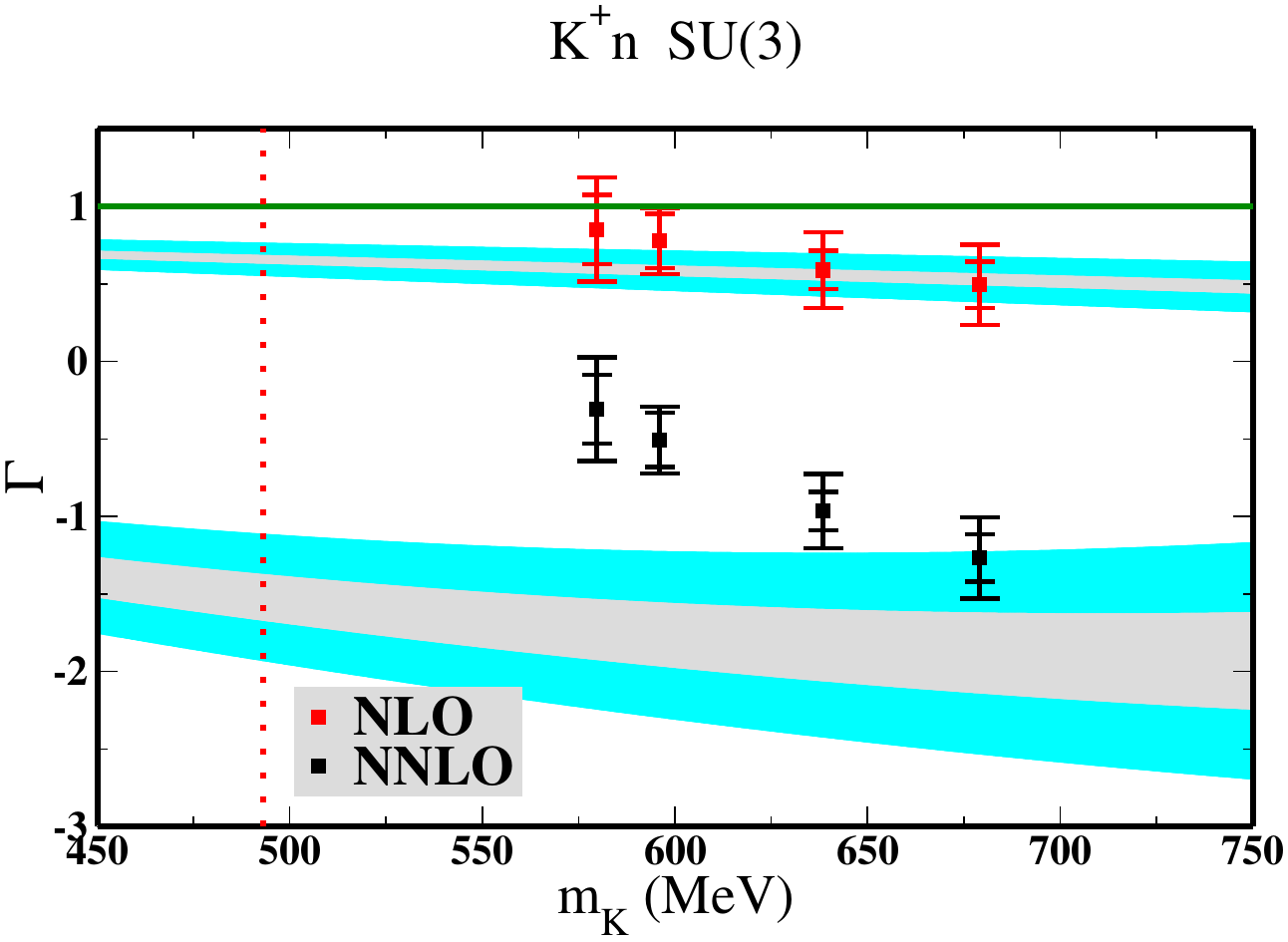}}
\caption{$\Gamma_{\rm NLO}$ and $\Gamma_{\rm NNLO}$ versus the
  meson masses for two of the meson-baryon processes in SU(3) HB$\chi$-PT. 
The line at $\Gamma=1$ is the LO curve, and dotted line is the 
physical meson mass. }
\label{fig:KostasNLONNLO}
\end{center}
\end{figure}

While pion-nucleon scattering is the best-studied meson-baryon
process, both theoretically and experimentally, its determination on
the lattice is computationally prohibitive since it involves
annihilation diagrams~\footnote{In recent work, the s-wave $\pi N$ phase-shift 
was extracted from the negative-parity single nucleon correlation 
function~\cite{Beane:2009kya}.
}. 
Combining the lowest-lying meson and baryon flavor octets, one can
form five meson-baryon elastic scattering processes that do not
involve annihilation diagrams: $\pi^+\Sigma^+$, $\pi^+\Xi^0$, $K^+ p$,
$K^+ n$, and ${\bar K}^0 \Xi^0$~\footnote{${\bar K}^0 \Sigma^+$ has
the same quantum numbers as $\pi^+\Xi^0$.}.  
Three of these processes
involve kaons and therefore are, in principle, amenable to an $SU(3)$
HB$\chi$-PT  analysis~\cite{Jenkins:1990jv}
for extrapolation. The remaining two processes involve pions
interacting with hyperons and therefore can be analyzed in conjunction
with the kaon processes in $SU(3)$ HB$\chi$-PT, or independently using
$SU(2)$ HB$\chi$-PT. 
An analysis of meson-baryon scattering using the mixed-action technology and 
resources was recently performed~\cite{Torok:2009dg}.
The scattering lengths of the five meson-baryon processes without
annihilation diagrams 
have been calculated to $\mathcal{O}(m_{\pi,K}^3)$ in $SU(3)$
HB$\chi$-PT~\cite{Liu:2006xja,Liu:2007ct}, and involve low-energy constants,
the $C$'s and the $h$'s, and the loop functions, $\mathcal{Y}$'s, 
which are given in
Ref.~\cite{Torok:2009dg}. 
The system of processes is found to be over-constrained, and 
multiple  fitting strategies
are possible, as discussed in Ref.~\cite{Torok:2009dg}.  
It is convenient to rewrite the
$\chi$-PT  formulas as polynomial expansions. 
For instance, 
in the case of $\pi^+\Xi^0$, the NLO and next-to-next-to-leading-order
(NNLO) polynomial expressions are 
\begin{eqnarray}
\Gamma_{\rm NLO}&&\equiv
-{4\pi f^2_\pi a_{\pi^+\Xi^0} \over m_\pi}
\left( 1 + {m_\pi\over M_\Xi}\right)\ =\ 
1-{ C_{01}} m_\pi 
\nonumber\\
\Gamma_{\rm NNLO}&&\equiv
-{4\pi f^2_\pi a_{\pi^+\Xi^0} \over m_\pi}
\left( 1 + {m_\pi\over M_\Xi}\right)
\ +\ {f^2_\pi\over m_\pi} \mathcal{Y}_{\pi^+\Xi^0}({ \Lambda_{\chi}})
\ =\ 1-{ C_{01}} m_\pi-8{ h_1(\Lambda_{\chi})} m_\pi^2 
\ \ \ .
\label{eq:gammadef}
\end{eqnarray}
The left-hand sides of these equations are given entirely
in terms of lattice-determined quantities, while the right-hand side provides a convenient
polynomial fitting function.
Figure~\ref{fig:KostasNLONNLO} shows the results of the Lattice QCD
calculation, along with the fits.
 The shift of the value of
$\Gamma$ from NLO to NNLO is dependent on the renormalization scale
$\mu$, and therefore with the choice $\mu=\Lambda_\chi$ one would
expect this shift to be perturbative if the expansion is
converging. The large shifts in $\Gamma$ from NLO to NNLO are
indicative of large loop corrections.
The low-energy constants (LECs) fit to the results of the 
lattice calculations are tabulated in
Ref.~\cite{Torok:2009dg}.  While the NNLO LECs $h_1$ and $h_{123}$
appear to be of natural size, the NLO LECs $C_0$ and $C_{01}$ are
unnaturally large. The extrapolated values of the five scattering
lengths are given in Table~\ref{tab:scattLpisig}. While the
$\pi^+\Sigma^+$ and $\pi^+\Xi^0$ scattering lengths appear to be
perturbative, the extrapolated kaon-baryon scattering lengths at NNLO
deviate by at least 100\% from the LO values.  The seemingly
inescapable conclusion is that the kaon-baryon scattering lengths are
unstable against chiral corrections in the three-flavor chiral
expansion, over the explored range of light-quark masses.
\begin{table}
\begin{center}
\begin{minipage}[t]{16.5 cm}
\caption{$SU(3)$ HB$\chi$-PT extrapolated scattering lengths. 
The first uncertainty is
  statistical, and the second is the statistical and systematic uncertainty
  added in quadrature. ``NLO (NNLO fit)'' indicates that
  the $C_1$ and $C_{01}$ from the NNLO fit to $\pi^+\Sigma^+$ and $\pi^+\Xi^0$
  have been used.
}
\label{tab:scattLpisig}
\end{minipage}
\begin{tabular}{|c|c|c|c|c|}
\hline
& & & & \\
Quantity & LO (fm) & NLO fit (fm) & NLO (NNLO fit) (fm) & NNLO (fm) \\
& & & & \\
\hline
& & & & \\
$a_{\pi\Sigma}$ & -0.2294 & -0.208(01)(03) & -0.117(06)(08) & -0.197(06)(08) \\
$a_{\pi\Xi}$ & -0.1158 & -0.105(01)(04) &  0.004(05)(11) & -0.096(05)(12) \\
$a_{Kp}$ & -0.3971 & -0.311(18)(44) &  0.292(35)(48) & -0.154(51)(63) \\
$a_{Kn}$ & -0.1986 & -0.143(10)(27) &  0.531(28)(68) &  0.128(42)(87) \\
$a_{K\Xi}$ & -0.4406 & -0.331(12)(31) &  0.324(39)(54) & -0.127(57)(70) \\
& & & & \\
\hline
\end{tabular}
%noalign{\smallskip\hrule}\cr}
\begin{minipage}[t]{16.5 cm}
\vskip 0.5cm
\noindent
\end{minipage}
\end{center}
\end{table}     

Given the poor convergence found in the three-flavor chiral expansion
due to the large loop corrections, it is natural to consider the
two-flavor theory with the strange quark integrated out. In this way,
$\pi\Sigma$ and $\pi\Xi$ may be analyzed in an expansion in
$m_\pi$. To $\mathcal{O}(m_\pi^3)$ in the two-flavor chiral expansion,
one has~\cite{Mai:2009ce}
\begin{eqnarray}
a_{\pi^+\Sigma^+}& = & \frac{1}{2 \pi}\frac{m_\Sigma}{m_\pi+m_\Sigma} 
\bigg[ -\frac{m_\pi}{f^2} + \frac{m_\pi^2}{f^2} {\mathrm{C}}_{\pi^+\Sigma^+} 
+ \frac{m_\pi^3}{f^2} h'_{\pi^+\Sigma^+} \bigg], \qquad
h'_{\pi^+\Sigma^+}=\frac{4}{f^2}\ell_4^r+ h_{\pi^+\Sigma^+} \ ,
\nonumber\\
a_{\pi^+\Xi^0} & = & \frac{1}{4 \pi}\frac{m_\Xi}{m_\pi+m_\Xi} 
\bigg[ -\frac{m_\pi}{f^2} + \frac{m_\pi^2}{f^2}{\mathrm{C}}_{\pi^+\Xi^0} 
+ \frac{m_\pi^3}{f^2} h'_{\pi^+\Xi^0} \bigg],\qquad
h'_{\pi^+\Xi^0}=\frac{4}{f^2}\ell_4^r + h_{\pi^+\Xi^0} \ ,
\label{eq:apixi2param}
\end{eqnarray}
where $\ell_4^r$ is the LEC which governs the pion mass
dependence of $f_\pi$~\cite{Colangelo:2001df}.  
\begin{figure}[ht]
\centerline{\includegraphics[width=0.4\linewidth]{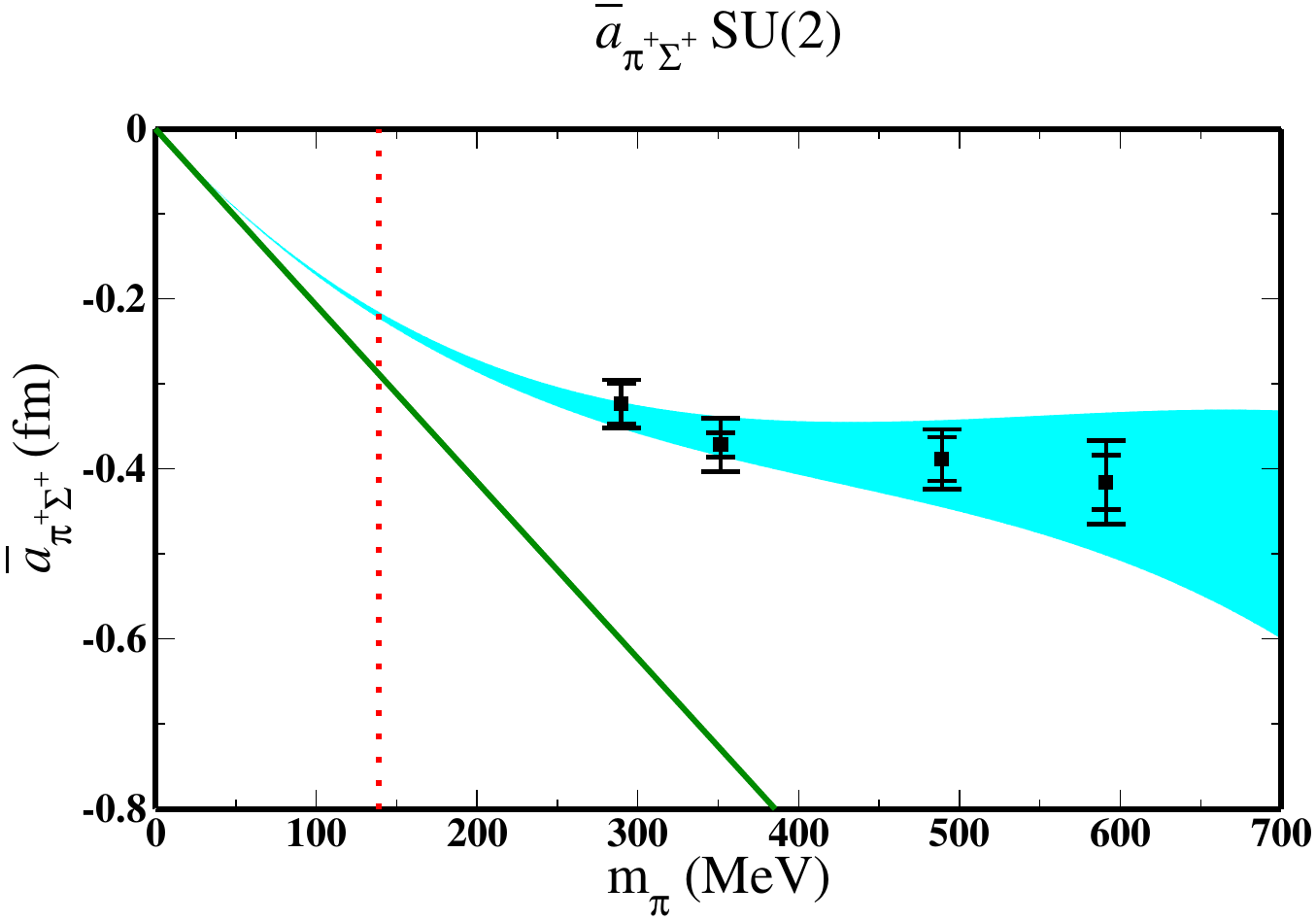}\ \ \ \
\includegraphics[width=0.4\linewidth]{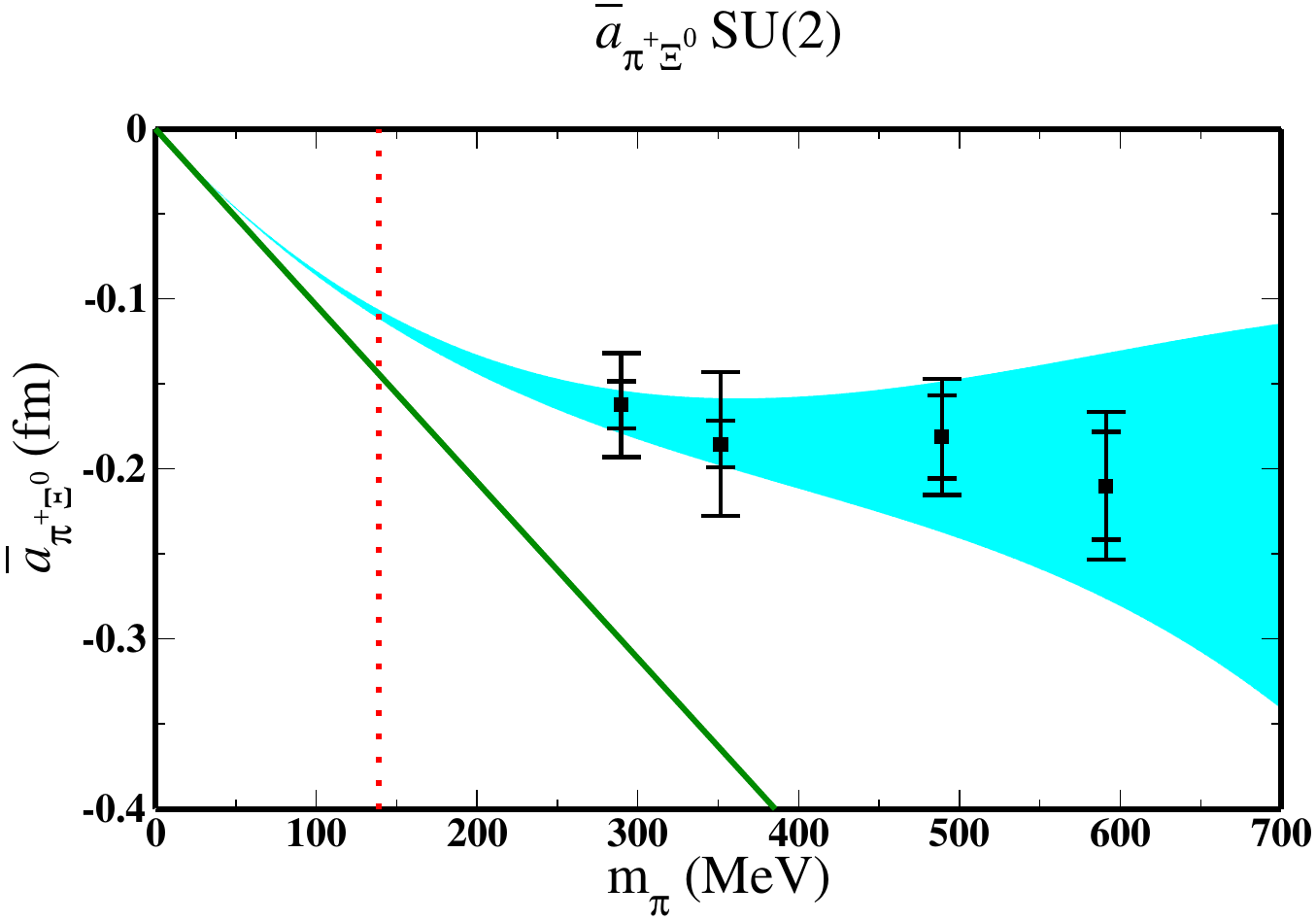}}
\caption{$\overline{a}$ versus the pion mass
for $\pi^+\Sigma^+$  and~$\pi^+\Xi^0$. The
  diagonal line is the LO curve, and the dotted line is the
  physical pion mass. The innermost error bar is the statistical uncertainty
  and the outermost error bar is the statistical and systematic uncertainty
  added in quadrature. The filled bands correspond to the fits to the LECs in the
  SU(2) case at NNLO.}
\label{fig:aSU2}
\end{figure}
Note that the chiral
logs have canceled, and in this form, valid to order
$m_\pi^3$ in the chiral expansion, the scattering lengths have a
simple polynomial dependence on $m_\pi$~\cite{Mai:2009ce}.  
Exploring the full 95\%
confidence interval error ellipse in the $h$-${\mathrm{C}}$ plane
yields pion-hyperon scattering lengths, extrapolated to the 
physical pion mass, of
\begin{eqnarray}
a_{\pi^+\Sigma^+}& = & -0.197 \pm 0.017~{\rm fm}
\ \ ,\ \ 
a_{\pi^+\Xi^0} \ =\  -0.096\pm 0.017~{\rm fm} 
\ \ \ ,
\label{eq:MP2}
\end{eqnarray}
and the scattering length versus the pion mass is shown in 
Figure~\ref{fig:aSU2}~\footnote{The bar denotes the scattering 
length rescaled by a kinematical factor~\cite{Torok:2009dg}.}.

The HB$\chi$-PT analyses of meson-baryon scattering 
support a general
observation about convergence in the three-flavor chiral expansion. 
As the pion masses considered in
the lattice calculation are comparable to the physical kaon mass, the
distinct convergence patterns of the two- and three-flavor chiral
expansions are suggestive that the breakdown in the
three-flavor case is not due to the relative largeness of the
strange-quark mass as compared to the light quark masses, but rather
due to some other enhancement in the coefficients of the loop
contributions, possibly related to a scaling with powers of $n_f$, the
number of flavors.

%%%%%%%%%%%%%%%%%%%%%%%%%%%%%%%%%%%%%%%%%%%%
\subsection{\it NN, YN and YY Interactions}

\noindent 
Perhaps the most studied and best understood of the two-hadron processes
are proton-proton and proton-neutron scattering.  In the S-wave, only two
combinations of spin and isospin are possible, a spin-triplet
isosinglet $np\ (^3S_1)$ and a spin-singlet isotriplet $pp\ (^1S_0)$.
At the physical pion mass, the scattering lengths in these channels
are unnaturally large and the $\siii-\diii$ coupled-channel contains a 
shallow bound
state, the deuteron, with a binding energy of $\sim 2.22~{\rm MeV}$.
These large scattering lengths and the shallow bound state are
described, in EFT, by the coefficient of the
momentum-independent four-nucleon operator having a non-trivial
fixed-point in its renormalization group flow
for the physical light-quark masses.  An
interesting line of investigation is the study of the scattering
lengths as a function of the quark masses to ascertain the sensitivity
of this fine-tuning to the QCD
parameters~\cite{Beane:2002vs,Beane:2002xf,Epelbaum:2002gb}.  The fine
tuning is not expected to persist away from the physical masses and we
expect present day (unphysical) Lattice QCD calculations to yield
scattering lengths that are natural-sized. More generally, it is
interesting to determine how the structure of nuclei depends upon the
fundamental constants of nature. In particular, we expect that any
nuclear observable is essentially a function of only five constants,
the length scale of the strong interactions, $\Lambda_{QCD}$, the
quark masses, $m_u$, $m_d$ and $m_s$, and the electromagnetic coupling
$\alpha_e$~\footnote{In the low-energy theory, the dependence on top, bottom
and charm quark masses is encapsulated in $\Lambda_{QCD}$.}.

The first study of baryon-baryon scattering with Lattice QCD was performed more
than a decade ago by Fukugita~{\it et
al}~\cite{Fukugita:1994na,Fukugita:1994ve}.  This calculation was
quenched and at relatively large pion masses, $m_\pi\gsim 550~{\rm
MeV}$.  Since this time, the dependence of the NN scattering lengths upon the
light-quark masses has been determined to various non-trivial orders
in the EFT expansion~\cite{Beane:2002vs,Beane:2002xf,Epelbaum:2002gb},
which is estimated to be valid up to $m_\pi\sim 350~{\rm
MeV}$. Therefore physical predictions of NN scattering parameters becomes
possible with Lattice QCD calculations that are performed
with pion masses less than $\sim350~{\rm MeV}$.

The NPLQCD collaboration performed the first $n_f=2+1$ QCD calculations
of nucleon-nucleon interactions~\cite{Beane:2006mx} and
hyperon-nucleon~\cite{Beane:2006gf} interactions at low-energies but with
unphysical pion masses, and  the nucleon-nucleon scattering lengths were found
to be of natural size. 
The fine-tunings at the physical values of the light-quark masses indicates
that 
Lattice QCD calculations with quark masses much
closer to the physical values (than today) 
are needed to extrapolate to the experimental
values. The results of the Lattice QCD calculation at the lightest
pion mass and the experimentally-determined scattering lengths at the
physical value of the pion mass were used to constrain the chiral
dependence of the scattering lengths from $m_\pi\sim 350~{\rm MeV}$
down to the chiral limit~\cite{Beane:2006mx}. 
However, these results suggest various
possible scenarios toward the chiral limit
which can only be resolved by way of Lattice QCD
calculations at lighter pion masses.  In contrast, very little is
known about the interactions between nucleons and hyperons from
experiment, and Lattice QCD calculations can provide the best
determinations of the corresponding scattering parameters and hence
determine the role of hyperons in neutron stars.

The energy eigenstates in the finite lattice volume are
classified by their global quantum numbers, baryon number, isospin,
third component of isospin, strangeness, total momentum, and behavior
under hyper-cubic transformations.  Six quark operators that are simple
products of three-quark baryon operators are (generally) used as sources for the
baryon-baryon correlation functions, but this is not a requirement.  
As a consequence,  the
baryon content of the interpolating operator is used to define the operator,
e.g. $n\Lambda(^3S_1)$, but this operator will, in principle,  couple
to all states in the volume with the quantum numbers $B=2$, $I={1\over
  2}$, $I_z=-{1\over 2}$, $s=-1$, and $^{2s+1}L_J =\; ^3S_1\ +\ ...$,
where the ellipses denote states with higher total angular momentum
that also project onto the $A_1$ irreducible representation of the
cubic-group~\footnote{ The spatial dimensions of the gauge field
  configurations that have been used to date for such calculations  
are identical, and as such the eigenstates of the QCD Hamiltonian
  can be classified with respect to their transformation properties
  under cubic transformations, $H(3)$, a subgroup of the group of
  continuous three-dimensional rotations, $O(3)$.  The two-baryon
  states that are calculated in this work all belong to the $A_1^+$
  representation of $H(3)$, corresponding to states with angular
  momentum $L=0,4,6,\ldots$\ .}.  Correlation functions for the nine
baryon-baryon channels shown in Table~\ref{tab:channels} have been
calculated, using both SS and SP correlators, as described in
Ref.~\cite{Beane:2009ky}.
\begin{table}
\begin{center}
\begin{minipage}[!ht]{16.5 cm}
\caption{
Baryon-baryon channels calculated in Ref.~\protect\cite{Beane:2009py}.
}
\label{tab:channels}
\end{minipage}
\begin{tabular}{|c|c|c|c|}
\hline
& & & \\
Channel   &   $I$   &  $I_z$  &   $s$   \\
& & & \\
\hline 
& & & \\
$pp$ ($^1S_0$) & $1$ & $+1$ & $0$   \\
$np$ ($^3S_1$) & $0$ & $0$ & $0$   \\
$n\Lambda$ ($^1S_0$) & ${1\over 2}$ & $-{1\over 2}$ & $-1$   \\
$n\Lambda$  ($^3S_1$) & ${1\over 2}$ & $-{1\over 2}$ & $-1$   \\
$n\Sigma^-$ ($^1S_0$) & ${3\over 2}$ & $-{3\over 2}$ & $-1$   \\
$n\Sigma^-$  ($^3S_1$) & ${3\over 2}$ & $-{3\over 2}$ & $-1$  \\
$\Sigma^-\Sigma^-$ ($^1S_0$) & $2$ & $-2$ & $-2$  \\
$\Lambda\Lambda$ ($^1S_0$) & $0$ & $0$ & $-2$   \\
$\Xi^-\Xi^-$ ($^1S_0$) & $1$ & $-1$ & $-4$   \\
& & & \\
\hline
\end{tabular}
%noalign{\smallskip\hrule}\cr}
\begin{minipage}[t]{16.5 cm}
\vskip 0.5cm
\noindent
\end{minipage}
\end{center}
\end{table}     
If the calculations were performed on gauge field configurations of
infinite extent in the time-direction, so that only forward
propagation could occur, some of the channels in
Table~\ref{tab:channels} could be analyzed by considering
contributions from a single scattering channel, e.g. $NN$,
$\Xi^-\Xi^-$, $\Sigma^-\Sigma^-$, $n\Sigma^-$, for which  a single,
well-separated ground state with these quantum numbers is expected. 
However, other
channels may require a multi-channel analysis, e.g. $n\Lambda$,
$\Lambda\Lambda$.  The $n\Lambda$ source will produce low-lying states
in the lattice volume that are predominately linear combinations of
the $n\Lambda$, $n\Sigma^0$ and $p\Sigma^-$ two-baryon states.  The
$\Lambda\Lambda$ source will produce low-lying states in the lattice
volume that are predominately linear combinations of the
$\Lambda\Lambda$, $\Sigma^{\pm,0}\Sigma^{\mp,0}$, and $N\Xi$
two-baryon states.

There are a number of ways to perform the statistical analysis of the
correlation functions and  determine the associated uncertainties.
One way is to use the Jackknife method to determine the uncertainty in
$p\cot\delta$ directly.  However, this is complicated by the
fact that ${\bf S}(x)$ in Eq.~(\ref{eq:energies}) is a singular function.  An
alternate method is to determine the value of $p_n^2$ and its
associated uncertainty from the two-baryon energy splitting from
Eq.~(\ref{eq:energieshift}), and then to propagate the central value and
$1\sigma$ uncertainties through Eq.~(\ref{eq:energieshift}) to determine
$p\cot\delta$. 

\begin{figure}[!ht]
\begin{center}
\begin{minipage}[t]{8 cm}
\centerline{\includegraphics[scale=0.65]{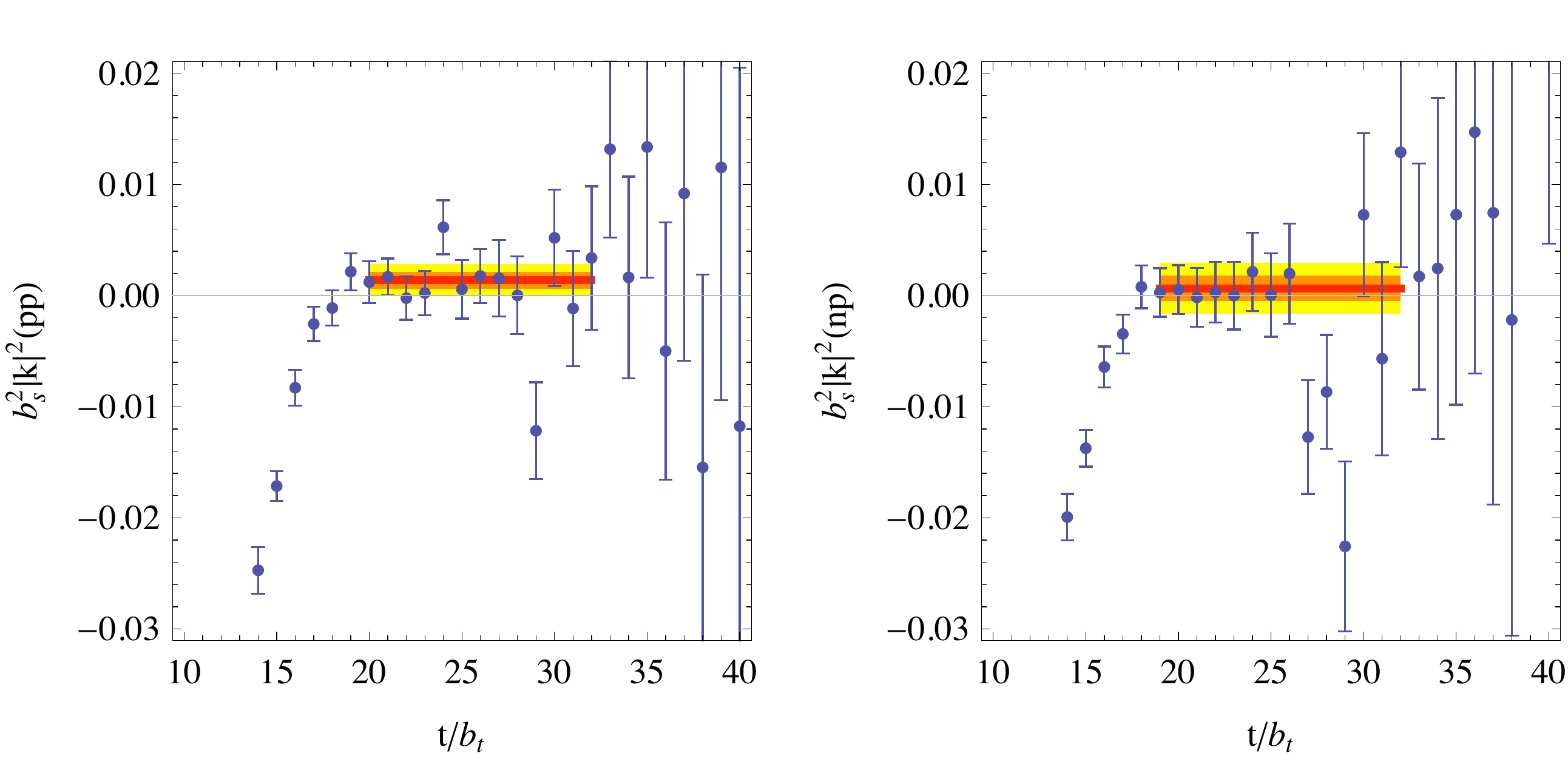}}
\end{minipage}
\begin{minipage}[t]{16.5 cm}
\caption{The left (right) panel is the effective $|{\bf k}|^2$ plot for the
    proton-proton $(^1S_0)$ (neutron-proton $(^3S_1)$)
channel and the fit to the
    plateau for the calculations on anisotropic clover gauge configurations 
in Ref.~\protect\cite{Beane:2009py}.  
\label{fig:NN-k2-BOT}}
\end{minipage}
\end{center}
\end{figure}

\begin{figure}[!ht]
\begin{center}
\begin{minipage}[t]{8 cm}
\centerline{\includegraphics[scale=0.35]{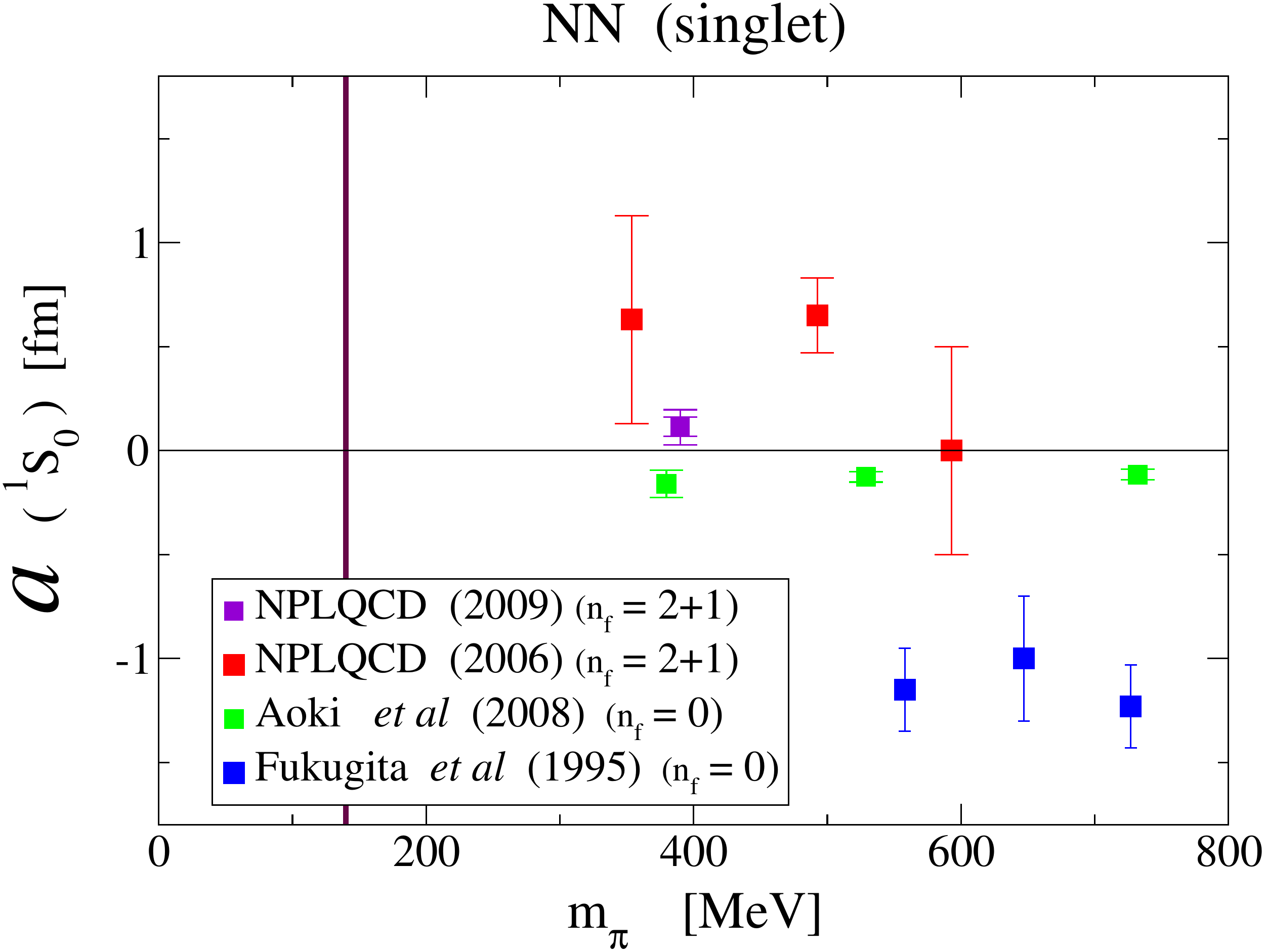}\includegraphics[scale=0.34]{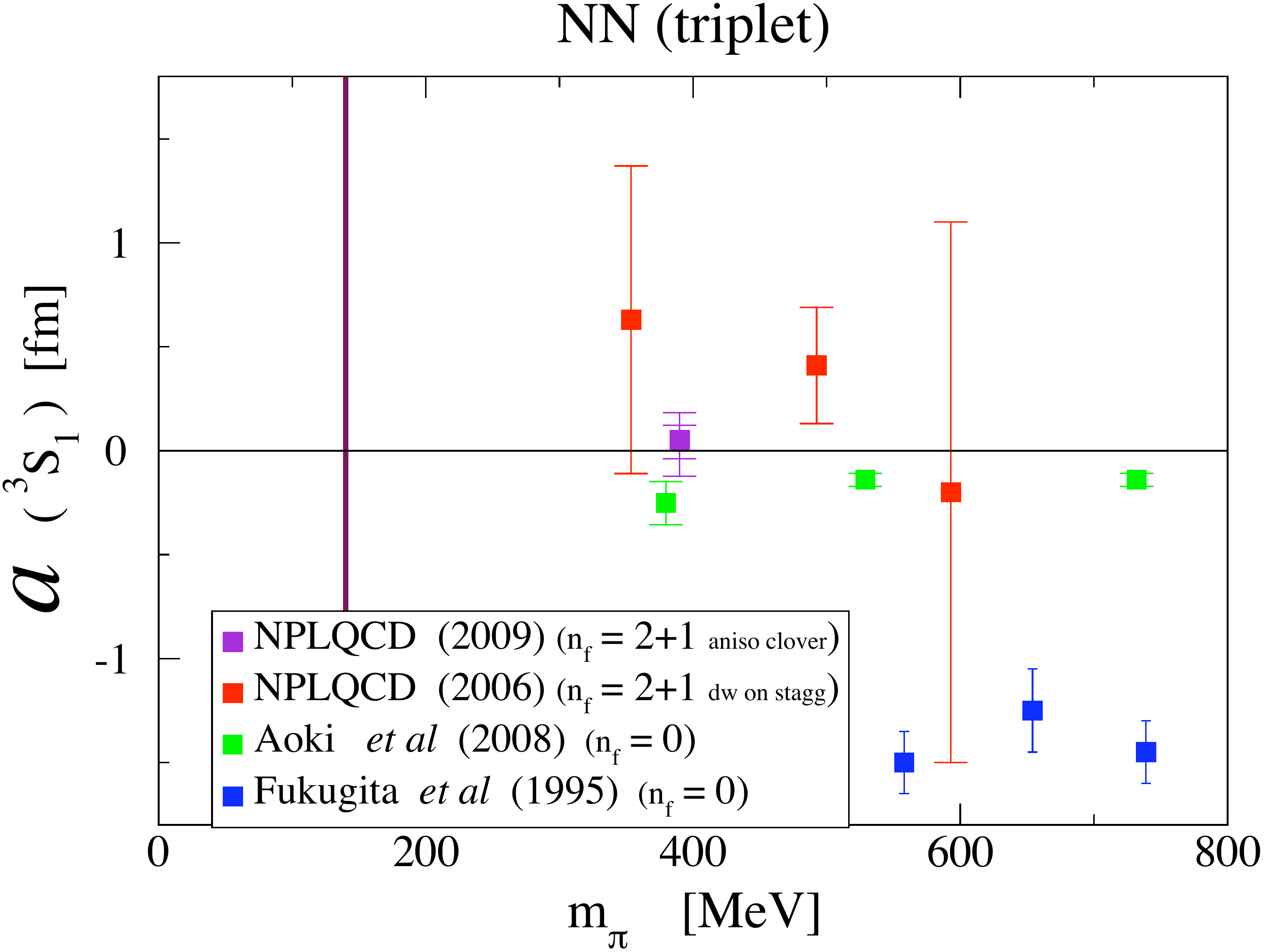}}
\end{minipage}
\begin{minipage}[t]{16.5 cm}
\caption{A compilation of the scattering lengths for NN scattering
    in the $^1S_0$ (left panel) and $^3S_1$ (right panel) calculated
    with Lattice QCD and with quenched Lattice QCD.  
The vertical lines correspond to
    the physical pion mass.
\label{fig:NN-ALL-LQCD}}
\end{minipage}
\end{center}
\end{figure}

An extensive exploration of
the impact of high-statistics on one-, two- and three-baryon correlation
functions on one ensemble of anisotropic clover gauge configurations
generated by the Hadron Spectrum
Collaboration was undertaken in
Refs~\cite{Beane:2009kya,Beane:2009gs,Beane:2009py}.  
A total of
$\gsim$440,000 sets of calculations were performed using $1200$ gauge
configurations of size $20^3\times 128$ with an anisotropy parameter
$\xi= b_s/b_t = 3.5$, a spatial lattice spacing of $b_s=0.1227\pm
0.0008~{\rm fm}$, and pion mass of $\mpi\sim 390~{\rm MeV}$.  The
ground state baryon masses (in lattice units) were extracted with
uncertainties that are at or below the $\sim 0.2\%$-level. 
Figure~\ref{fig:NN-k2-BOT} shows the effective $|{\bf k}|^2$ 
plot for both the proton-proton and neutron-proton channels.
Both channels exhibit plateaus in $|{\bf k}|^2$.  
While the plateau in the proton-proton channel differs from
zero at the $\sim 1$-$\sigma$ level, 
the plateaus in both 
channels are consistent with zero.  
We conclude that at this value of
the pion mass, the interactions between nucleons produce a small
scattering length in both channels compared to the naive estimate of
$m_\pi^{-1}\sim 0.5~{\rm fm}$.   
A summary of all Lattice QCD calculations of NN scattering is shown in
Figure~\ref{fig:NN-ALL-LQCD}.  
The results calculated 
on the anisotropic clover gauge configurations
are consistent with those obtained with
mixed-action Lattice QCD~\cite{Beane:2006mx}.  It is interesting to
note that the results of quenched calculations~\cite{Aoki:2008hh}
yield scattering lengths that are consistent within uncertainties with
the fully-dynamical $n_f=2+1$ values.
\begin{figure}[!htb]
\begin{center}
\begin{minipage}[t]{8 cm}
\centerline{\includegraphics[scale=0.7]{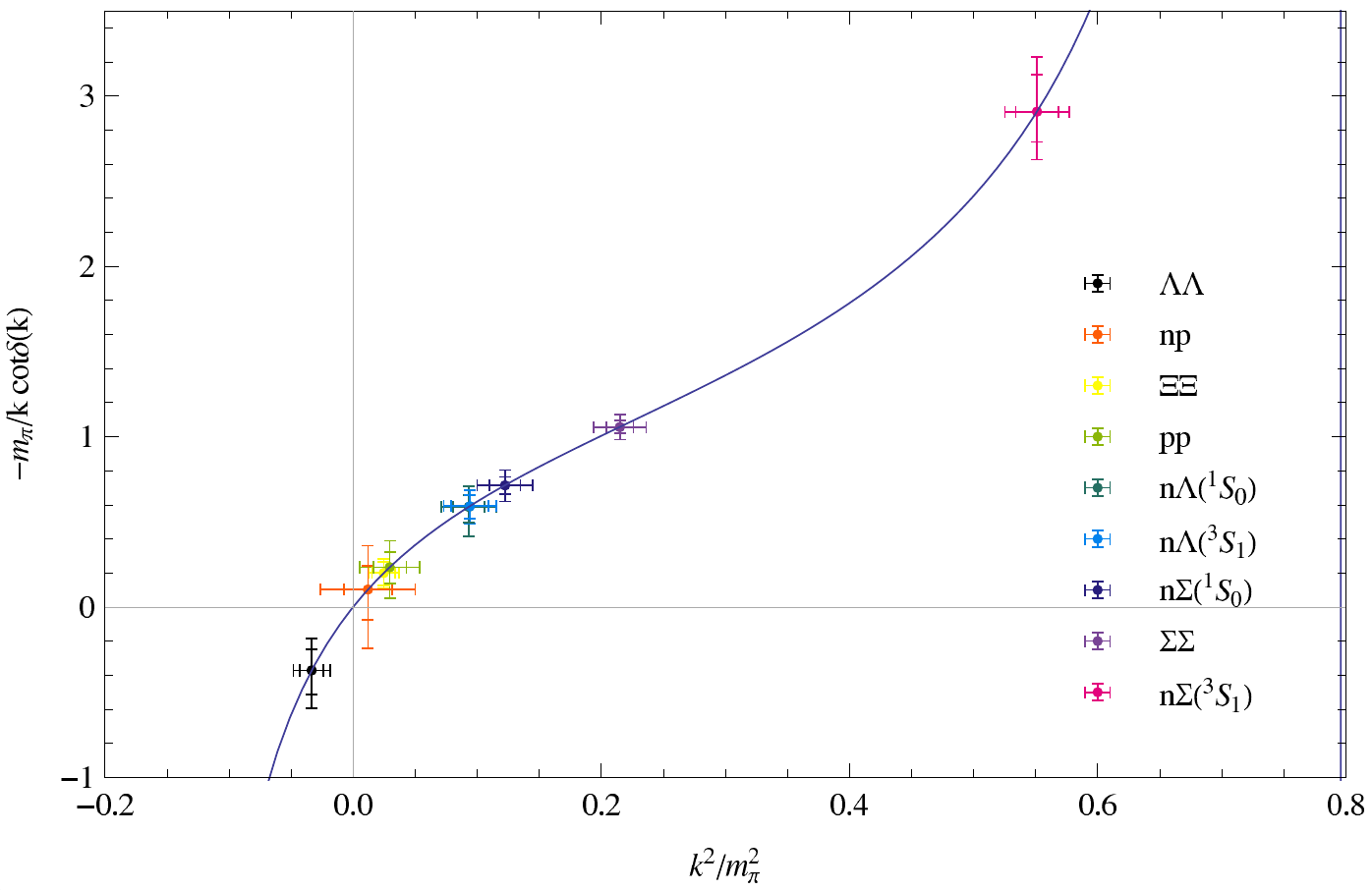}}
\end{minipage}
\begin{minipage}[t]{16.5 cm}
\caption{Baryon-baryon interactions extracted from calculations at 
$m_\pi\sim 390~{\rm MeV}$ and in a lattice volume of 
$20^3\times 128$~\cite{Beane:2009py}.
The top-most point of the
plot-legend corresponds to the left-most point on the plot, 
and the bottom- most point of the plot-legend corresponds
to the right-most point on the plot. The other points are ordered accordingly.
\label{fig:BB}}
\end{minipage}
\end{center}
\end{figure}
Figure~\ref{fig:BB} shows a summary of the high-statistics  
results for baryon-baryon
interactions 
obtained on the anisotropic clover gauge configurations.  The
$\Lambda\Lambda$ channel is found to be negatively shifted in energy which may
signal that the lowest-state is in fact a bound-state, but
calculations in a larger volume are required before more definitive
conclusions can be drawn. 
The calculations 
constitute an order of magnitude jump forward in the volume of 
output for this type of calculation.  The two additional lattice
volumes in which we are presently performing calculations at this pion
mass will allow for a systematic exploration of the volume dependence
of the scattering amplitude for all of the two-hadron systems.  In
principle, this will enable a separation of scattering and bound
states. However, calculations at this pion mass should be viewed to be only a
proof of principle and a testing ground for new analysis techniques.
In order for lattice calculations to provide meaningful constraints 
on interactions at the physical pion mass through
chiral extrapolation, the pion mass must be substantially reduced down
toward the physical value ($m_\pi\sim 140~{\rm MeV}$) while
maintaining the integrity of the calculation (i.e. small enough
lattice spacings and large enough volumes).

%%%%%%%%%%%%%%%%%%%%%%%%%%%%%%%%%%%%%
\section{Few-Body Physics}
\noindent
Computing resources and lattice algorithms have reached a stage where
it is now possible to consider more complicated hadronic observables
such as those in the baryon number, $B>2$ sector --- the domain of
nuclei. Here there are many observables 
about which very little (or nothing) is known
experimentally or theoretically, but which 
are phenomenologically
important to nuclear structure and interactions and to nuclear
astrophysics. Systems containing more than two  mesons are of
phenomenological interest in a number of areas from heavy ion
collisions at RHIC, to the equation of state of neutron
stars. In the last few years, there has been
a concerted effort to understand and develop the techniques needed for
studying systems of a few or many hadrons.

%%%%%%%%%%%%%%%%%%%%%%%%
\subsection{\it Multi-Meson Interactions}

\label{sec:npi}

As discussed previously, it has been known for many years how to exploit
the volume dependence of the eigen-energies of two-hadron systems to
extract infinite volume scattering phase shifts~\cite{Luscher:1986pf}
provided that the effective range of the interaction, $r$ is small
compared to the spatial extent of the lattice volume 
(since $r\sim m_\pi^{-1}$ for most
interactions, this constraint is $m_\pi\ L \gg1$). In recent works,
this has been extended to systems involving $n>2$ 
bosons~\cite{Beane:2007qr,Tan:2007bg,Detmold:2008gh} and 
$n=3$ fermions~\cite{Luu:2008fg} in the situation where the 
relevant scattering length,
$a$, is small compared to the spatial extent of the lattice.  
By performing a perturbative EFT 
calculation in a finite
volume, the ground state energy of a system of $n$ bosons was computed
in Refs.~\cite{Beane:2007qr,Tan:2007bg,Detmold:2008gh}. The
shift in energy of $n$ bosons of mass $M$ from the non-interacting
system is
\begin{eqnarray}
 \Delta E_n &=&
  \frac{4\pi\, \abar}{M\,L^3}\Choose{n}{2}\Bigg\{1
-\left(\frac{\abar}{\pi\,L}\right){\cal I}
+\left(\frac{\abar}{\pi\,L}\right)^2\left[{\cal I}^2+(2n-5){\cal J}\right]
\nonumber 
\\&&\hspace*{2cm}
-
\left(\frac{\abar}{\pi\,L}\right)^3\Big[{\cal I}^3 + (2 n-7)
  {\cal I}{\cal J} + \left(5 n^2-41 n+63\right){\cal K}\Big]
\nonumber
\\&&\hspace*{2cm}
+
\left(\frac{\abar}{\pi\,L}\right)^4\Big[
{\cal I}^4 - 6 {\cal I}^2 {\cal J} + (4 + n - n^2){\cal J}^2 
+ 4 (27-15 n + n^2) {\cal I} \ {\cal K}
\nonumber\\
&&\hspace*{4cm}
+(14 n^3-227 n^2+919 n-1043) {\cal L}\ 
\Big]
\Bigg\}
\nonumber\\
&&
+\ \Choose{n}{3}\left[\ 
{192 \ \abar^5\over M\pi^3 L^7} \left( {\cal T}_0\ +\ {\cal T}_1\ n \right)
\ +\ 
{6\pi \abar^3\over M^3 L^7}\ (n+3)\ {\cal I}\ 
\right]
\nonumber\\
&&
+\ \Choose{n}{3} \ {1\over L^6}\ \overline{\overline{\eta}}_3^L\ 
\ \ + \ {\cal O}\left(L^{-8}\right)
\ \ \ \ ,
\label{eq:energyshift}
\end{eqnarray}
where the parameter $\abar$ is related to the scattering length, $a$,
and the effective range, $r$, by
\begin{eqnarray}
a
& = & 
\overline{a}\ -\ {2\pi\over L^3} \overline{a}^3 r \left(\ 1 \ -\
  \left( {\overline{a}\over\pi L}\right)\ {\cal I} \right)\ \ .
\label{eq:aabar}
\end{eqnarray}
The geometric constants, ${\cal I},\ {\cal J},\ {\cal K},\ {\cal L},\,
{\cal T}_{0,1}$, that enter into Eq.~(\ref{eq:energyshift}) are
defined in Ref.~\cite{Detmold:2008gh} and $^nC_m$ are the binomial 
coefficients.  The
three-body contribution to the energy-shift given in
Eq.~(\ref{eq:energyshift}) is represented by the parameter
$\overline{\overline{\eta}}_3^L$ (see Ref.~\cite{Detmold:2008gh}).

Lattice QCD calculations of these energy shifts allow for an extraction of
the parameters $\abar$ and $\overline{\overline{\eta}}_3^L$. To
determine the energy shifts, the multi-meson correlation functions 
(specifying to the multi-pion system)
\begin{eqnarray}
C_n(t) 
 & \propto & \left\langle 
\left(\sum_{\bf x} \pi^-({\bf x},t)
\right)^n
\left( 
\phantom{\sum_x\hskip -0.2in}
\pi^+({\bf 0},0)
\right)^n
\right\rangle\
 \ \ ,
\label{eq:Cnfun}
\end{eqnarray}
are calculated.
On a lattice of infinite temporal extent,\footnote{Effects of temporal
  (anti-)periodicity are discussed in Ref.~\cite{Detmold:2008yn}.} the
combination
\begin{eqnarray}
G_n(t) 
&  \equiv & { C_n(t) \over \left[\ C_1 (t)\ \right]^n }
\ \stackrel{t\to\infty}{\longrightarrow}\ {\cal B}_0^{(n)}\ e^{- \Delta E_n\ t}
\ \ \ ,
\label{eq:Gnlarget}
\end{eqnarray}
allows for an extraction of the ground-state energy shift, $\Delta E_n$, which can
then be used as input into Eq.~(\ref{eq:energyshift}) to extract the scattering
and interaction parameters.
To compute the $(n!)^2$ Wick contractions in Eq.~(\ref{eq:Cnfun}),
the correlation function can be written as
\begin{eqnarray}
C_n(t) 
 & \propto & 
\langle \ \left(\ \overline{\eta} \Pi \eta\ \right)^n \ \rangle
\ \propto \ 
\varepsilon^{\alpha_1\alpha_2..\alpha_n\xi_1..\xi_{12-n}}\ 
\varepsilon_{\beta_1\beta_2..\beta _n\xi_1..\xi_{12-n}}\ 
\left(\Pi\right)_{\alpha_1}^{\beta_1} \left(\Pi\right)_{\alpha_2}^{\beta_2} 
.. \left(\Pi\right)_{\alpha_n}^{\beta_n} 
\ \ \ ,
\nonumber\\
\Pi
& = &  \sum_{\bf x} \ S({\bf x},t;0,0) \   S^\dagger({\bf x},t;0,0)
 \ \ ,
\label{eq:Cnfungrassman}
\end{eqnarray}
where $S({\bf x},t;0,0)$ is a light-quark propagator.  
The object
(block) $\Pi$ is a $N\times N$  bosonic
time-dependent matrix where $N=12$ 
($N=N_S\times N_C$ with $N_S=4$-spin and $N_C=3$-color), 
and $\eta_\alpha$ is a twelve component
Grassmann variable. 
Further simplifications are possible resulting in the
correlation functions being written in terms of traces of powers of $\Pi$.
As an example, the
contractions for the $3$-$\pi^+$ system give
\begin{eqnarray}
C_3(t) & \propto & 
{\rm tr_{C,S}}\left[ \Pi \right]^3
\ -\  3\  {\rm tr_{C,S}}\left[ \Pi^2 \right] {\rm tr_{C,S}}\left[\Pi\right]
\ +\  2\  {\rm tr_{C,S}}\left[ \Pi^3 \right]
\ \ \ ,
\label{eq:threePiCorrelator}
\end{eqnarray}
where the traces, ${\rm tr_{C,S}}$, are over color and spin indices.
Contractions for $n\le N$ $\pi^+$'s  are given explicitly in
Ref.~\cite{Detmold:2008fn}.

%%%%%%%%%%%%%%%%%%%%%%%%%
\subsection{\it Two- and Three- Body Interactions}
\label{sec:two-three-body}
\begin{figure}[!ht]
  \centering
\begin{minipage}[!ht]{8 cm}
\centerline{  \includegraphics[scale=1.1]{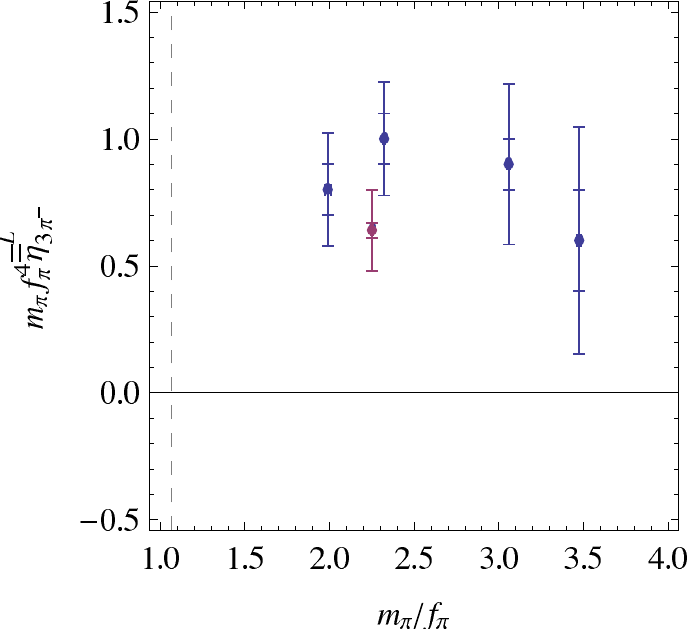}\\ \hspace*{1cm}\\
  \includegraphics[scale=1.1]{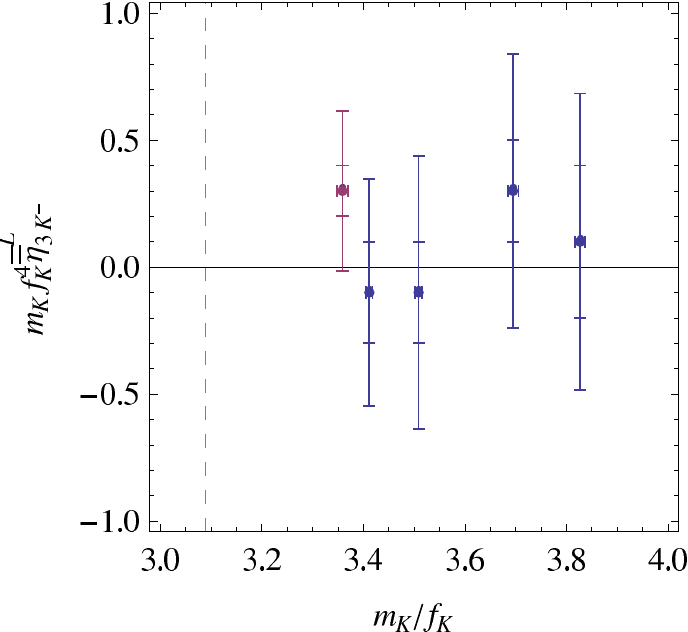}}
\end{minipage}
\begin{minipage}[t]{16.5 cm}
  \caption{Three pion (left-panel) and kaon (right-panel) 
interactions determined
    from the MILC coarse (lighter, blue) and fine (darker, magenta) 
lattices plotted as
    a function of the dimensionless ratios $m_\pi/f_\pi$ and $m_K/f_K$
    respectively. }
  \label{fig:three}
\end{minipage}
\end{figure}

\noindent
The NPLQCD collaboration has 
performed mixed-action Lattice QCD calculations of
the $n\leq 12$ pion and kaon
correlation functions~\cite{Detmold:2008yn,Detmold:2008fn,Beane:2007es}. 
In order to
correctly calculate these correlators for large $n$, very high
numerical precision is necessary and the ARPREC arbitrary precision
numerical library~\cite{ARPREC} was used. 
By performing a correlated
fit to the effective energy differences extracted from these
calculations,  the two- and three-body
interactions were determined~\cite{Detmold:2008yn,Detmold:2008fn,Beane:2007es},
and  the two-body interactions
agree with those extracted from the two-body sector 
alone~\cite{Beane:2007xs}. The resulting three-body interactions are
displayed in Figure~\ref{fig:three}. The $3$-$\pi^+$ interaction is found
to be repulsive with a magnitude consistent with the expectation from
naive dimensional analysis (tree-level $\chi$-PT). 
In contrast, the three $K^+$ interaction
is consistent with zero within somewhat larger uncertainties.

%%%%%%%%%%%%%%%%%%%%%%%%%%%%%%%%%%%%%%%
\subsection{\it Mixed Species Meson Systems}
\label{sec:mixed-species-meson}

In the case of a mixed system comprising $n$ mesons of one type and
$m$ mesons of a second type, the results summarized in Section~\ref{sec:npi}
have been generalized by Smigielski and 
Wasem~\cite{Smigielski:2008pa}, working to ${\cal O}(1/L^6)$. The energy
shift of the interacting system from the free system depends on three
two-body interaction parameters and four three-body interaction
parameters. As might be expected, the full form of the energy shift is
cumbersome and we refer the interested reader to the original
paper. 
These expressions are currently being used to analyze LQCD calculations of the 
energies of
systems of pions and kaons~\cite{BJS}. Correlation functions are calculated
in a way that is similar to the way that the multi-pion correlation 
functions are calculated.

%%%%%%%%%%%%%%%%%%%%%%%%%%%%%%

\subsection{\it Contractions for Large ($N>12$) Systems of Mesons}
\label{sec:contr-large-n12}

In order to go beyond $n=12$ meson systems, or to study systems of
many different types of mesons, it is necessary to make use of more
efficient methods of performing the contractions of quark
fields. While better than the na\"ive factorial construction (which
scales as $n!^2$), the construction used in Section~\ref{sec:npi} scales
poorly to large numbers of mesons, behaving at best as $n!^{1/2}$
(provided the matrix $\Pi$ is generalized to give a nonzero
result). In Ref.~\cite{Detmold:2010au}, a recursive method for performing these
contractions was developed that allows the extension of the study of
meson systems to larger $n$ and also greatly simplifies the
contractions required for systems of many different species of
mesons. We will summarize the construction by first considering the
recursive approach to the contractions for a single species of meson
before reporting the general case.

By rescaling the correlation functions of the single-species, single-source
system considered previously as 
\begin{eqnarray}
  C_{n\pi^+}(t)   & = &   (-)^n\ n!\ \langle\ R_n\ \rangle
  \ \ \ \ ,
  \label{eq:npiDET}
\end{eqnarray}
(the angle brackets denote a trace over spin and color indices) it is
straightforward to show that the objects, $R_n$ satisfy an ascending
recursion 
\begin{eqnarray}
  R_{n+1} & = & \langle\ R_n\ \rangle\ \Pi\ -\ n\ R_n\ \Pi \, .
  \label{eq:npipRECURSION}
\end{eqnarray}
with the initial condition that $R_1=\Pi$ as defined in
Eq.~(\ref{eq:Cnfungrassman}).  To see how
this works we explicitly construct the first few terms:
\begin{eqnarray}
  R_2 & = & \langle\ R_1\ \rangle \Pi \ -\ R_1\ \Pi
  \ =\ \langle\ \Pi\ \rangle\ \Pi \ -\ \Pi^2
  \nonumber\\
  \langle \ R_2\ \rangle & = &
 \langle\ \Pi\ \rangle^2 \ -\ \langle\ \Pi^2\ \rangle
  \ \ ,
\nonumber\\
  R_3 & = & \langle\ R_2\ \rangle\  \Pi \ -\ 2\ R_2\ \Pi
  \ =\  
  \langle\ \Pi\ \rangle^2 \ \Pi
  \ -\ \langle\ \Pi^2\ \rangle\ \Pi
  \ -\ 2\ \langle\ \Pi\ \rangle\ \Pi^2 
  \ +\ 2\ \Pi^3\ 
  \nonumber\\
  \langle\ R_3 \ \rangle 
  & = & 
  \langle\ \Pi\ \rangle^3
  \ -\ 3\ \langle\ \Pi^2\ \rangle\ \  \langle\ \Pi\ \rangle\
  \ -\ 2\ \langle\ \Pi^3\ \rangle\ 
  \ \ ,
  \label{eq:pipR3}
\end{eqnarray}
in agreement with Eq.~(\ref{eq:threePiCorrelator}).
Descending recursions also exist.

A correlation function for a system composed of $n_{ij}$
mesons of the $i^{\rm th}$ species from the $j^{\rm th}$ source at
$({\bf y}_j,0)$, where $0\le i\le k$ and $0\le j\le m$, is of the form
\begin{eqnarray}
  && C_{\bf n}(t) \ = \  
  \Bigg\langle\ 
  \left(\ \sum_{\bf x}\ {\cal A}_1 ({\bf x},t)\ \right)^{{\cal N}_1} \ 
  ...
  \left(\ \sum_{\bf x}\ {\cal A}
    _k ({\bf x},t)\ \right)^{{\cal N}_k} \
  \nonumber\\
  &&
  \left( \phantom{\sum_{\bf x}}\hskip -0.22in
    {\cal A}_1^\dagger({\bf  y_1},0) \right)^{n_{11}} ...\  
  \left( \phantom{\sum_{\bf x}}\hskip -0.22in
    {\cal A}_1^\dagger({\bf  y_m},0) \right)^{n_{1m}} ...\  
  \left( \phantom{\sum_{\bf x}}\hskip -0.22in
    {\cal A}_k^\dagger({\bf  y_1},0) \right)^{n_{k1}} ...\  
  \left( \phantom{\sum_{\bf x}}\hskip -0.22in
    {\cal A}_k^\dagger({\bf  y_m},0) \right)^{n_{km}}  
  \Bigg\rangle
  \ ,
  \label{eq:mk}
\end{eqnarray}
where ${\cal N}_i = \sum_j\ n_{ij}$ is the total number of mesons of
species $i$, and the subscript in $C_{\bf n} (t)$ labels the number of
each species from each source,
\begin{eqnarray} {\bf n} & = &\left(
    \begin{array}{cccc}
      n_{11}&n_{12}&...&n_{1m}\\
      \vdots &\vdots &\vdots &\vdots \\
      n_{k1} & n_{k2}&...&n_{km}
    \end{array}
  \right)
  \ \ \ .
  \label{eq:nvecdef}
\end{eqnarray}
The ${\cal A}_i({\bf y},t)$ denotes a quark-level interpolating 
operator ${\cal A}_m
({\bf x},t) = \overline{q}_m({\bf x},t) \ \gamma_5\ u ({\bf x},t)$, and
it is straightforward to show that~\cite{Detmold:2010au} 
\begin{eqnarray}
  && C_{\bf n}(t) \ = \  
  \left(\ \prod_i\ {\cal N}_i!\ \right)\ 
  \left\langle\ \prod_{i,j} 
  \left(\ \overline{\eta}\ P_{ij}\ \eta\ \right)^{n_{ij}}\ \right\rangle
\ =\ 
  (-)^{\overline{\cal N}}\ 
  {\left(\ \prod_i\ {\cal N}_i!\right)\  \ \left(\ \prod_{i,j}\ n_{ij}!
    \right)\ 
    \over \overline{\cal N}!}
  \ 
  \langle\ T_{  {\bf n } }\ \rangle
  \ \ \ ,
  \label{eq:mkGRASS}
\end{eqnarray}
where the $\eta$ are $m\times N$-component Grassmann variables, and
the $P_{ij}$ are $\overline{N}\times\overline{N}$ dimensional
matrices, where $\overline{N}=m\times N$, which are generalizations of
the $\Pi$ defined in Eq.~(\ref{eq:Cnfungrassman}) with an additional
species index, $i$. They are defined as
\begin{eqnarray}
  P_{ij} & = & 
  \left(
    \begin{array}{c|c|c|c}
      0&0&...&0 \\
      \hline
      \vdots & \vdots & ... & \vdots \\
      \hline
      \left(A_i\right)_{j1}(t)&\left(A_i\right)_{j2}(t)&\ \  ... \ \ & \left(A_i\right)_{jm}(t) \\
      \hline
      0&0& ... & 0\\
      \hline
      \vdots & \vdots & ... & \vdots \\
      \hline
      0&0& ... & 0
    \end{array}
  \right)
\ \ ,
\nonumber\\
  \left(\ A_i\ \right)_{ab} & = & 
  \sum_{\bf x}\ S({\bf x},t;{\bf y}_b,0)\ S_i^\dagger ({\bf x},t;{\bf
    y}_a,0)\ 
  \ \ \ ,
  \label{eq:Sijdef}
\end{eqnarray}
The $ \left(\ A_i\ \right)_{ab}$
are $N\times N$ dimensional matrices, one for each flavor, $i$, and
pair of source indices, $a$ and $b$.  
$\overline{\cal N} = \sum_i\ {\cal N}_i$ is the total number of
mesons in the system, with $\overline{\cal N} \le \overline{N}$.  The
$T_{{\bf n} }$ 
defined implicitly in Eq.~(\ref{eq:mkGRASS})
satisfy the recursion relation
\begin{eqnarray}
  T_{  {\bf n}+{\bf 1}_{rs} }
  & = & 
  \sum_{i=1}^k\ 
  \sum_{j=1}^m\ 
  \ 
  \langle\ T_{ {\bf n}  + {\bf 1}_{rs} - {\bf 1}_{ij}  }\ \rangle\ P_{ij}
  \ -\
  \overline{\cal N}\  T_{  {\bf n}  + {\bf 1}_{rs} - {\bf 1}_{ij} }\ \ P_{ij}
  \ \ \ \ ,
  \label{eq:genRECUR}
\end{eqnarray}
where 
\begin{eqnarray} 
{\bf 1}_{ij} \ & = &\left(
    \begin{array}{cccc}
      0 & 0& \cdots & 0\\
      \vdots & \vdots  & \hdots \  1 \ \hdots & \vdots \\
      0 & 0&\cdots&0
    \end{array}
  \right)
  \ \ \ ,
  \label{eq:nplusvecdef}
\end{eqnarray}
and where the non-zero value is in the $(i,j)^{\rm th}$ entry.
Defining ${\cal U}_j = \sum_i \ n_{ij}$ to be the number of mesons
from the $j^{\rm th}$ source, it is clear that the correlation
function vanishes when ${\cal U}_j>N$ for any source $j$.

These recursion relations (and those for the simpler systems also
studied in Ref.~\cite{Detmold:2010au}) allow for the calculation of
arbitrarily large systems of mesons. 
As formulated above, the systems
are restricted to contain quarks of one flavor but
anti-quarks of any number of flavors~\footnote{This restriction is not
necessary, but relaxing it results in much more complex sets of
recursions.} or vice versa. 
Importantly the above algorithm scales linearly with
the total number of mesons in the system.

%%%%%%%%%%%%%%%%%%%%%%
\subsection{\it Pion and Kaon Condensation}
\label{sec:pion-kaon-cond}

\begin{figure}[!ht]
   \centering
\begin{minipage}[!ht]{8 cm}
\centerline{\includegraphics[scale=0.9]{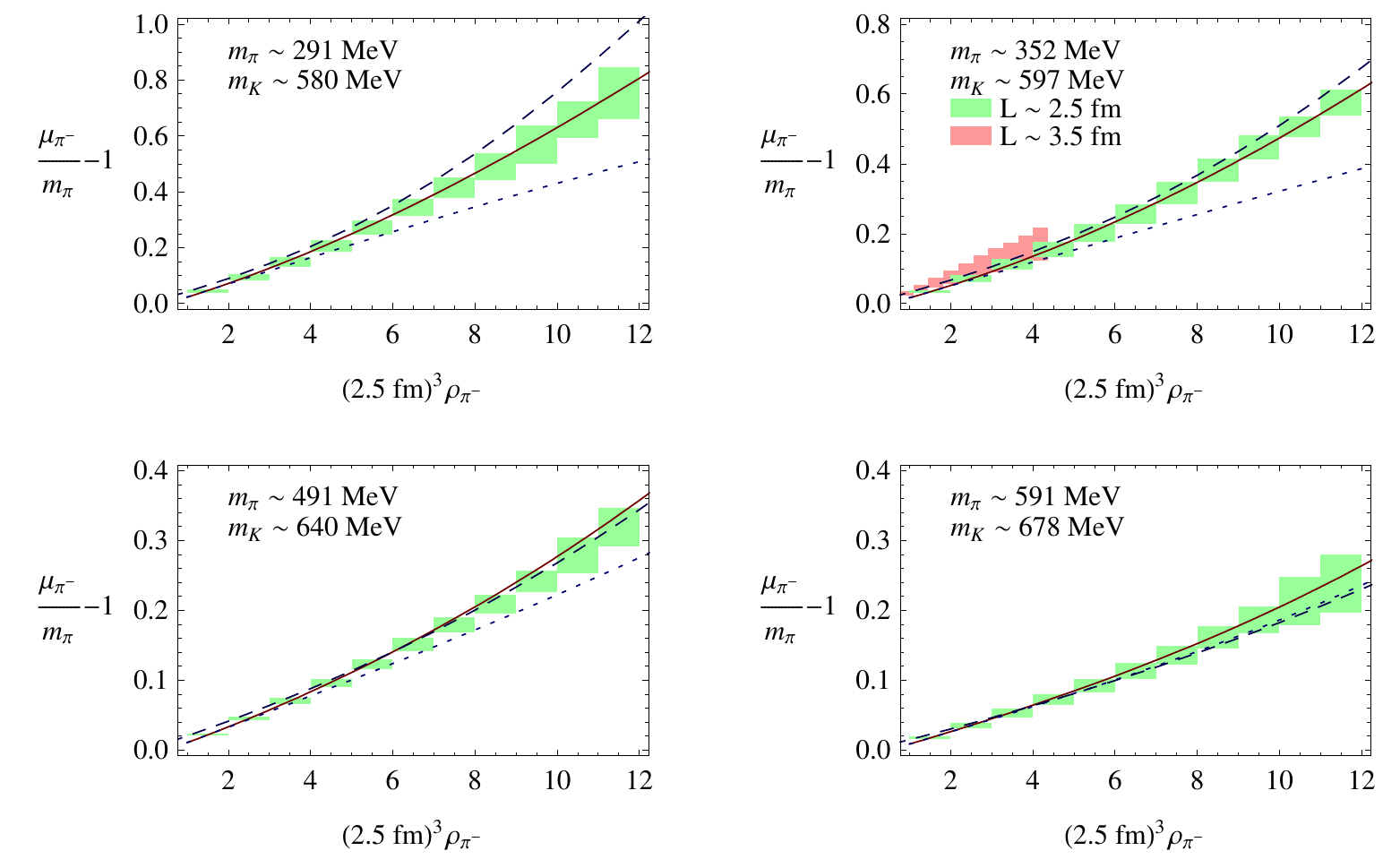}}
\end{minipage}
\begin{minipage}[t]{16.5 cm}
   \caption{The dependence of the
     isospin chemical potential on the pion 
     density. The curves
     correspond to the predictions of tree level $\chi$-PT 
(dashed)~\protect\cite{Son:2000xc}, the energy shift of 
Eq.~(\ref{eq:energyshift})
     (solid) and without the three-body interaction (dotted).}
   \label{fig:Istr}
\end{minipage}
 \end{figure}

\begin{figure}[!ht]
   \centering
\begin{minipage}[!ht]{8 cm}
\centerline{\includegraphics[scale=0.9]{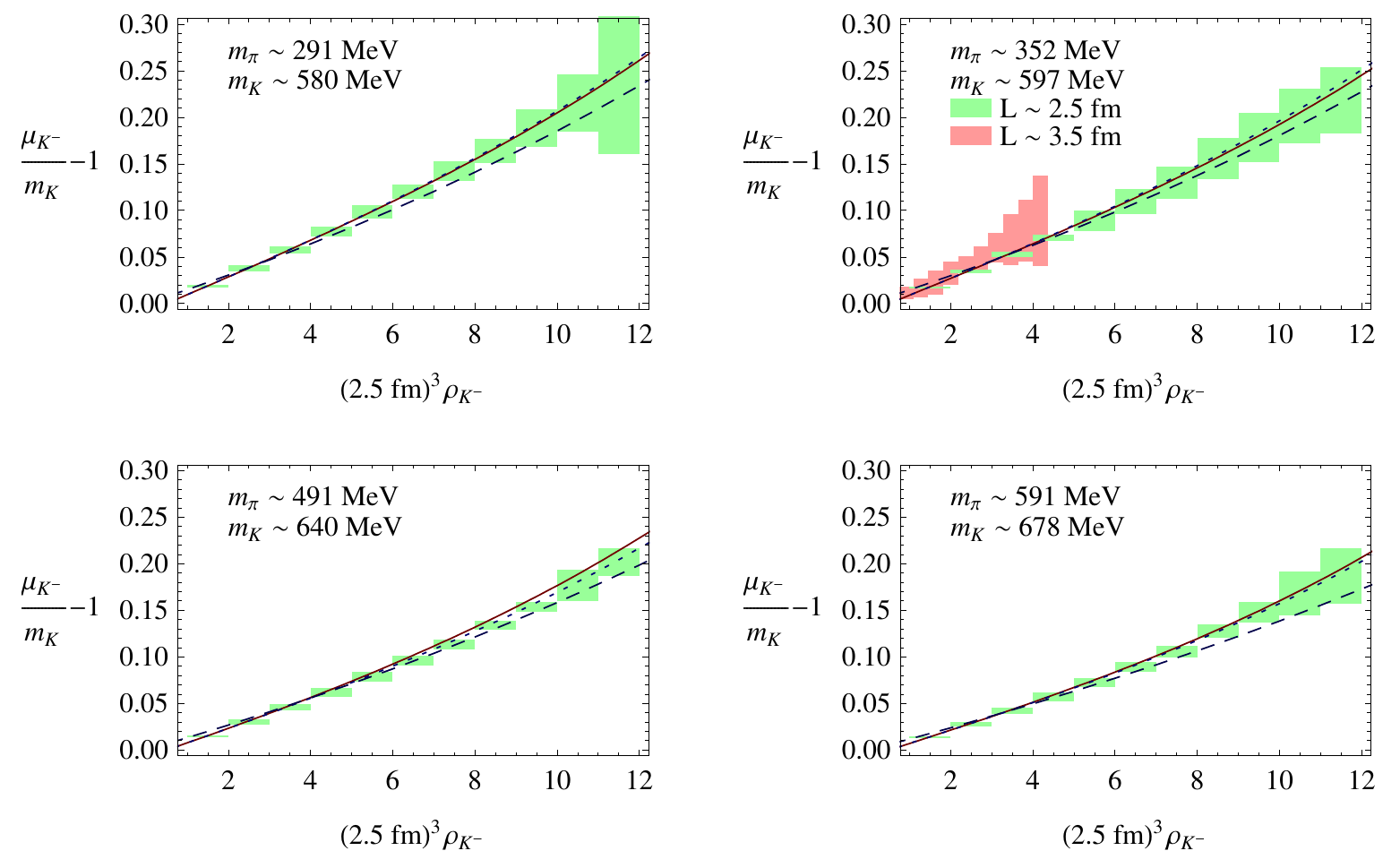}}
\end{minipage}
\begin{minipage}[t]{16.5 cm}
   \caption{The dependence of the
     strangeness chemical potential on the kaon
     density. The curves
     correspond to the predictions of tree level chiral perturbation
     theory (dashed)~\protect\cite{Son:2000xc}, the energy shift of 
Eq.~(\ref{eq:energyshift})
     (solid) and without the three-body interaction (dotted).}
   \label{fig:str}
\end{minipage}
 \end{figure}

The ground state of a generic system of many bosons with repulsive
interactions is a Bose-condensed phase. The QCD systems of pions or
kaons discussed above 
form a Bose-Einstein condensate of fixed third-component 
of isospin, $I_z$, and strangeness, $s$. It is of
significant theoretical and phenomenological interest to investigate
the properties of such systems. Theoretical efforts have used 
LO $\chi$-PT
to investigate the phase diagram at
low chemical potential~\cite{Son:2000xc} and it is important to assess
the extent to which these results agree with QCD. In neutron stars, it
is possible that it is energetically favorable to (partially-)
neutralize the
system with a condensate of $K^-$ mesons, and an isospin chemical
potential resulting from the isospin asymmetry of the colliding nuclei
may play an important role in the study of the QCD phase diagram at
RHIC. Numerical calculations provide  a probe of
the dependence of
the energy on the pion  or kaon density, and thereby allow for 
an extraction of the
chemical potential via a finite difference. The results 
from mixed-action calculations are shown in
Figure~\ref{fig:Istr} and Figure~\ref{fig:str}. 
Also shown are the predictions of tree-level $\chi$-PT, 
which are in remarkably good agreement.
This is encouraging for studies of kaon condensation in
neutron stars where, typically, tree level $\chi$-PT
interactions  amongst kaons, and between kaons and 
baryons~\cite{KaplanNelson}, are assumed.

%%%%%%%%%%%%%%%%%%%%%%%%%%%%
\subsection{\it Hadronic Medium Effects}
\label{sec:hadr-medi-effects}

The hadronic medium formed in the multi-meson calculations described
above is naturally expected to influence the properties of other
hadrons that interact with it. A phenomenologically relevant example
is the passage of a $J/\Psi$ meson through medium as $J/\Psi$
suppression is taken to be a key signal for the formation of a
new state of matter at high temperatures, 
such as the quark-gluon plasma~\cite{Matsui:1986dk}. In
Ref.~\cite{Detmold:2008bw}, a simplified version of this scenario was
considered, and the screening of the static quark potential (the
potential between an infinitely massive quark--anti-quark pair) by a
Bose-Einstein condensate of $\pi^+$'s was computed. 
In order to extract the shift in the
static potential caused by the presence of the medium, the following
ratio of correlation functions was calculated
\begin{eqnarray}
G_{n,W}(R,t_\pi,t_w,t) & = & 
{ C_{n,W} (R,t_\pi,t_w,t)\over C_n(t_\pi,t) \ C_W (R,t_w,t)
}
\ \ ,
\label{eq:medrat}
\end{eqnarray}
where $C_n$ is defined in Eq.~(\ref{eq:Cnfun}) and 
\begin{eqnarray}
  \label{eq:corrs}
C_W (R,t_w,t) &=&   \Big\langle 0 \Big| 
\sum_{{\bf y},|{\bf r}|=R}
{\cal W}\left({\bf y}+{\bf r}, t;{\bf y},t_w\right)
\Big|0\Big\rangle
\ \ ,
\nonumber\\
C_{n,W} (R,t_\pi,t_w,t) &=&   \Big\langle 0 \Big| 
\Big[\sum_{\bf x} {\pi^-}({\bf x},t) {\pi^+}(0,t_\pi)\Big]^n
\sum_{{\bf y},|{\bf r}|=R}
{\cal W}\left({\bf y}+{\bf r}, t; {\bf y},t_w\right)
\Big|0\Big\rangle
\ \ ,
\end{eqnarray}
and ${\cal W}\left({\bf y},t_0; {\bf y}+{\bf r}, t\right)$ is the
Wilson-loop operator formed from products of gauge links joining the
vertices at $({\bf y},t_0)$, $({\bf y}+{\bf r},t_0)$, $({\bf y}+{\bf
  r},t)$ and $({\bf y},t)$. At large Euclidean time, this ratio falls
exponentially with a scale $\delta V(R,n)$ that is the shift in static
potential for separation $R=|{\bf r}|$ and isospin density
$\rho=\rho_0 n$ (where $\rho_0=1/V\sim 0.064\ {\rm fm}^{-3}$).
Further details are discussed in Ref.~\cite{Detmold:2008bw}.

\begin{figure}[!t]
  \centering
\begin{minipage}[t]{8 cm}
 \centerline{ \includegraphics{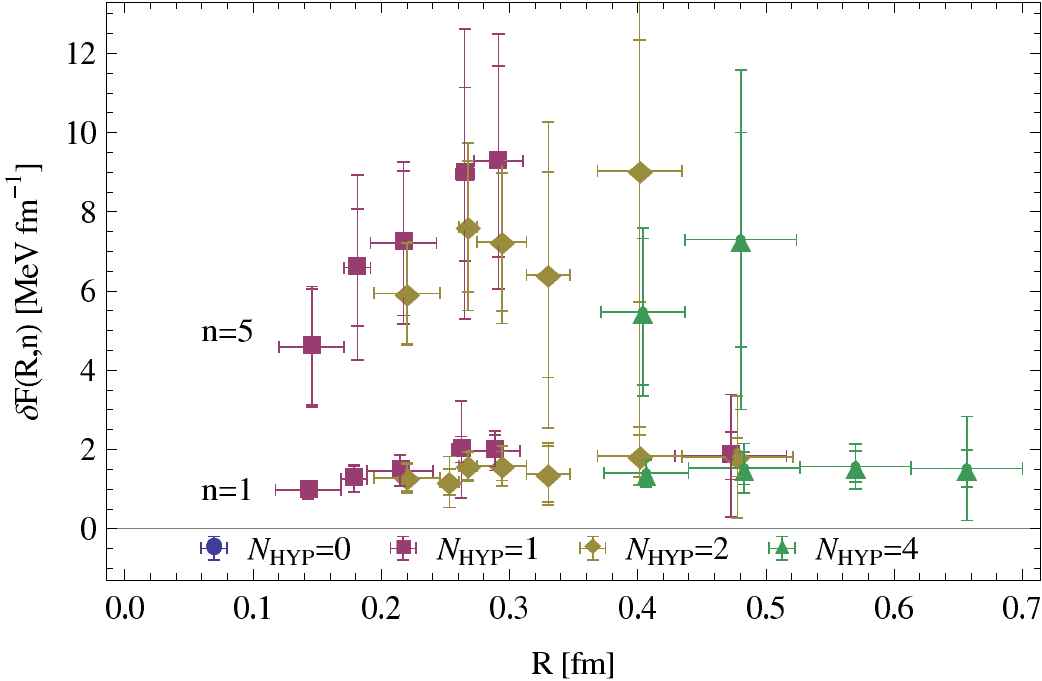} }
\end{minipage}
\begin{minipage}[t]{16.5 cm}
  \caption{The in-medium contribution to the radial $Q\overline{Q}$
    force, $\delta F(R,n)$, at a pion number density of $\rho_0$ and
    $5\rho_0$.The inner uncertainty associated with each point is
    statistical, while the outer is the statistical and systematic
    uncertainties combined in quadrature.}
\label{fig:pionscreen}
\end{minipage}
\end{figure}

As expected from the weak nature of pion interactions with isoscalar
objects, the overall screening effect is small. In the region where
the force between the static quark and anti-quark is constant, $F\sim
1\ {\rm GeV \ fm^{-1}}$, the shift is ${\cal O}({\rm MeV\  fm^{-1}})$. The
shift in the force was found to vary linearly with the isospin density
of the hadronic system as shown in Figure~\ref{fig:pionscreen}. This is
consistent with a simple interpretation in terms of a dielectric in
the volume of the color flux tube between the quark--anti-quark pair.
Further calculations extending this study to dynamical charm quarks
are underway. To directly connect with phenomenology, the more
difficult problem of finite baryon density must be confronted.

%%%%%%%%%%%%%%%%%%
\subsection{\it Three-Baryon Systems}
\noindent
The first significant steps towards the calculation of the properties of 
nuclei directly
from QCD were taken by the NPLQCD collaboration~\cite{Beane:2009gs}
and the PACS-CS collaboration~\cite{Yamazaki:2009ua}
during 2009.  
The NPLQCD collaboration performed a $n_f=2+1$ calculation  
of the correlation function with the quantum
numbers of the strangeness $s=-4$, baryon number $B=3$ system, which is labeled
as ``$n\Xi^0\Xi^0$'' for convenience, 
and also of the correlation function with the quantum numbers 
of the triton (or $^3$He),
at a pion mass of $m_\pi\sim 390~{\rm MeV}$~\cite{Beane:2009gs}. 
Further, the PACS-CS collaboration performed a quenched calculation of 
the system with the quantum numbers of the triton and of the 
$\alpha$-particle at a pion mass of 
$m_\pi\sim 800~{\rm MeV}$~\cite{Yamazaki:2009ua}.

The NPLQCD collaboration calculation used the product of single baryon sources
and sinks for the source and sink of the three-baryon correlation functions.
In the $n\Xi^0\Xi^0$-channel they were the product of two $\Xi$ and one
nucleon source or sink, with the spin quantum numbers constructed to produce a
state with total $J=0$ and third component of isospin $I_z={1\over 2}$.
In
the triton channel, the product of three nucleon interpolating  operators were
used to construct an operator with $J={1\over 2}$ and with total isospin 
$I={1\over 2}$.
\begin{figure}[!tb]
\begin{center}
\begin{minipage}[t]{8 cm}
\centerline{\includegraphics[scale=0.45]{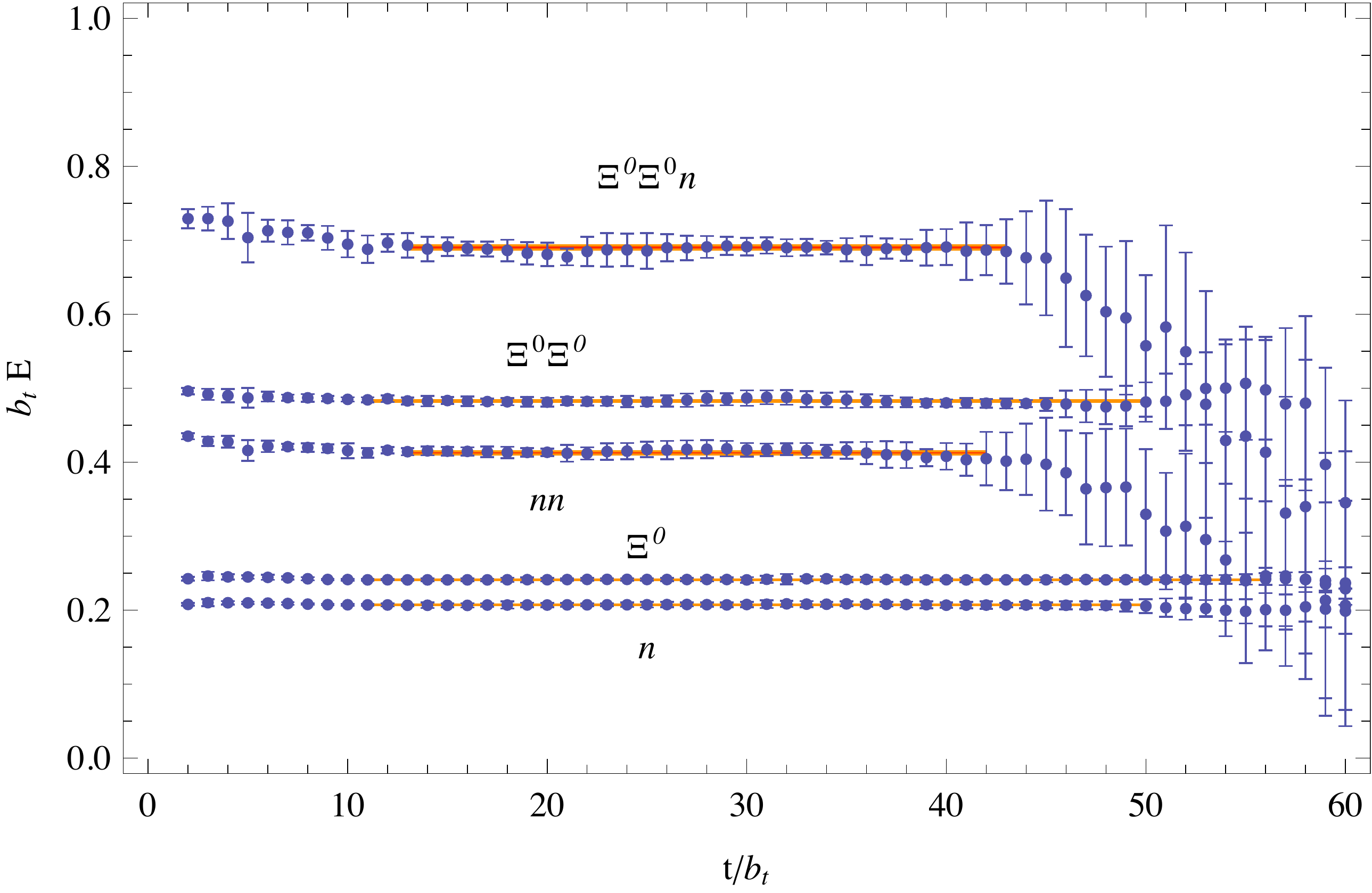}}
\end{minipage}
\begin{minipage}[t]{16.5 cm}
\caption{
The EMPs for the nucleon, 
$\Xi$, $nn$, $\Xi^0\Xi^0$ and
$n\Xi^0\Xi^0$ systems obtained by the NPLQCD 
collaboration~\protect\cite{Beane:2009gs}.
\label{fig:nXiXi}}
\end{minipage}
\end{center}

\end{figure}
\begin{figure}[tb]
\begin{center}
\begin{minipage}[t]{8 cm}
\centerline{\includegraphics[scale=0.3]{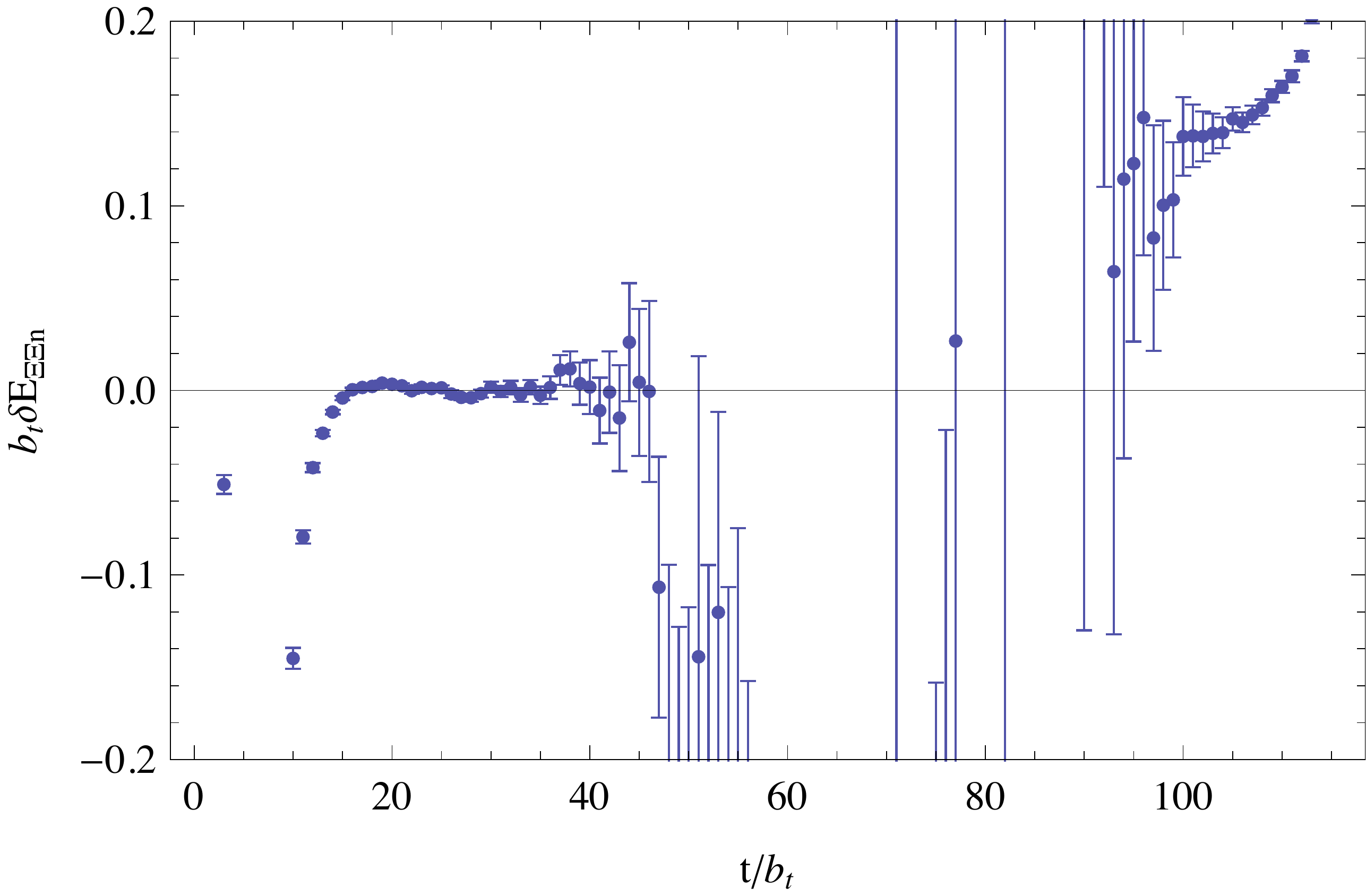}\ \ \includegraphics[scale=0.3]{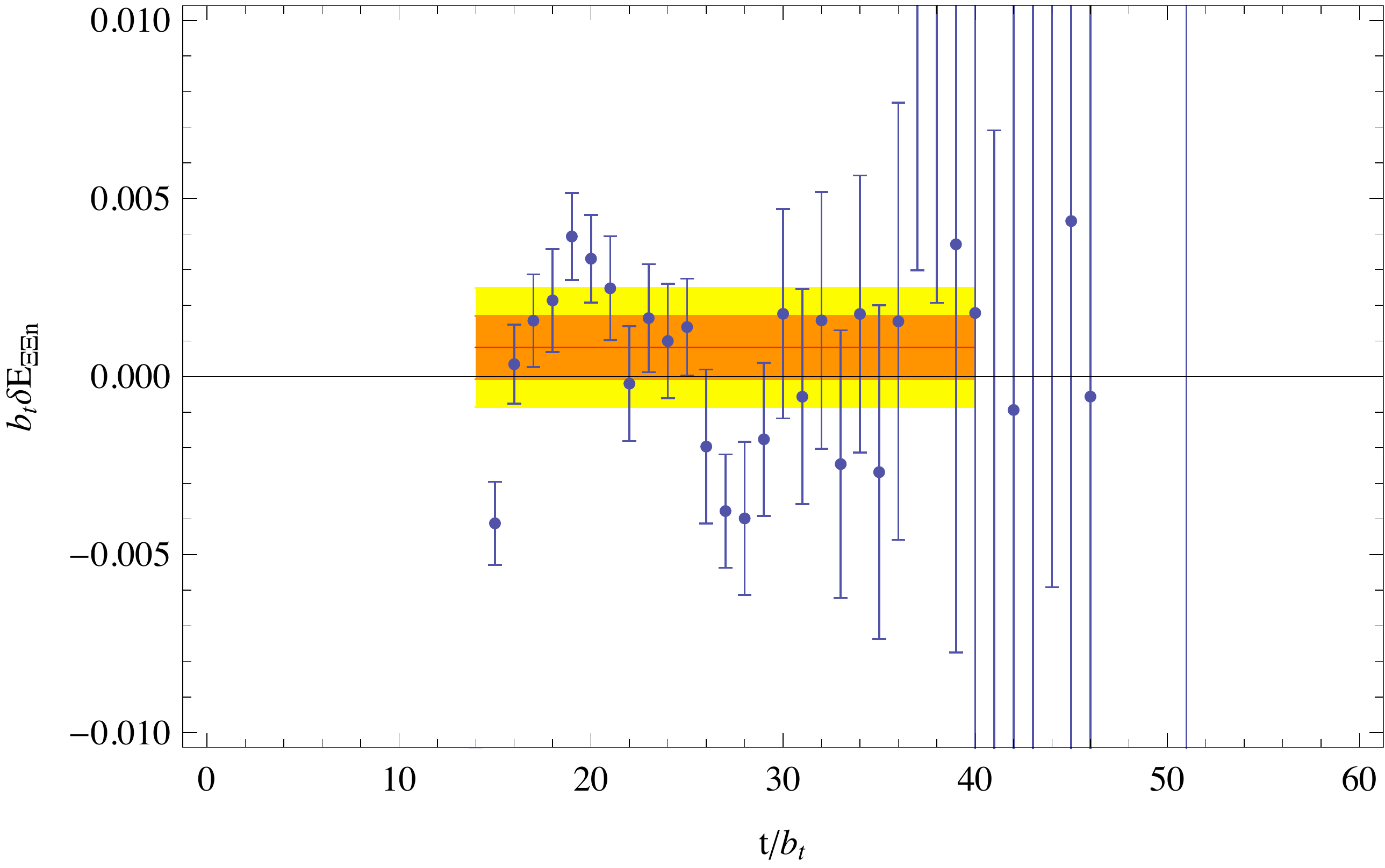}}
\end{minipage}
\begin{minipage}[t]{16.5 cm}
\caption{
The EMP of the difference between the energy of the
$n\Xi^0\Xi^0$ channel and its constituent hadrons~\protect\cite{Beane:2009gs}.
The right panel is an enlargement of part of the left panel.
\label{fig:nXiXiDIFF}}
\end{minipage}
\end{center}

\end{figure}
The correlation function in the $n\Xi^0\Xi^0$-channel, shown in
Figure~\ref{fig:nXiXi}, 
yields an energy-splitting that is consistent with zero within the
uncertainties of the calculation, as shown in Figure~\ref{fig:nXiXiDIFF},
\begin{eqnarray}
\delta E_{\Xi^0\Xi^0 n} & = & 4.6\pm 5.0\pm 7.9\pm4.2~{\rm MeV}
\ \ ,\ \ \chi^2/{\rm dof} \ =\ 2.0
\ \ \ ,
\label{eq:fitval}
\end{eqnarray}
It is very encouraging that the uncertainty in the
energy-shift per baryon is $\sim 3~{\rm MeV}$, which is smaller than
the binding-energy per nucleon in typical nuclei, $B \sim 8~{\rm
  MeV}$, and not significantly larger than the binding-energy per
nucleon in the deuteron or triton at the physical values of the
light-quark masses.  The single energy-level fit to the EMP in
Figure~\ref{fig:nXiXiDIFF} has a $\chi^2/{\rm dof} \ =\ 2.0$,
indicating that there may be additional structure in the correlation
function.  Including a second energy-level shifted by $\Delta E\sim
-0.004$ lattice units might provide a better description of the EMP,
and would be consistent with the lower-energy state,
$\Xi^0\Lambda\Lambda$, that is expected to contribute to the four
low-lying eigenstates in the lattice-volume.  However, enhanced
statistics are required to determine if this is, in fact, the case.

At present, unlike the situation in multi-meson
systems~\cite{Beane:2007es,Detmold:2008fn,Detmold:2008yn}, the
analytical tools are not in place to use the above energy shift and
those of the associated two-baryon systems to extract the parameters
describing the relevant two- and three-body interactions. While the
volume dependence of the simplest three-fermion systems has been
studied in Ref.~\cite{Luu:2008fg}, the mixing we expect between four
closely spaced states significantly complicates the situation.

The NPLQCD collaboration 
focused on the state(s) that couples to the $\Xi^0\Xi^0
n$ interpolating-operator simply due to limited computational
resources and the expectation that the
large strange quark content would lead to a clean signal.  
In the past it has been the case that gauge field
generation has required the majority of LQCD resources, but this is no
longer true for precise calculations of baryonic observables.  The
resources 
that are presently
required to perform the large number of calculations
required for nuclear systems is significantly greater than that
required for gauge field generation.  This situation will improve as
more effort is put into algorithmic improvements for contractions, in
the same way that the use of deflation and
other techniques have greatly reduced the resources required for
propagator generation. Work in this direction is in progress. 
Given the observed behavior of the signal-to-noise ratio, which will be
discussed in 
Section~\ref{sec:noise-correlations},
identification of the ground state in 
systems of four and five baryons is anticipated in the near future.

\begin{figure}[tb]
\begin{center}
\begin{minipage}[t]{8 cm}
\centerline{\includegraphics[scale=0.75]{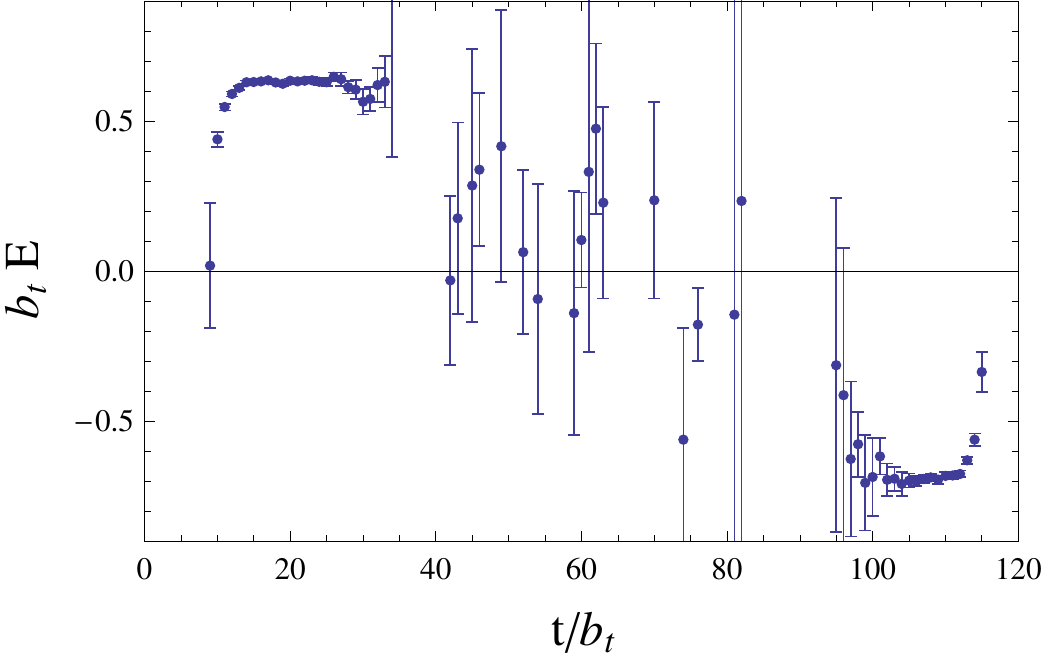}\ \ 
\includegraphics[scale=0.75]{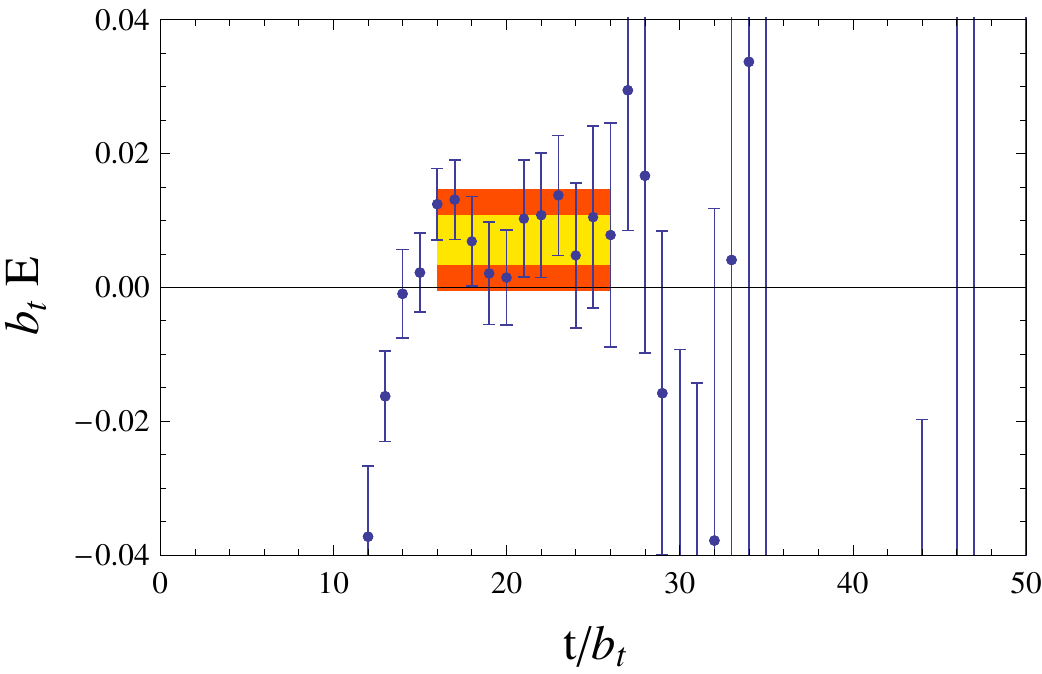}}
\end{minipage}
\begin{minipage}[t]{16.5 cm}
\caption{
The EMP for the correlation function with the quantum numbers
of the triton (left panel), and for the difference between the triton and
its constituents (right panel)~\protect\cite{Beane:2009gs}.
\label{fig:triton}}
\end{minipage}
\end{center}

\end{figure}

The number of contractions that must be performed in order to calculate the
correlation function in the triton channel is significantly greater than the
number in the $\Xi^0\Xi^0 n$, and therefore a smaller number of contractions
was possible with the computational resources available to NPLQCD.
The results of the calculations performed by the NPLQCD collaboration of the correlation
function in the channel with the quantum numbers of the triton are shown 
in Figure~\ref{fig:triton}.
A clear plateau is found for the ground-state of the system in the lattice
volume in the window of time-slices where the signal-to-noise ratio is not
degrading exponentially. However, the uncertainty remains too large to determine
if the ground state corresponds to a positively or negative shifted state,
\begin{eqnarray}
\delta E_{pnn} & = & 40\pm 21\pm 38~{\rm MeV}
\ \ ,\ \ \chi^2/{\rm dof} \ =\ 0.97
\ \ \ ,
\label{eq:fitvaltriton}
\end{eqnarray}
corresponding to $\sim 13~{\rm MeV}$-per-baryon.
However, the fact that a plateau in the triton channel has been observed at a
relatively light pion mass ($m_\pi\sim 390~{\rm MeV}$) is a very encouraging
step forward.

Unfortunately, none of the calculations that have been performed to 
date  for systems of three or more baryons
have
managed to calculate energy-shifts that exceed $5$-$\sigma$ deviations from zero,
and, as such,  compelling calculations 
are yet to be performed.
However, such calculations are expected in the near future.

At the physical pion mass, the expected energy eigenvalues 
of two nucleons in a
lattice volume with $m_\pi L\gg 1$
have been determined~\cite{Beane:2003da}.
Recently,  there have been efforts undertaken to determine the expected energy
eigenvalues of systems composed of three
nucleons~\cite{Luu:2008fg,Kreuzer:2009jp,HammerKreuzer:2010:triton,Epelbaum:2010xt,Lee:2010:triton}
in cubic volumes with periodic boundary conditions using EFT.
It is found that the triton binding energy  is less sensitive
to a finite-volume than the deuteron binding energy, which can be understood
in terms of the spatial extent of the respective wavefunctions.
These works are the beginning of a series of theoretical calculations 
that should be performed to provide a guide in determining the  
Lattice QCD calculations that 
should be performed to best extract the parameters in the 
EFT which will then allow for a description
of more complex processes of relevance to nuclear physics.

%%%%%%%%%%%%%%%%%
\subsection{\it Four-Baryon Interaction}

\begin{figure}[tb]
\begin{center}
\begin{minipage}[t]{8 cm}
\centerline{\includegraphics[scale=0.4]{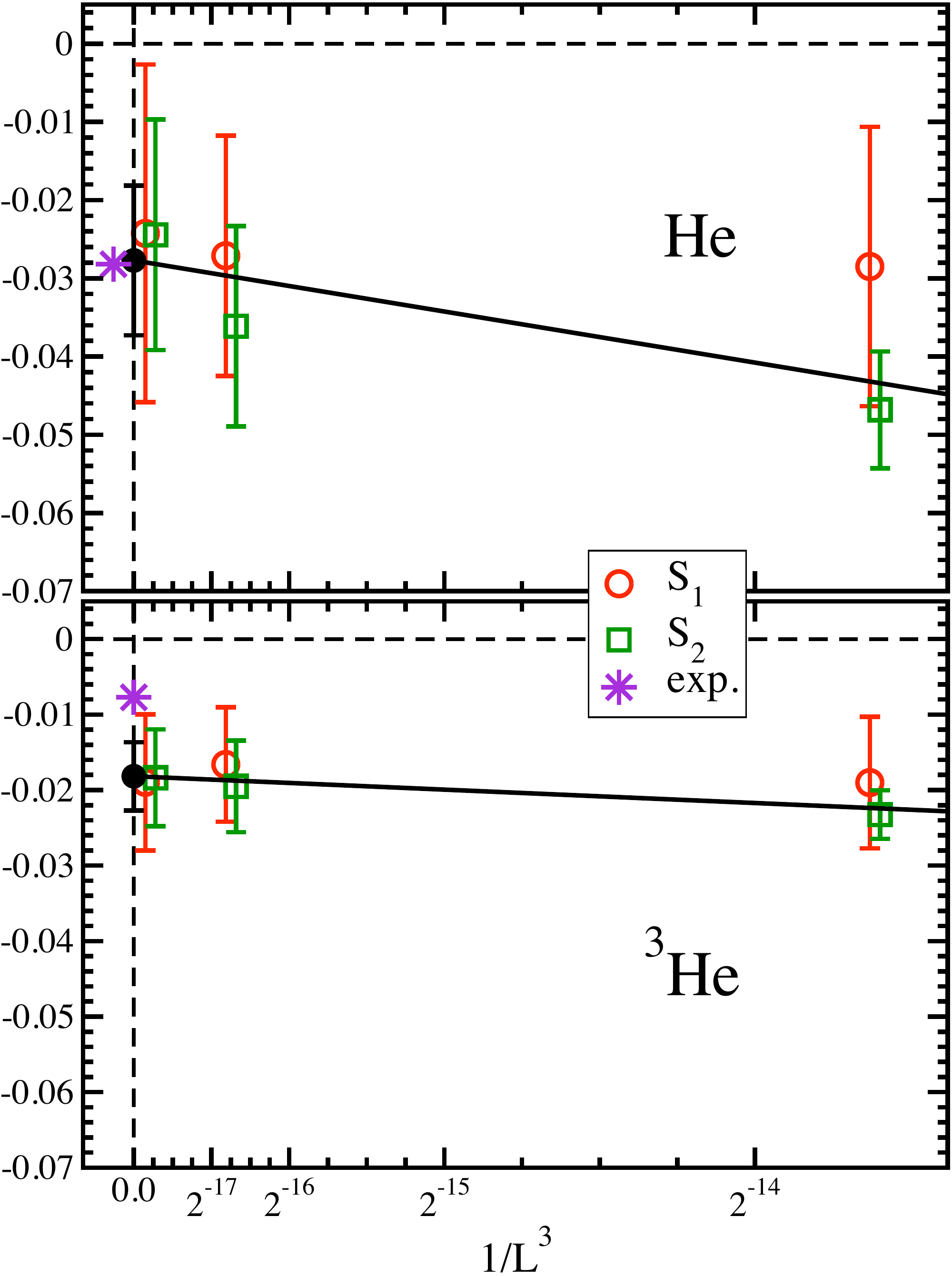}}
\end{minipage}
\begin{minipage}[t]{16.5 cm}
\caption{
The quenched results 
for the binding energies (in lattice units)
obtained by the PACS-CS collaboration in the triton
channel and the
channel with the quantum numbers of 
the $\alpha$-particle~\cite{Yamazaki:2009ua}. 
The pion mass in these calculations is
$m_\pi\sim 800~{\rm MeV}$.
\label{fig:alpha}}
\end{minipage}
\end{center}

\end{figure}

In late 2009,
the PACS-CS collaboration performed the first quenched calculation of 
a four-baryon correlation function~\cite{Yamazaki:2009ua} 
in the channel that will contain the 
$\alpha$-particle ($^4$He nucleus) when performed at the
physical pion mass.
The pion mass used in these calculations is large, $m_\pi\sim 800~{\rm MeV}$,
and sea quark effects are ignored.
Nonetheless this is a very important step towards calculating the 
properties
and interactions of nuclei.  
The PACS-CS collaboration results are shown in
Figure~\ref{fig:alpha}.
One hopes to see significantly improved
statistics 
in the near future for such an important quantity.
%

%%%%%%%%%%%%%%%%%%%%%%%%%%%%%%%%%
\section{The Signal-to-Noise Ratio in Baryon Correlation Functions}
\label{sec:noise-correlations}
\noindent
Until recently  it had been lore that the signal-to-noise ratio in baryon
correlation functions degrades exponentially with 
the baryon number.   The correlation functions exhibit this behavior at
large times~\cite{Lepage:1989hd}, 
and if this were the only part of the correlation functions that
could be calculated with precision, the calculation of quantities of importance to
nuclear physics would require exponentially more computing resources than those
of importance to particle physics.   However, during the last year it has been
realized that such exponential degradation of the signal-to-noise ratio
is absent at intermediate times over an interval that is dictated by the
structure of the source and sink of the correlation 
function~\cite{Beane:2009ky,Beane:2009gs,Beane:2009py}.
A consequence of this behavior  is that it is possible to extract
information about multi-baryon systems where the only additional computational
resources required are those to perform the contractions.

In the case of a source that has the
quantum numbers of a single positive parity nucleon, the correlation
function 
on an ensemble of gauge field configurations with infinite temporal extent
has the form~\cite{Lepage:1989hd}
\begin{eqnarray}
  \langle \theta_{N}(t)\rangle &=&
  \sum_{\bf x}\ 
  \Gamma_+^{\beta\alpha}
  \ \langle 0 | \ N^\alpha({\bf x},t) \overline{N}^{\beta} ({\bf 0},0)
  \ |0\rangle
  \ \rightarrow\ Z_N \ e^{-M_N t}
  \ \ ,
  \label{eq:Gfunproton}
\end{eqnarray}
where $N^\alpha({\bf x},t)$ is an interpolating field 
(composed of three quark operators) that has
non-vanishing overlap with the nucleon, $\Gamma_+$ is a positive
energy projector, and the angle brackets indicate statistical
averaging over calculations on an ensemble of configurations.  The
variance of this correlation function is given by
\begin{eqnarray}
  {\rm N}\  \sigma^2 & \sim & 
  \langle \theta^{\dagger}_N(t)
  \theta_N(t)\rangle  - \langle \theta_{N}(t) \rangle^2 \nonumber \\
  & = & 
  \sum_{\bf x,y} \Gamma_+^{\delta\alpha}\Gamma_+^{\gamma\beta\dagger}\ 
  \langle 0|\ N^\alpha({\bf x},t) \overline{N}^{\beta}({\bf y},t) N^\gamma({\bf 0},0)
  \overline{N}^{\delta}({\bf 0},0) 
  \ |0 \rangle\ \ -\ \langle \theta_{N}(t) \rangle^2
  \nonumber\\
  & \rightarrow & 
  Z_{N\overline{N}} e^{-2 M_N t} - Z_N^2  e^{-2M_N t}
  \ +\ 
  Z_{3\pi}\  e^{-3 m_\pi t}\ +\ ...
  \ \ \stackrel{t\to\infty}{\rightarrow} \ Z_{3\pi}\  e^{-3 m_\pi t}
  \ \ ,
  \label{eq:GGdaggerfunproton}
\end{eqnarray}
where all interaction energies have been neglected, and N is the
number of (independent) calculations.  
At large times, the noise-to-signal ratio has the form, 
as argued by Lepage~\cite{Lepage:1989hd},
\begin{eqnarray} {\sigma\over\overline{x}} & = & {\sigma (t)\over
    \langle \theta(t) \rangle } \sim {1\over \sqrt{\rm N}} \ e^{\left(
      M_N - {3\over 2} m_\pi\right) t} \ \ .
  \label{eq:NtoSproton}
\end{eqnarray}
More generally, for a system of $A$ nucleons, the noise-to-signal
ratio behaves as
\begin{eqnarray} {\sigma\over\overline{x}} & & \sim {1\over \sqrt{\rm
      N}} \ e^{A \left( M_N - {3\over 2} m_\pi\right) t} \ \
  \label{eq:NtoSnucleus}
\end{eqnarray}
at large times.

The various
``Z-factors'', such as $Z_{3\pi}$, depend upon the details of the
sources and sinks interpolators that are used.  
For the calculations 
performed by the NPLQCD collaboration,
the projection onto zero-momentum final state nucleons,
introduces a $1/\sqrt{\rm Volume}$ suppression of the amplitudes of
the various components (except for $N\overline{N}$) in addition to
color and spin rearrangement suppressions that exists independent of
the spatial structure of the source. As a consequence, an interval of
time slices exists at short times (the ``Golden Window'') in which the
variance of the correlation function is dominated by the terms in
Eq.~(\ref{eq:NtoSproton}) that behave as $\sim e^{-2 M_N t} $.  In
this window, the signal-to-noise ratio of the single baryon
correlation function is independent of time.  Further, the
signal-to-noise ratio does not degrade exponentially faster in
multi-baryon correlation functions than in single-baryon correlation
functions in the ``Golden Window''.  

The finite temporal extent introduces backward propagating states
(thermal states) into the correlation functions which lead to
exponentially worse signal-to-noise ratios at large
times~\cite{Beane:2009ky,Beane:2009gs,Beane:2009py}. These contributions are
suppressed by at least $\exp(m_\pi T)$, however, they  can cause
complications. We note that the impact of these
states can be mitigated by working at
larger temporal extents and exponentially large computational
resources are not required.

With the high statistics calculations that have been performed, 
the behavior of the signal-to-noise ratio has been carefully
examined, and it was  found to be useful to form the effective noise-to-signal
plot~\cite{Beane:2009ky}.  On each time
slice, the quantity
\begin{eqnarray} {\cal S}(t) & = & {\sigma (t)\over\overline{x}(t)} \
  \ \ ,
  \label{eq:stondefn}
\end{eqnarray}
is formed, from which the energy governing the exponential behavior (the
signal-to-noise energy-scale)
can be extracted via
\begin{eqnarray}
  E_{\cal S}(t;t_J) & = & 
  {1\over t_J}\ \log\left({ {\cal S}(t+t_J) \over {\cal S}(t)}\right)
  \ \ \ .
  \label{eq:Estondefn}
\end{eqnarray}
For a correlation function that is dominated by a single state with a
corresponding variance correlation function dominated by a single
energy scale, the quantity $E_{\cal S}(t;t_J)$ will be independent of
both $t$ and $t_J$.

The signal-to-noise ratio in the one- and two-nucleon sector has
the simplest structure as only up and down quarks appear in the
interpolating operators.  In the single nucleon sector, it is expected  that
the energy scales $E_S \sim 0$, $M_N-{3\over 2} m_\pi$, and others,
contribute to the signal-to-noise ratio.  At  times when the
nucleon correlation function is in the ground state, and the variance
correlation function is dominated by the nucleon-antinucleon state,
$E_S = 0$ should dominate the signal-to-noise ratio. At large
times the variance correlation function is dominated by
the 3-pion state and $E_S=M_N-{3\over 2} m_\pi$
should dominate.  This is modified by the finite temporal 
direction~\cite{Beane:2009ky} as the hadrons produced by the sources of the
correlation function and the variance correlation function can
propagate forward and backward in time. 
\begin{figure}[tb]
\begin{center}
\begin{minipage}[t]{8 cm}
\centerline{\includegraphics[scale=0.5]{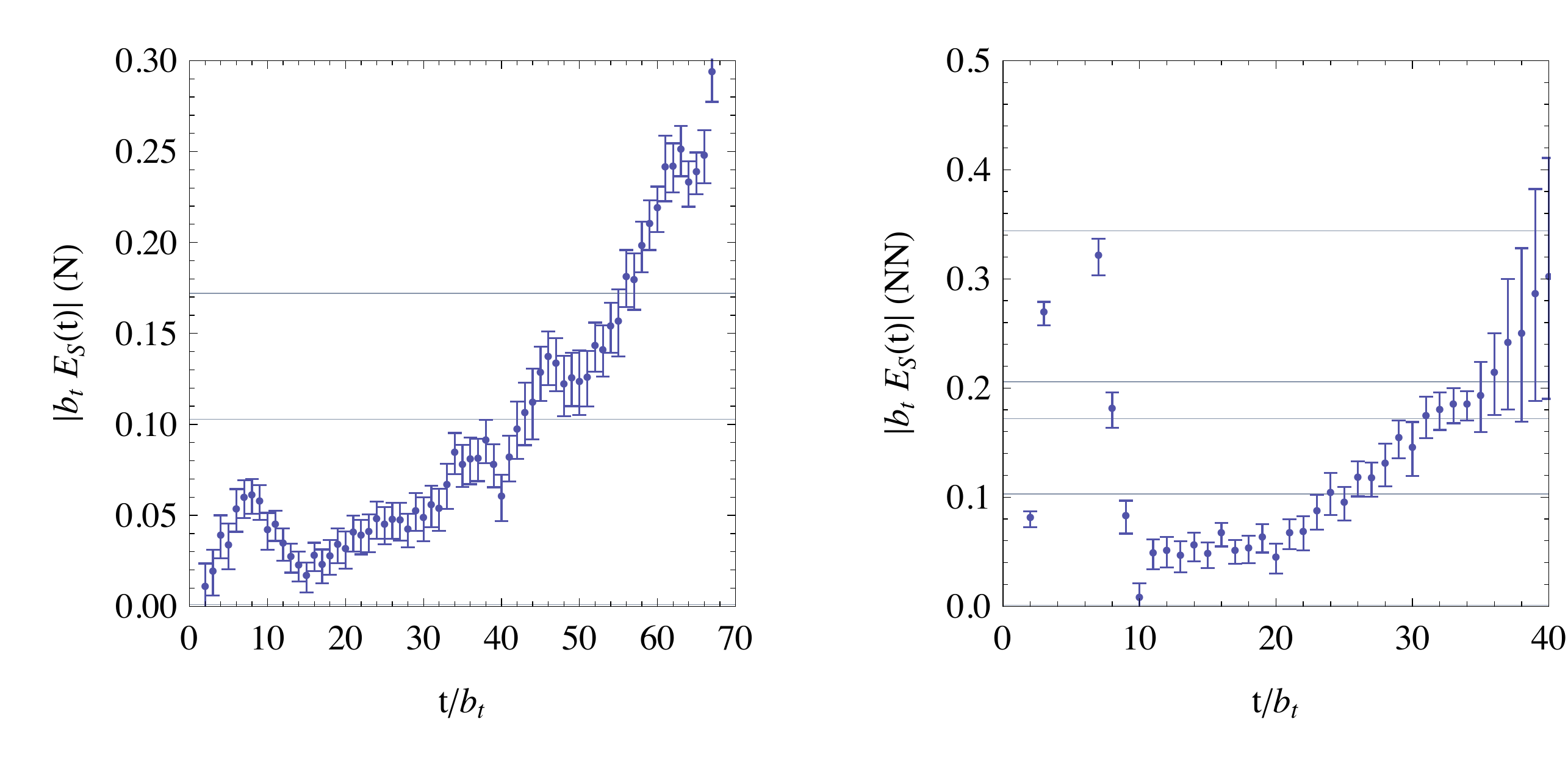}}
\end{minipage}
\begin{minipage}[t]{16.5 cm}
\caption{
The energy scale of the signal-to-noise ratio in the
    nucleon (left panel) and proton-proton (right panel) correlation
    functions, defined in Eq.~(\protect\ref{eq:Estondefn}), 
obtained on 
$20^3\times 128$ anisotropic clover gauge field 
configurations~\protect\cite{Beane:2009ky,Beane:2009gs,Beane:2009py}.  
The horizontal lines in the left panel correspond to
    $E_s=0$, $M_N-{3\over 2} m_\pi$ and $M_N-{1\over 2} m_\pi$, while
    those in the right panel correspond to $E_s=0$, $M_N-{3\over 2}
    m_\pi$, $M_N-{1\over 2} m_\pi$, $2 M_N-3 m_\pi$, $2 M_N- m_\pi$.
\label{fig:NN-noise}}
\end{minipage}
\end{center}
\end{figure}
The calculated energy scale of the signal-to-noise ratio of the single
nucleon correlation function 
obtained on $20^3\times 128$ anisotropic clover gauge field 
configurations~\cite{Beane:2009ky,Beane:2009gs,Beane:2009py}
is shown in Figure~\ref{fig:NN-noise}.  It
exhibits behavior that is consistent with expectations, and exceeds
the long-time behavior expected from the Lepage argument at
approximately time-slice $t=50$ due to the temporal boundary
conditions.

The structure of the two-nucleon variance correlation function has
significantly more structure than that for one nucleon.  On
configurations with infinite temporal extent, the proton-proton
correlation function is of the form (neglecting interactions between
the hadrons)
\begin{eqnarray}
  \langle \theta_{NN}(t)\rangle &=&
  \sum_{\bf x, y}\ 
  \Gamma_+^{\alpha\gamma\beta\rho}
  \ \langle 0 | \ N^\alpha({\bf x},t) N^\gamma({\bf y},t)
  \overline{N}^{\beta}({\bf 0},0) 
  \overline{N}^{\rho} ({\bf 0},0)
  \ |0\rangle \nonumber\\
  & \rightarrow & Z_{NN} \ e^{-2 M_N  t} + \ldots,
  \label{eq:GfunNN}
\end{eqnarray}
and the variance correlation function has the form
\begin{eqnarray} {\rm N}\ \sigma^2 &\sim& \langle
  \theta^{\dagger}_{NN}(t)
  \theta_{NN}(t)\rangle  - \langle \theta_{NN}(t) \rangle^2 \nonumber \\
  &= & \sum_{\bf x,y,z,w} \Gamma_+^{\alpha\rho\delta\psi}
  \Gamma_+^{\beta\eta\gamma\zeta\dagger}\ \langle 0|\ N^\alpha({\bf
    x},t) N^\rho({\bf y},t) \overline{N}^{\beta}({\bf z},t)
  \overline{N}^{\eta}({\bf w},t)\;\times \nonumber\\
  &&\qquad \quad\qquad\qquad\qquad \overline{N}^{\delta}({\bf 0},0)
  \overline{N}^{\psi}({\bf 0},0) N^\gamma({\bf 0},0) N^\zeta({\bf
    0},0) \ |0 \rangle\ -\ \langle \theta_{NN}(t) \rangle^2
  \nonumber\\
  & \rightarrow & Z_{NN\overline{N}\overline{N}} e^{-4 M_N t} -
  Z_{NN}^2 e^{-4 M_N t} \ +\ Z_{3\pi N\overline{N}}\ e^{-(2 M_N + 3
    m_\pi) t} + Z_{6\pi}\ e^{-6 m_\pi t}+\ldots
  \nonumber\\
  & \rightarrow & Z_{6\pi}\ e^{-6 m_\pi t}.
  \label{eq:GGdaggerfunNN}
\end{eqnarray}
Therefore, we anticipate finding energy scales of approximately
$E_s=0, M_N-{3\over 2} m_\pi$ and $2 M_N- 3m_\pi$ in the
signal-to-noise ratio on gauge field configurations of infinite
temporal extent, see Figure~\ref{fig:NN-noise}.  
The temporal boundary conditions imposed in the
present calculation introduce additional energy scales due to the
backward propagating states.
\begin{figure}[tb]
\begin{center}
\begin{minipage}[t]{8 cm}
\centerline{\includegraphics[scale=0.5]{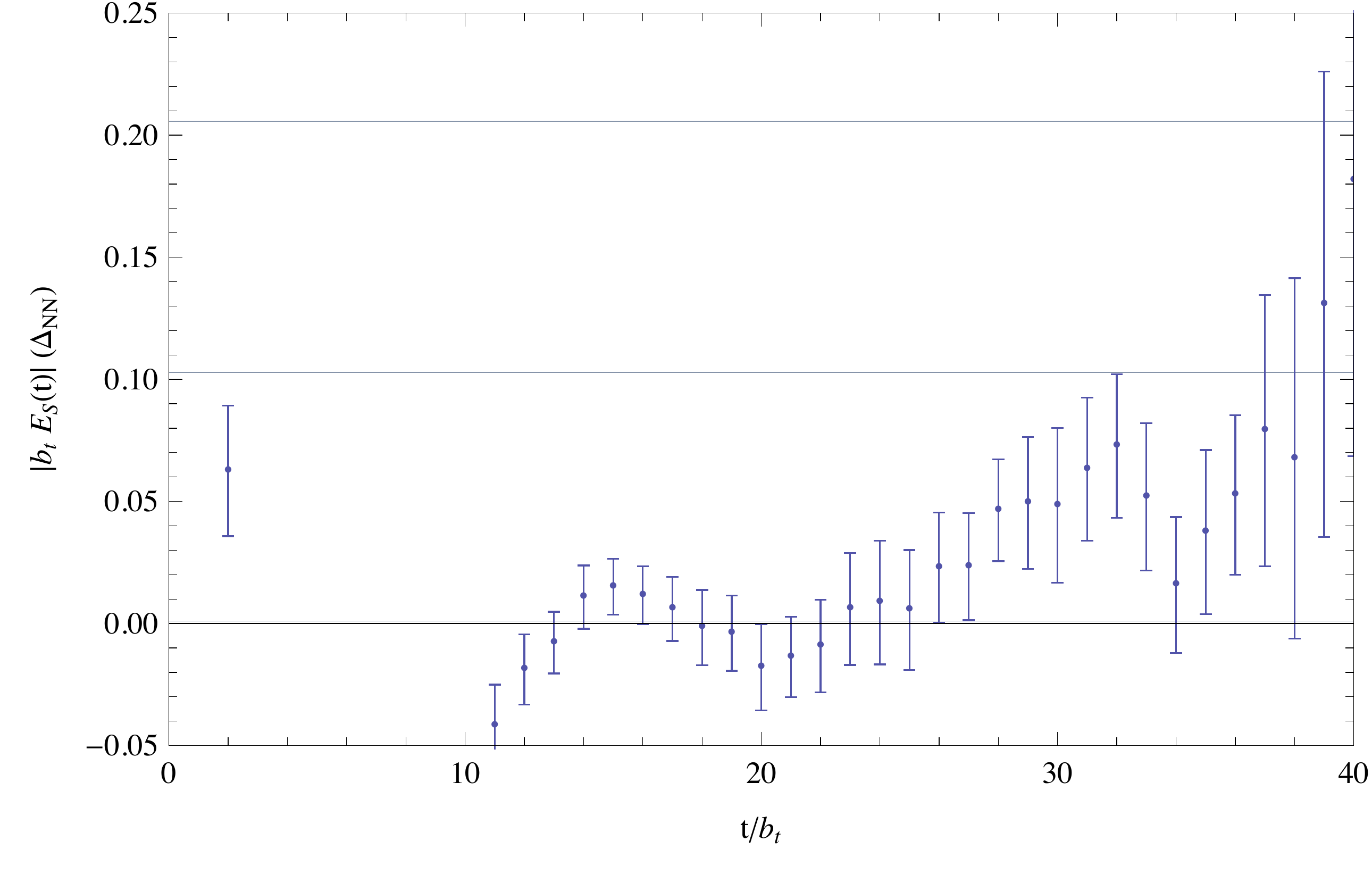}}
\end{minipage}
\begin{minipage}[t]{16.5 cm}
\caption{
The energy scale of the signal-to-noise ratio, defined
    in Eq.~(\protect\ref{eq:Estondefn}), in the ratio of correlation
    functions that produces the shift in energy between two
    interacting protons and two isolated protons, 
obtained on 
$20^3\times 128$ anisotropic clover gauge field 
configurations~\protect\cite{Beane:2009ky,Beane:2009gs,Beane:2009py}.  The
    horizontal lines in the left panel correspond to $E_s=0$,
    $M_N-{3\over 2} m_\pi$ and $2 M_N-3 m_\pi$. 
\label{fig:DIFF-NN-noise}}
\end{minipage}
\end{center}
\end{figure}
Figure~\ref{fig:DIFF-NN-noise} shows the energy scale associated with
the signal-to-noise ratio for the ratio of correlation functions that
provides the energy splitting between two interacting protons and two
isolated protons from which the $p\cot\delta(p)$ is extracted,
obtained on 
$20^3\times 128$ anisotropic clover gauge field 
configurations~\protect\cite{Beane:2009ky,Beane:2009gs,Beane:2009py}.  
It is
clear that the energy scale of the energy splitting is significantly
less than for the individual energies, and is consistent with zero
throughout much of the Golden Window of time slices.  This indicates
that the signal-to-noise ratio associated with the energy splitting in
the proton-proton sector, and hence the scattering parameters and
bound-state energies, are time independent, and therefore do not
degrade exponentially with time.  This is an exceptionally important
result, as it means that the extraction of NN, and more generally,
multi-nucleon interactions, does not require an exponentially large
number of calculations  for each relevant correlation function.

The signal-to-noise ratio in the $\Xi\Xi$ sector is noticeably
larger than in the NN and the YN sectors. 
It is likely that the improved signal-to-noise behavior in the
$\Xi\Xi$ sector is due to a reduced overlap of the source onto the
multi-meson intermediate states in the variance correlation function
compared to purely baryonic intermediate states.  Such a reduction is
expected based on the fact that the volume occupied by multiple
$\Xi$'s is smaller than that of multiple nucleons, and serves to
extend the Golden Window beyond its range in nucleon correlation
functions.

As discussed previously, the correlation functions for states with the
quantum numbers of the triton (or $^3{\rm He}$) and $\Xi^0\Xi^0 n$ have been
calculated, 
with well-defined plateaus being observed in the EMPs
in both channels.
As the largest number of calculations have been performed in
the  $\Xi^0\Xi^0 n$ channel, the signal-to-noise ratio of its correlation function
has been calculated~\cite{Beane:2009gs}. 
\begin{figure}[tb]
\begin{center}
\begin{minipage}[t]{8 cm}
\centerline{\includegraphics[scale=0.5]{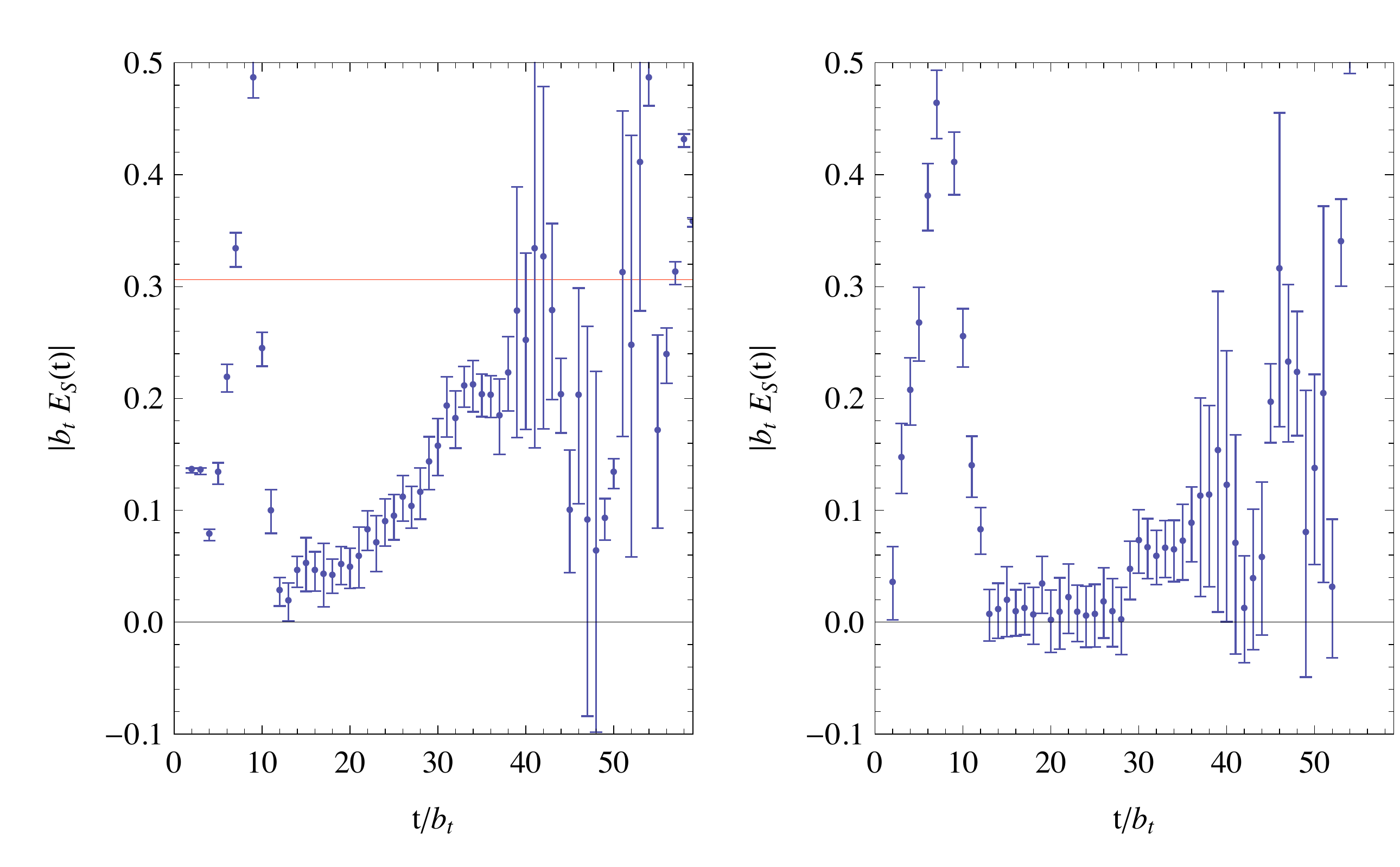}}
\end{minipage}
\begin{minipage}[t]{16.5 cm}
\caption{
The left panel shows the energy-scales
    associated with the signal-to-noise ratio for the $\Xi^0\Xi^0 n$
    correlation function~\cite{Beane:2009gs}, as defined in
    Eq.~(\protect\ref{eq:Estondefn}).  The horizontal line corresponds
    to $m_N + 2 m_\Xi - 2 m_\eta - {5\over 2} m_\pi $, the asymptotic
    energy-scale in a lattice with infinite temporal extent.  The
    right panel shows the difference between the signal-to-noise
    energy scales of the diagonalized $\Xi^0\Xi^0 n$ correlation
    function and that of the nucleon and twice that of the $\Xi$
    correlation function. 
\label{fig:XiXin-noise}}
\end{minipage}
\end{center}
\end{figure}
In Figure~\ref{fig:XiXin-noise} we show the  energy-scales
    associated with the signal-to-noise ratios for the $\Xi^0\Xi^0 n$
    correlation function as a function of time-slice~\cite{Beane:2009gs}.
In particular, the right-panel shows that the difference of energy-scales
between the three-baryon system and its constituents is consistent with zero
within the Golden Window of time-slices.
It is clear from this figure that the current computational technology used by
NPLQCD would allow for the calculation of systems with $A>3$ 
using algorithms
that allow for the multi-baryon contractions to be performed in a
reasonable amount of computer time.

\begin{figure}[tb]
\begin{center}
\begin{minipage}[t]{8 cm}
\centerline{\includegraphics[scale=0.45]{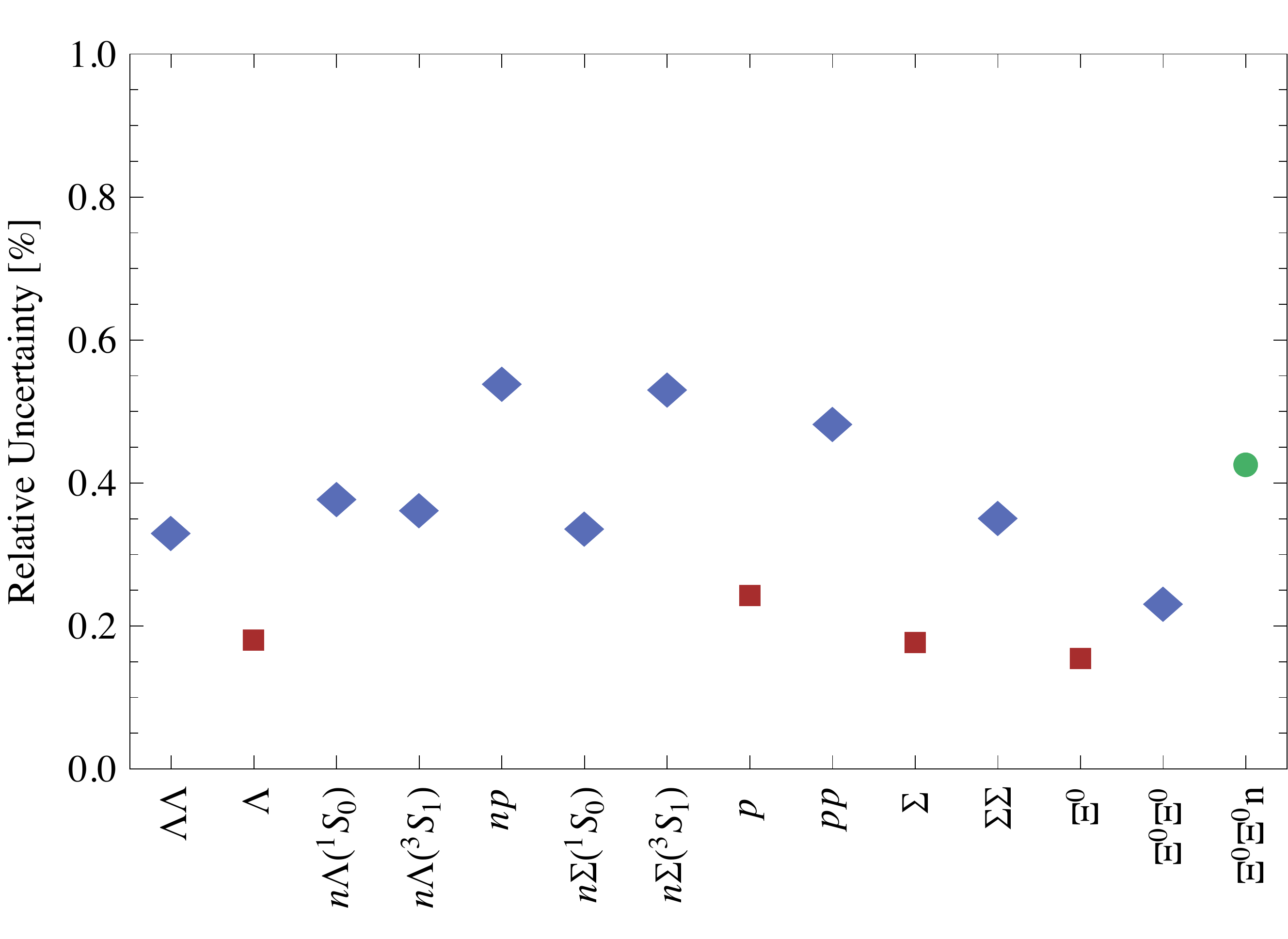}}
\end{minipage}
\begin{minipage}[t]{16.5 cm}
\caption{
The relative uncertainties in the
    extraction of one-, two- and three-baryon ground state energies
obtained in Refs.~\protect\cite{Beane:2009ky,Beane:2009gs,Beane:2009py}.
\label{fig:uncertainty}}
\end{minipage}
\end{center}
\end{figure}
In Figure~\ref{fig:uncertainty}, the  relative uncertainties in the
    extraction of one-, two- and three-baryon ground state energies
obtained in Refs.~\cite{Beane:2009ky,Beane:2009gs,Beane:2009py} are shown.
It is somewhat remarkable from the numerical standpoint that the relative
uncertainty per baryon is essentially independent of baryon number when
determined within the Golden Window of time-slices.  With the high-statistics
calculations performed by the NPLQCD collaboration, an uncertainty of
approximately $6~{\rm MeV}$ per baryon has been obtained, which is at the
threshold for performing meaningful calculations for nuclear physics (once the
pion mass is reduced to its physical value).

It is clear that the Golden Window of time-slices that has been uncovered in 
Refs.~\cite{Beane:2009ky,Beane:2009gs,Beane:2009py}, which  provides a way to
evade the need for exponentially large computational resources to calculate
multi-baryon systems, requires careful  source and sink optimization.
This involves two considerations  to make optimal use of available resources:
\begin{enumerate}
\item
maximal overlap of the interpolating operators
onto the baryon states 
\item 
minimum overlap of the interpolating operators onto the mesonic states in the
correlation function dictating the variance of the baryon correlation
functions
\end{enumerate}

To close this section, we make the comment that it is highly desirable to
develop algorithms that explicitly eliminate the exponential degradation of the
signal-to-noise ratio in generic correlation functions.
While this is an obvious comment to make, such algorithms do not yet exist,
however, there have been encouraging developments.
It has been shown, by L\"uscher and Weisz~\cite{Luscher:2001up},
that a multi-level scheme, involving the iterative 
sequential update of gauge fields and averaging the products of gauge links,
leads to large Wilson-loops that have statistical uncertainties that are
exponentially smaller than those resulting from the usual method of evaluation
(direct evaluation of the Wilson loop on each member of the ensemble of gauge
configurations). 
This multi-level scheme has been applied to the calculation of the Yang-Mills
spectrum by inserting projectors into the transfer-matrix that restrict the
quantum numbers of the states that can propagate forward in 
time~\cite{DellaMorte:2008jd,DellaMorte:2009rf}.  Such
projections lead to an exponential improvement in the time-dependence of the
signal-to-noise ratio.
It is conceivable that analogous projections can be developed for $n_f=2+1$
calculations, for instance producing gauge field configurations which do not
permit the propagation of single pions, thereby eliminating the long-time
correlations in the light-quark propagators, and thereby exponentially reducing
the cancellations that occur in forming the correlation functions for one or more
baryons.
There has also been progress in the calculation of the properties of
spin-systems by using cluster-algorithms where the partition functions is split
into the sum of contributions that individually do not suffer from the sign
problem, e.g. Ref.~\cite{Chandrasekharan:1999vz}.  
However, this algorithm has yet to be transcribed into Lattice QCD calculations.

%%%%%%%%%%%%%%%%%%%%%%%%%%%%%%%%%%%%%%%%
\section{Optimization through the Analysis of Simulated  Calculations}
\label{sec:fakedata}
\noindent
Large computational resources are required to perform Lattice QCD calculations,
especially 
of quantities that impact nuclear physics.
As such, it is important to 
simulate the results of Lattice QCD calculations in order to 
estimate the volumes, lattice spacings and pion masses of calculations that will 
optimize the physics output with fixed resources.
As an example, consider the results of a set of simulated
calculations on an ensemble of
gauge field configurations with spatial dimensions $L\sim 12.3~{\rm fm}$ at the
physical pion mass ($m_\pi L\sim 8.7$).  The experimentally determined 
nucleon-nucleon $\siii-\diii$ coupled-channels scattering amplitude 
is used 
to determine the two energy-levels
in the finite lattice volume that are below the inelastic threshold 
(set by $|{\bf p}|<m_\pi/2$)
using the L\"uscher relation in Eq.~(\ref{eq:energies}). 
In particular, a scattering length of $a^{\rm input} = 5.425~{\rm fm}$,
an effective range of $r^{\rm input}=1.749~{\rm fm}$, 
and a nucleon mass of $M_N^{\rm input}=939~{\rm MeV}$
are used,
which produce a
deuteron bound by $B=2.214~{\rm MeV}$ when d-wave interactions and higher order
terms in the effective range expansion are ignored. 
Simulated results are then generated with $10\%$, $5\%$ and $1\%$ 
uncertainties by 
randomly distributing the centroid of the energy-level by the corresponding
amount, and assigning the corresponding uncertainty.  
\begin{table}
\begin{center}
\begin{minipage}[!ht]{16.5 cm}
\caption{Simulated calculations of the shifts of the 
two lowest-lying energy levels of two
  nucleons in a finite lattice volume with spatial extent $L\sim 12.3~{\rm fm}$
  generated from the physical $^3S_1$ scattering amplitude (at the physical values 
of the light-quark masses) at varying levels of
  precision. 
For simplicity, the precision of the two calculated energy levels are taken to
be equal. 
}
\label{tab:fake}
\end{minipage}
\begin{tabular}{|c|c|c|}
\hline
& & \\
{\rm Precision Level of Energy Shift} & {\rm Bound State Energy} (MeV)& ${\rm 1^{st} \ Continuum
  Level}$ (MeV) \\
& & \\
\hline
& & \\
$0\%$ & $-3.147$ & $4.005$ \\
$1\%$ & $-3.111\pm 0.031$ & $4.015\pm 0.040$ \\
$5\%$ & $-2.95\pm 0.16$ & $4.24\pm 0.20$ \\
$10\%$ & $-2.66\pm 0.31$ & $3.65\pm 0.40$ \\
& & \\
\hline
\end{tabular}
%noalign{\smallskip\hrule}\cr}
\begin{minipage}[t]{16.5 cm}
\vskip 0.5cm
\noindent
\end{minipage}
\end{center}
\end{table}     
The  deuteron bound state energy in this finite lattice volume~\footnote{The
  deuteron becomes more tightly bound in the finite lattice volume due to the
  exclusion of momentum modes by the periodic boundary conditions imposed in
the spatial directions.  It is interesting to note that if instead of a volume
with periodic boundary conditions, the nucleons were confined with a harmonic
oscillator potential, as is common-place in nuclear structure calculations, the
deuteron binding energy would be reduced in magnitude due to the positive energy
contributions from the potential.}
is
$E_D^{(L)}=-3.147~{\rm MeV}$, 
and the lowest
lying continuum state appears at $E^{(L)}=+4.005~{\rm MeV}$.
Given the simulated calculations shown in Table~\ref{tab:fake}, 
the goal is to extract the deuteron binding energy and the 
scattering parameters at the 
varying levels of precision.

\begin{figure}[tb]
\begin{center}
\begin{minipage}[t]{8 cm}
\centerline{\includegraphics[scale=0.65]{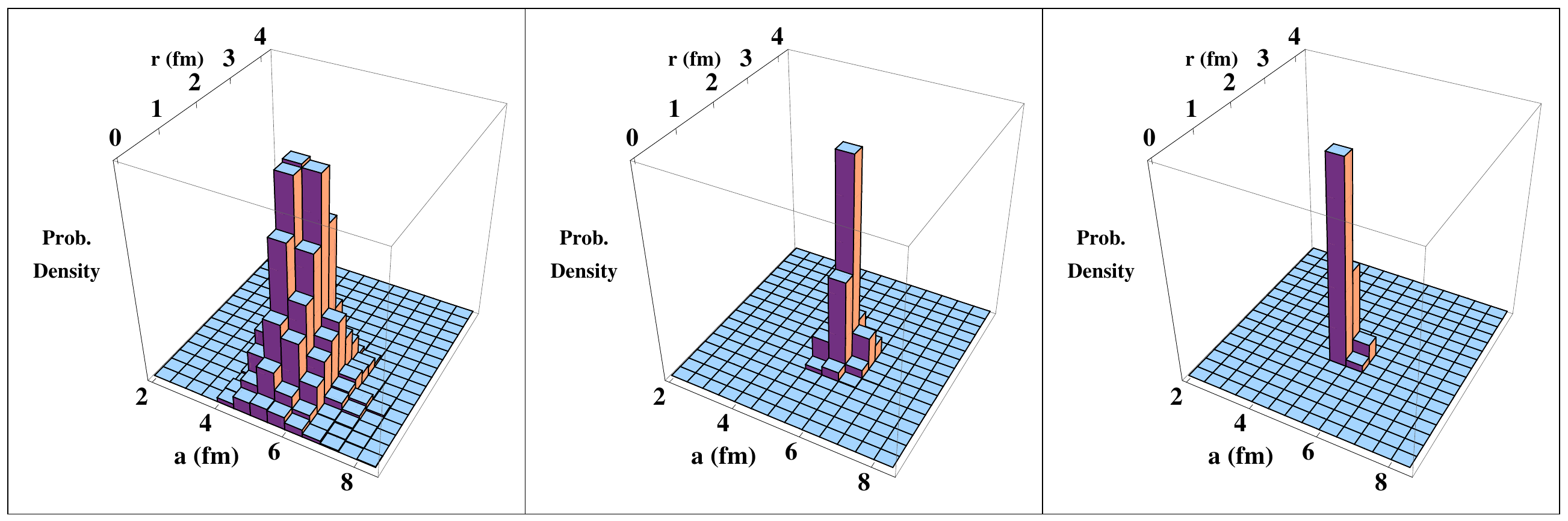}}
\end{minipage}
\begin{minipage}[t]{16.5 cm}
\caption{
The scattering parameters extracted from Monte-Carlo fitting to the simulated
calculations of $p\cot\delta$ shown in Table~\protect\ref{tab:fake}.
The left panel corresponds to $10\%$ precision, the middle panel to $5\%$
precision and the right panel to $1\%$ precision.
\label{fig:FakeaVr}}
\end{minipage}
\end{center}
\end{figure}
\begin{figure}[tb]
\begin{center}
\begin{minipage}[t]{8 cm}
\centerline{\includegraphics[scale=0.65]{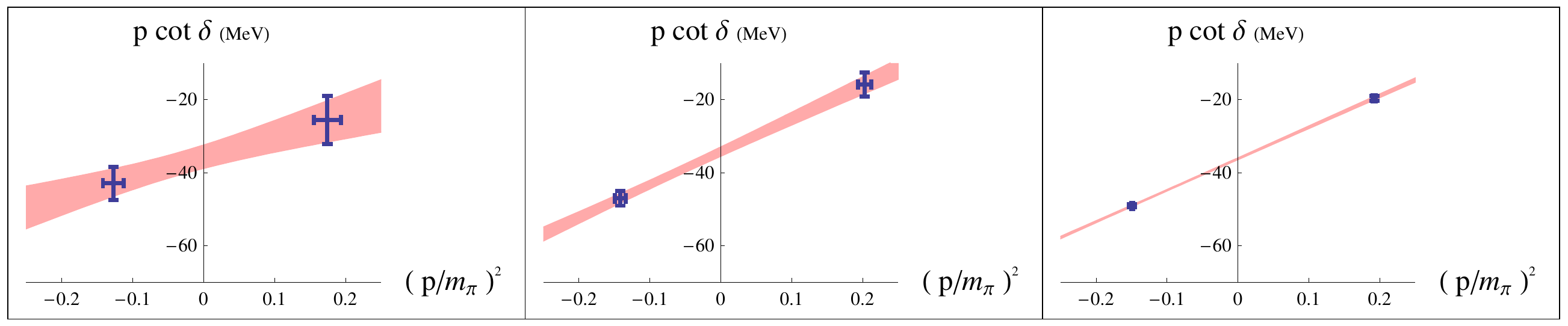}}
\end{minipage}
\begin{minipage}[t]{16.5 cm}
\caption{
The 1-$\sigma$ determination of the $p\cot\delta$ function (shaded region)
extracted from the simulated calculations
in Table~\protect\ref{tab:fake} (shown as points with uncertainties).
The left panel corresponds to $10\%$ precision, the middle panel to $5\%$
precision and the right panel to $1\%$ precision.
\label{fig:FakePCOT}}
\end{minipage}
\end{center}
\end{figure}
\begin{figure}[tb]
\begin{center}
\begin{minipage}[t]{8 cm}
\centerline{\includegraphics[scale=0.65]{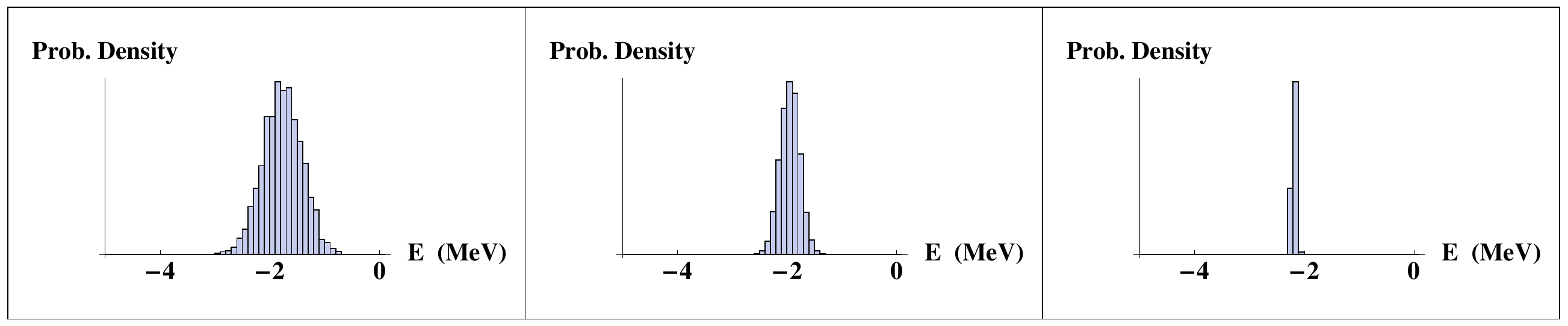}}
\end{minipage}
\begin{minipage}[t]{16.5 cm}
\caption{
The deuteron binding energy extracted from Monte-Carlo fitting to the simulated 
calculations in Table~\protect\ref{tab:fake}.
The left panel corresponds to $10\%$ precision, the middle panel to $5\%$
precision and the right panel to $1\%$ precision.
\label{fig:Fakedeuteronbinding}}
\end{minipage}
\end{center}
\end{figure}

In order to propagate the correlated uncertainties associated with the
extraction of the scattering parameters, we perform a Monte-Carlo analysis
of this simulated  set of calculations.  For each value of $p^2$ and of 
$p\cot\delta$ a
value is randomly drawn from a normal-distribution with mean and standard
deviation set by the quantity and its uncertainty~\footnote{
It should be noted that the uncertainties associated 
with the simulated calculations
in Table~\ref{tab:fake} are uncorrelated (between the $p^2$ and $p\cot\delta$
for each energy-level).
However, this will not be the situation in actual Lattice QCD 
calculations where $p\cot\delta$ is determined from $p^2$ using
Eq.~(\ref{eq:energies}). 
For this exercise we treat them as independent (which tends to increase
calculated uncertainties).}.  
A two-parameter fit is performed in this simplified analysis,
terminating the energy expansion of  $p\cot\delta$ at the effective
range parameter, providing a pair of fit values to the scattering length and
effective range, $\{a_i,r_i\}$.  
The distributions of these extracted pairs are shown in the lego plots in 
Figure~\ref{fig:FakeaVr}.
These ensembles of extractions are  then used to
generate ranges of values of  $p\cot\delta$ for any  $p^2$ (using the
functional form), from which  means
and standard deviations can be determined.  The results of which are shown in 
Figure~\ref{fig:FakePCOT}.
For each pair $\{a_i,r_i\}$,
the effective range expression can be used to determine the deuteron binding
energy,  the distribution of which is shown in 
Figure~\ref{fig:Fakedeuteronbinding}.
The results of the analysis of the simulated  
calculations are shown in Table~\ref{tab:fakeResults}
\begin{table}
\begin{center}
\begin{minipage}[t]{16.5 cm}
\caption{Results of the analysis of the simulated  calculations in Table~\protect\ref{tab:fake}.
}
\label{tab:fakeResults}
\end{minipage}
\begin{tabular}{|c|c|c|c|}
\hline
& & & \\
{\rm Precision Level} & {\rm Deuteron Binding Energy} (MeV) & {\rm Scattering Length}
(fm) & {\rm Effective Range} (fm) \\
& & & \\
\hline
& & & \\
$0\%$ & $-2.214$ (input) & $5.425$ (input) & $ 1.749$ (input)\\
$1\%$ & $-2.180 (35)$ & $5.447 (38)$  & $ 1.737 (32)$\\
$5\%$ & $-1.94 (18)$ & $5.77 (22)$  & $ 1.82 (16)$\\
$10\%$ & $-1.77 (35)$ & $5.58 (51)$  & $ 1.14 (42)$\\
& & & \\
\hline
\end{tabular}
%noalign{\smallskip\hrule}\cr}
\begin{minipage}[t]{16.5 cm}
\vskip 0.5cm
\noindent
\end{minipage}
\end{center}
\end{table}     

A number of 
conclusions can be drawn from the analysis of simulated  calculations:
\begin{enumerate}
\item 
Determining the shifts in the
lowest two energy eigenvalues in a lattice with $L\sim
  12~{\rm fm}$ with precision exceeding $\sim 10\%$ is sufficient to determine
  the deuteron binding energy with a precision that exceeds  the
  5-$\sigma$-level.  
For the physical binding energy of the deuteron, $10\%$ uncertainty in the
energy shift corresponds to a $\sim 0.01\%$ uncertainty in the total energy.
This is approximately an order of magnitude more precise than present
calculations.
\item 
The precision of the scattering length extraction is somewhat better than
  the deuteron binding energy, while that of the effective range is somewhat
  worse.
This is consistent with the fact that the scattering length enters at
LO in the momentum expansion of the scattering length, while the
effective range enters at NLO.
\item 
Having calculations on both sides of $p^2=0$, combined with the fact that the
magnitude of   
the deuteron binding energy increases as the lattice volume decreases, allows
  for a much more precise determination of the scattering parameters and
  deuteron binding energy than would be the case with calculations of the
  deuteron binding energy alone on two lattice ensembles with different
  volumes.
This is because an interpolation is required to reach the physical
deuteron binding energy, as opposed to an extrapolation that would be required
for single-level (bound-state) calculations in multiple lattice volumes.
\item
A NNLO analysis, including the extraction of the
shape-parameter, $r_1$ requires calculating three energy eigenvalues in one
lattice-volume, or calculations on a  number of lattice volumes.  
Given the experimentally
observed smallness of the shape parameters contributing to nucleon-nucleon
scattering, high-precision calculations of the energy eigenvalues will be
required for such an extraction.
\end{enumerate}

%%%%%%%%%%%%%%%%%%%%%%%%%%%%%%%%%
\section{Other Lattice QCD Efforts in Low-Energy  Nuclear Physics}
\label{sec:others}
\noindent
The content of this review is primarily focused on the work of the NPLQCD
collaboration, and its efforts to extract the properties and interactions
of nuclei from lattice QCD calculations.  
However, this agenda has been taken up by other collaborations also.
As discussed previously, the HALQCD collaboration has performed both quenched
and dynamical Lattice QCD calculations of baryon-baryon correlation functions
from which non-local,
energy-dependent and interpolating operator dependent baryon-baryon 
potentials are extracted~\cite{Ishii:2006ec,Nemura:2008sp}.
Unfortunately these potentials contain no more (rigorous) information than the location of
the energy-levels in the lattice volume, and the scattering parameters that can
be derived using the L\"uscher relation. 
Important work has been recently performed
by the PACS-CS collaboration~\cite{Yamazaki:2009ua}. 
Exploratory quenched Lattice QCD calculations of the $^3$He and $^4$He
correlation functions strongly suggest that both nuclei are bound
for pions with a mass of $m_\pi\sim 800~{\rm MeV}$.

An exciting development that has taken place during the last couple of years
is the exploration of nuclei with $n_f=1$ Lattice QCD at strong 
coupling~\cite{deForcrand:2009dh,Fromm:2009xw}.  In this limit there is no
Yang-Mills contribution to the action, and instead of integrating out the
fermionic fields to leave an integral over the gauge fields that is evaluated by
Monte-Carlo, the gauge links are integrated out to leave a integration over the
fermion fields which can also be performed in closed-form.  The partition
function becomes a weighted sum over configurations of dimers and self-avoiding
baryon loops.  
The nucleon-nucleon potential that is found, is qualitatively consistent with
the modern phenomenological nucleon-nucleon potentials.
Further, the binding energy
of larger nuclei up to $^{12}$C  derived in this limit,  
were found to be qualitatively described by the first two terms in the
von Weis\"acker formula.

An area of Lattice QCD calculations that is peripherally related to
multi-nucleon systems, but is having increased overlap with the study of
hadronic interactions is the calculation of the excited states of baryons 
and mesons.  At heavier pion masses, where many of the excited states of the
nucleon and strange baryons are stable against strong decay, it is somewhat
straightforward to calculate their spectrum with lattice QCD. The excited
states correspond to exponentially decaying states in the correlation functions
created by interpolating operators with the correct quantum numbers.  Enormous
effort has been put into successfully refining the basis of interpolating operators and
developing techniques with which to cleanly separate multiple excited states
from the ground-state and other excited states, for example 
Ref.~\cite{Bulava:2009jb,Mahbub:2010jz,Gattringer:2008vj}.
The situation becomes vastly more complex as the pion mass approaches the
physical value and the resonances become unstable to strong decay.  The
energy eigenstates in the lattice volume are the baryon ground-state and the
scattering states with the same quantum numbers.  The excited states correspond
to the energies for which the scattering phase-shift is at $\delta_0 = {\pi\over
  2}$, and the energy interval over which the phase-shift is ``near''
$\delta_0$ dictates the width. Given that the location of the scattering states
in the volume cannot be dictated in the generation of gauge field
configurations, calculations must be performed in multiple volumes.  Further,
given that $m_\pi L \gsim 2\pi$ must be satisfied by calculations in order for the
wavefunctions to be asymptotic at the boundary of the lattice, the location of
multiple excited states in the lattice volume must be precisely 
determined, e.g. Ref.~\cite{Gockeler:2008kc}.  
This is an exceptionally challenging problem, but progress is being made in
extracting the properties of the $\rho$-meson and the $\Delta$-resonance.

An area that is not Lattice QCD, but which makes use of techniques that were
developed for Lattice QCD calculations and impacts low-energy nuclear physics
is Lattice Nuclear EFT.
Low-energy nuclear structure and reactions are determined
in the low-energy EFT using a space-time lattice, and
performing the path integral.
The conventional power-counting and
perturbative expansion that are employed successfully for particle physics
observables fail in the nuclear context as  nuclei are nonperturbative objects.
After many years of development,
EFT's for nuclear physics are reasonably well developed, and recently progress
has been made in calculating S-matrix elements in systems involving more than
two nucleons by latticizing the EFT's~\cite{Epelbaum:2010xt,Lee:2008fa}.

%%%%%%%%%%%%%%%
\section{Resource Requirements and the Next Decade}
\begin{figure}[tb]
\begin{center}
\begin{minipage}[t]{8 cm}
\centerline{\includegraphics[scale=0.55]{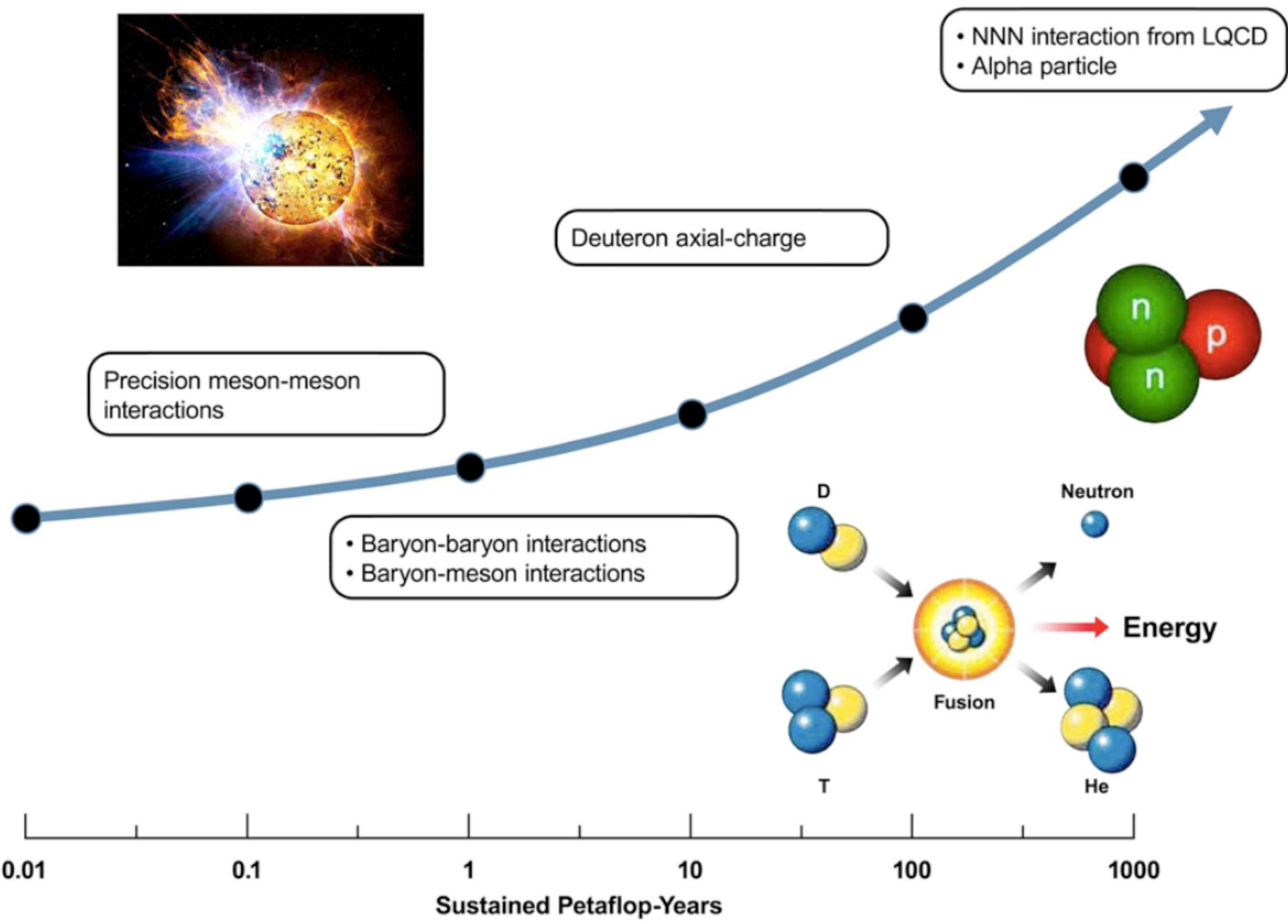}}
\end{minipage}
\begin{minipage}[t]{16.5 cm}
\caption{
Estimates of the resources required to complete calculations of importance
to nuclear physics~\protect\cite{ExascaleReport:2009}. 
Except for the quantities indicated as requiring exa-scale
resources, 
the resource requirements are for 
calculations performed in the isospin limit and without
electroweak interactions. 
\label{fig:BBresources}}
\end{minipage}
\end{center}
\end{figure}

\noindent
A number of workshops focusing on the science need for exa-scale
computing resources sponsored  by the US Department of Energy were held during
2009.
One of the workshops, {\it Forefront Questions in Nuclear Science and the Role of
  Computing at the Extreme Scale}~\cite{ExascaleReport:2009}, 
established the need for exa-scale computing
resources in order for the main goals of the field of nuclear physics to be accomplished.
One of the major goals of the field that requires exa-scale computing
resources is the calculation of nuclear forces from QCD using Lattice QCD,
and Figure~\ref{fig:BBresources} presents 
an overview of
current estimates of these requirements.

As discussed in a previous review~\cite{Beane:2008dv}, 
a complete calculation of the nucleon-nucleon
scattering amplitude, and the hyperon-nucleon and hyperon-hyperon
scattering amplitudes 
(including multiple lattice spacings, volumes and light quark masses)
will require sustained peta-scale resources, as shown in 
Figure~\ref{fig:BBresources}.
The same is true for the meson-baryon interactions.  
It is estimated that sustained-sub-peta-scale-year resources are
required to perform high-precision calculations of meson-meson  scattering-phase shifts,
extrapolated to the physical pion mass (in the isospin limit), including the
contributions from disconnected diagrams to the iso-singlet $\pi\pi$ channel.

It is currently estimated that sustained peta-scale resources are required in
order to calculate the matrix elements of electroweak operators, such as those
determining neutrino-induced breakup of the deuteron, in the few-nucleon
sector.
Great progress is being made in computing single hadron matrix elements of such
operators, such as the isovector axial-current matrix element in the nucleon,
$g_A$~\cite{Khan:2006de,Bratt:2010jn,Yamazaki:2009zq}.  
While the extrapolation to the physical pion mass, and to infinite
volume remain the subject of discussions in the community, relatively rapid
progress is being made.  The calculations of matrix elements of operators that
receive contributions from disconnected diagrams remain difficult with
currently available resources, but will be completed with peta-scale resources.
A significant  uncertainty
in the experimentally determined  properties of neutrinos comes from
the uncertainties in weak matrix elements between nuclear states.
Such  uncertainties in few-nucleon systems should be reduced within
the next decade with anticipated  Lattice QCD calculations, as indicated in 
Figure~\ref{fig:BBresources}.

Despite the first Lattice QCD calculations of three-
(dynamical~\cite{Beane:2009gs}) 
and
four-baryon (quenched~\cite{Yamazaki:2009ua}) systems
appearing this year, it is estimated that exa-scale computing resources will be
required to extract the nuclear interactions among three-nucleons and
determine the spectrum of the $\alpha$-particle.
Given that the three-nucleon interaction is relatively imprecisely known when
compared with the two-nucleon interactions, this calculation will have
significant impact upon nuclear structure and reaction calculations.  
The three-baryon interactions between strange and non-strange baryons will be
calculable at the same time with the same level of precision
with minimal additional resource requirements.

The current discussions regarding exa-scale computing facilities suggests that
it may be possible to see such resources deployed sometime around 
2018~\cite{ExascaleReport:2009}.
Clearly, such resources are required for the calculation of quantities of
central importance to the nuclear physics program. 
During the next decade the field will develop the ability to perform low-energy 
strong interaction calculations with quantifiable uncertainty estimation.
Lattice QCD  will supplement the present ability of the 
experimental nuclear physics program to
determine quantities with certain precision to processes that cannot be accessed
experimentally.

%%%%%%%%%%%%%%%%%%%%%%%%%%%%%%
\section{Conclusion}
 
A central goal of the field of nuclear physics is to establish a framework with
which to perform high-precision calculations, with quantifiable uncertainties,
of strong-interaction processes occurring under a broad range of conditions.
Quantum chromodynamics was established as the underlying theory of the strong
interactions during the 1970's, however, nuclear physics is 
the regime of QCD in which its defining property of 
asymptotic freedom is hidden by the vacuum and by 
the phenomenon of confinement.
Lattice QCD, in which the QCD path integral is evaluated numerically, is the
only known way to perform rigorous  QCD calculations of low-energy
strong interaction processes.
With the research and development into high-performance computing, nuclear
physics,  quantum field theory, applied
mathematics, and numerical algorithms  that has taken
place over the last few decades, the field of nuclear physics is entering into
an era in which Lattice QCD will become a quantitative tool in much the same
way that experiments are, but with a different scope and different range of
applicability.
Rapid progress is currently being made in the calculation of the interactions
among nucleons and, more generally, among the low-lying baryons. 
Present day Lattice QCD calculations are being performed at
pion masses larger than the physical pion mass, but as exploratory 
calculations are
now being performed at the physical pion mass, the interactions among  baryons
will be known from QCD at the physical light-quark masses within the next
several years (if computational resources devoted to these calculations
continue to increase as they have during the last decade).

Determining the three-body and higher-body interactions among nucleons and
hyperons directly from QCD will be a remarkable achievement for Lattice QCD,
and will provide crucial input into the calculations of the structure and
interactions of light nuclei.  During the last year the first calculations of
three- and four- nucleon systems were reported.  Such calculations are
presently difficult, but are primarily limited by the available computational
resources and not by conceptual or formal issues.  It is anticipated that the
three-nucleon interactions will be calculated with high-precision with Lattice
QCD during the next decade.
Beyond the three-body systems, we expect that some of the 
properties and
interactions of light nuclei will be calculable with Lattice QCD during the 
same
time-period.

The dream of being able to perform reliable
calculations of the interactions among  multiple nucleons and 
hyperons, and of the
structure and reactions of light-nuclei,  directly from QCD
is starting to be realized. 
The path forward is clear, and the next decade will be a truly remarkable 
period for nuclear physics.

\vskip 0.5 in

{\it We would like to thank 
Paulo Bedaque,
Robert Edwards,
Balint Joo,
David Kaplan,
Huey-Wen Lin, 
Tom Luu, Assumpta Parre\~no, Aaron Torok, Andre Walker-Loud
for
  their contributions to our understanding of this area of nuclear physics.
}

\vskip 0.5 in
{\it
The work of Silas Beane was
supported in part by the National Science Foundation CAREER grant No.
PHY-0645570.
The work of William Detmold 
was supported in part by the U.S.~Department~of Energy grants
DE-AC05-06OR23177 (JSA), DE-FG02-04ER41302, Outstanding
Junior Investigator (OJI) grant DE-SC0001784, and by the Jeffress
Memorial Trust.  
The work of Kostas Orginos was supported in part by
the U.S.~Department~of Energy grants DE-AC05-06OR23177 (JSA),  
DE-FG02-04ER41302,  Outstanding
Junior Investigator (OJI) grant DE-FG02-07ER41527,
by the NSF grant
CCF-0728915, and by the Jeffress
Memorial Trust grant J-813.
The work of Martin Savage was supported in part by the 
U.S.~Department~of Energy under grant DE-FG03-97ER4014. 

Computational work described in this review was performed in part with
resource allocations to the NPLQCD collaboration from 
NERSC (Office of Science of the U.S. Department of Energy), 
the Institute for Nuclear Theory, Centro
Nacional de Supercomputaci\'on (Barcelona, Spain), Lawrence Livermore
National Laboratory, the National Science Foundation through
the Teragrid, and the Thomas Jefferson National Accelerator
Facility and Fermi National Accelerator Laboratory through the
USQCD collaboration under {\it The Secret Life of a Quark}, a
U.S. Department of Energy SciDAC project.  
}

%%%%%%%%%%%%%    References  %%%%%%%%%%%%%%%%%

\end{document}